\documentclass[twoside,12pt]{article}
\usepackage{epsfig}
\usepackage{graphicx,amsfonts,amsmath,amssymb,color,epstopdf,setspace,float,cite}
\usepackage[vcentermath]{youngtab}

\def\Journal#1#2#3#4{{#1} {#2} (#4) #3 }

\def\NPA{{\em Nucl. Phys.} A}

\def\PLB{{\em Phys. Lett.} B}

\def\PRL{\em Phys. Rev. Lett.}

\def\PREP{\em Phys. Rep.}

\def\PRC{{\em Phys. Rev.} C}

\newcommand{\be}{\begin{equation}}
\newcommand{\ee}{\end{equation}}
\newcommand{\bea}{\begin{eqnarray}}
\newcommand{\eea}{\end{eqnarray}}

\newcommand{\SU}[1]{\ensuremath{\mathrm{SU}( #1 )}}
\newcommand{\Un}[1]{\ensuremath{\mathrm{U}( #1 )}}
\newcommand{\SO}[1]{\ensuremath{\mathrm{SO}( #1 )}}
\newcommand{\On}[1]{\ensuremath{\mathrm{O}( #1 )}}
\newcommand{\Spn}[1]{\ensuremath{\mathrm{Sp}( #1 )}}
\newcommand{\SpR}[1]{\ensuremath{\mathrm{Sp}( #1,\mathbb{R} )}}
\newcommand{\CG}[3]{\ensuremath{\langle#1;#2|\,#3\rangle}}

\newcommand{\RedCG}[3]{\ensuremath{\langle#1;#2\|#3\rangle}}

\newcommand{\ket}[1]{\ensuremath{\left| #1 \right\rangle}}

\newcommand{\half}{\ensuremath{\textstyle{\frac{1}{2}}}}

\newcommand{\betb}{\begin{tabular}{p{4.0cm}p{9.0cm}}}
\newcommand{\entb}{\end{tabular}}

\newcommand{\ho}{\ensuremath{\hbar\Omega}}
\newcommand{\ph}[1]{\ensuremath{#1}p-\ensuremath{#1}h}
\newcommand{\Nmax}[2]{\ensuremath{\langle#1\rangle #2}}
\begin{document}

\title{ \vspace{1cm} Symmetry-guided large-scale shell-model theory}
\author{Kristina D. Launey$^1$, Tomas Dytrych$^{1,2}$ ,  and Jerry P. Draayer$^1$\\
\\
$^1$Department of Physics and Astronomy, Louisiana State University,\\
 Baton Rouge, LA 70803, USA\\
$^2$Nuclear Physics Institute, 250 68 \v{R}e\v{z}, Czech Republic
}
\maketitle
\begin{abstract} 
In this review, we present a symmetry-guided strategy that utilizes exact as well as partial symmetries for enabling a deeper understanding of and advancing {\it ab initio} studies for determining the microscopic structure of atomic nuclei.
These symmetries expose physically relevant degrees of freedom that, for  large-scale calculations with QCD-inspired interactions,  allow  the model space size to be reduced  through a very structured selection of the  basis states to physically relevant subspaces. This can guide explorations of simple patterns in nuclei and how they emerge from first principles, as well as extensions of the theory beyond current limitations toward heavier nuclei and larger model spaces. This is illustrated for the {\it ab initio} symmetry-adapted no-core shell model (SA-NCSM) and two significant underlying  symmetries, the symplectic \SpR{3} group and its deformation-related \SU{3} subgroup.  We review the broad scope of nuclei, where these symmetries have been found to play a key role --  from the light $p$-shell systems, such as $^{6}$Li, $^{8}$B, $^{8}$Be, $^{12}$C, and $^{16}$O, and $sd$-shell nuclei exemplified by $^{20}$Ne, based on first-principle explorations; through  the Hoyle state in $^{12}$C and enhanced collectivity in intermediate-mass nuclei, within a no-core shell-model perspective; up to strongly deformed species of the rare-earth and actinide regions, as investigated in earlier studies. A complementary  picture, driven by  symmetries dual to \SpR{3}, is also discussed. We briefly review symmetry-guided techniques that prove useful in various nuclear-theory models, such as  Elliott model, {\it ab initio} SA-NCSM, symplectic model, pseudo-\SU{3} and pseudo-symplectic models, {\it ab initio} hyperspherical harmonics method, {\it ab initio} lattice effective field theory, exact pairing -plus- shell model approaches, and cluster models, including  the resonating-group method.  Important implications of these approaches that have deepened our understanding of  emergent phenomena in nuclei, such  as  enhanced collectivity, giant resonances, pairing, halo, and clustering, are discussed,  with  a focus on  emergent patterns in the framework of the {\it ab initio} SA-NCSM with no {\it a priori} assumptions. 
\end{abstract}
\eject
\tableofcontents

\section{Introduction}
%\begin{itemize}
%\item Background
%\item Experimental signatures of symmetries and emergent phenomena 
%\item Emergent symmetries from {\it ab initio} large-scale calculations
%\item Role of exact and near symmetries in symmetry-guided many-body theory
%\end{itemize}

Major progress in the development of realistic inter-nucleon interactions \cite{WiringaSS95,Machleidt01,EntemM03,Epelbaum06}
along with the utilization of massively parallel computing resources (e.g., see ~\cite{LangrSTDD12}) have placed
\textit{ab initio} large-scale simulations at the frontier of nuclear structure explorations. Several {\it ab initio} nuclear-theory
approaches have been recently advanced, including  Green's function Monte Carlo (GFMC) \cite{WiringaS98, PieperVW02}, no-core shell model (NCSM) \cite{NavratilVB00,BarrettNV13,MarisVN13,NavratilQSB09,BaroniNQ13} together with NCSM with a core \cite{LisetskiyBKNSV08} and importance truncation NCSM \cite{RothN07}, coupled-cluster method (CC) \cite{WlochDGHKPP05,Hagen2008},  lattice effective field theory \cite{EpelbaumKLM11}, in-medium SRG \cite{TsukiyamaBS11,BognerHHSBCLR14}, symmetry-adapted  no-core shell model (SA-NCSM) \cite{DytrychLMCDVL_PRL12}, Monte Carlo NCSM \cite{AbeMOSUV12}, and self-consistent Green's function \cite{CipolloneBN13}. {\it Ab initio} approaches build upon a `first principles' foundation. They provide a long-missing link that bridges from the many-particle nucleus down to the fundamental blocks, namely, the properties of only two or three nucleons often tied to  chiral symmetry-breaking patterns dictated by the underlying Quantum Chromodynamics (QCD). And while this bridge transforms the nuclear problem into a computational- and data-intensive challenge, it empowers {\it ab initio} models with two invaluable features: (1) a universal character essential for modeling the coexistence of  diverse nuclear substructures and (2) predictive capabilities vital for descriptions of experimentally inaccessible nuclear species far off the valley of stability. As such nuclei are often found key to understanding processes in extreme environments, from stellar explosions to the interior of nuclear reactors or fusion capsules, first-principle nuclear models have been  and will be demonstrating a tremendous impact for advancing the frontiers in multiple branches of physics such as astrophysics, neutrino physics, and applied physics \cite{WiringaP02, HayesNV03,MarisSV10,NavratilQ12,HagenHJMP12, CipolloneBN13, BaccaBLO13, HupinQN14,EpelbaumKLLMR14,RedondoQNH14,McGrawHill14}. 

While the predictive capability is an essential feature of {\it ab initio} theories, especially in regions inaccessible to experiment, the need for such theories goes beyond solely achieving accurate results. Their main purpose is  to advance our understanding of strongly interacting systems based on the nature of the strong force and of the way this force governs the complex nuclear dynamics that often displays striking simplicities. Indeed, models restricted in their interactions and model spaces can mimic simple patterns and can be misleading.
One of the most striking recent examples is related to the low-lying $0^+$ nuclear states in experimental excitation spectra that for a long time have been regarded, and hence, modeled, as vibrations, and only recently the different mechanism of shape coexistence has been suggested  \cite{RoweTW06, Kulp08,RoweW2010book}. Shape coexistence has been found to occur in many nuclei across the entire mass surface \cite{HeydeW11} --  it has been argued that it probably occurs in nearly all nuclei \cite{Wood16}.
And while vibrational spectra are often associated with spherical nuclei, the fact that a nucleus has a zero quadrupole moment in its $0^+$ ground state does not  imply that it is spherical in {\it its intrinsic frame}; for quantum mechanical reasons, it appears spherical in lab frame. With an expanding body of experimental evidence, it is becoming evident that non-zero deformation is far more widespread than zero deformation and that even nuclei that are spherical in their ground state have low-lying deformed excited states \cite{HeydeW11}.
In fact, first-principle calculations in the SA-NCSM \cite{DytrychLMCDVL_PRL12} (see also Sec. \ref{SANCSMspectra}) have unveiled that even the lightest of nuclei, exemplified by $^6$Li, in their ground state exhibit considerable collectivity, as seen by the dominance of prolate deformed configurations in the $^6$Li wave function. At the same time, these calculations  closely reproduce the nearly vanishing $^6$Li ground-state ($gs$) quadrupole moment of  $Q (1^+_{gs}) = -0.0818(17)\,e\mathrm{fm}^2$ \cite{Tilley2002}. This quadrupole moment, an $L=2$ operator, is in fact attributed to the considerable contribution  of $L=0$ configurations ($\sim 87\%$) to the ground state. 
Hence,  {\it ab initio} theory opens the path to explain from first principles simple patterns, revealed amidst experimental data, while providing deeper understanding of  emergent phenomena, such  as  enhanced collectivity, giant resonances, pairing, halo, and clustering, in a plethora of nuclei from stable to unstable, without {\it a priori} assumptions. 

In this review, we present a symmetry-guided strategy that utilizes exact as well as partial symmetries for enabling a deeper understanding of and advancing {\it ab initio} studies for determining the microscopic structure of atomic nuclei. These symmetries naturally provide a physically relevant basis that, for  large-scale calculations,  allows  the model space size to be reduced  through a very structured selection of the  basis states to physically relevant subspaces and can guide extensions of the theory beyond current limitations.  This is crucial, as model space dimensionality and associated computational resource demands grow combinatorially with the number of particles and the spaces in which they primarily reside (so-called, ``scale explosion"), thereby limiting the number of active particles that could be handled or precluding microscopic descriptions of largely deformed spatial structures. 

We focus on the symplectic \SpR{3} symmetry and its deformation-related \SU{3} subgroup, which underpins the Elliott model \cite{Elliott58, Elliott58b}, the symplectic model \cite{RosensteelR77,Rowe85} and the {\it ab initio} SA-NCSM \cite{DytrychLMCDVL_PRL12}, and review the broad scope of nuclei, where these symmetries have been found to play a key role -- from the lightest $p$-shell systems of $^{6}$Li and $^{8}$Be, through $sd$-shell intermediate-mass nuclei, up  to strongly deformed nuclei of the rare-earth and actinide regions (see also the review Ref. \cite{LauneyDDSD15}). That \SU{3} plays a key role tracks with the seminal work of Elliott \cite{Elliott58, Elliott58b}, and is further reinforced by the fact that \SU{3} underpins the microscopic symplectic model~\cite{RosensteelR77,Rowe85}, which provides a theoretical framework for understanding deformation-dominated collective phenomena in atomic nuclei~\cite{Rowe85, DraayerWR84} and which naturally contains low-energy shape coexisting excitations 
%but do not exhibit low-energy collective quadrupole vibrational excitations 
\cite{RoweW2010book}. 
 We also discuss complementary symmetries that provide alternative reorganization (classification) of the model space. We review important applications of various microscopic approaches built upon these classification schemes, with a focus on open-core shell-model theory (for a major class of valence shell models and their significant role for heavy nuclei, see the reviews \cite{Brown01,CaurierMNPZ05, KooninDL97, AlhassidMNO12}). While  many of these approaches have adopted  simple inter-nucleon interactions, their expansion  to manage realistic interactions and large-scale model spaces  is feasible with current massively parallel computing resources. The convergence of results with larger model space sizes in such {\it ab initio} theory tracks with the symmetry of choice. For a near symmetry,  convergence is fast and nuclear states can be described by a small number of symmetry-adapted basis states.  If the symmetry is largely broken,  the eigensolutions converge slowly and require  many basis states.  

Symmetries underpin orderly patterns in nuclear dynamics. As mentioned above, experimental evidence supports formation of deformation  and rotational patterns, including the  dominance of large deformation in low-lying nuclear states, as suggested by enhanced $E2$ transitions and large quadrupole moments.
As shown in the reviews \cite{Rowe85, Rowe96}, the dominance of large deformation and, by duality, low spin  (see Sec. \ref{spin} and \ref{def}) 
has been demonstrated by symmetry-guided theoretical studies, with further recognition of a new simple structure in nuclei, associated with  \SpR{3} symplectic symmetry (see Sec. \ref{symplectic}). In this review, we discuss how such highly structured orderly patterns emerge from first-principle  investigations  \cite{DytrychSBDV_PRL07,DytrychDSBV09} starting with  bare nucleon-nucleon ($NN$) interactions. Remarkably, the outcome  of these studies has revealed that typically only one or two symplectic irreducible representations (irreps), also referred to as ``vertical cones" of many-particle basis states, suffice to represent a large fraction of each of the {\it ab initio} wave functions of $^{12}$C and $^{16}$O, typically  in excess of about 80\% of the physics. Such a symplectic pattern has been also observed in {\it ab initio} SA-NCSM  results for light and intermediate-mass nuclei using symmetry-adapted \SU{3}-scheme basis states \cite{DytrychLMCDVL_PRL12}.  Implications of these studies to understanding the nature of nuclear dynamics and symmetry-guided applications are reviewed in Sec. \ref{appSANCSM}.

\section{Shell-model theory}
%U(N), SU(3): 3D harmonic oscillator\\
%
%
In its most general form, the nuclear shell model (SM) \cite{BrussardG77,Shavitt98,BarrettNV13,CaurierMNPZ05}, a many-body ``configuration interaction" (CI) method, solves the many-body Schr\"odinger equation for $A$ particles,
\begin{equation}
H \Psi(\vec r_1, \vec r_2, \ldots, \vec r_A) = E \Psi(\vec r_1, \vec r_2, \ldots, \vec r_A),
\label{ShrEqn}
\end{equation}
for which the interaction and basis configurations are adopted as follows.
\begin{description}
\item[{\rm \em Interaction} ] The intrinsic non-relativistic nuclear plus Coulomb interaction Hamiltonian is defined as:
\begin{equation}
H = T_{\rm rel} + V_{NN}  + V_{3N} + \ldots + V_{\rm Coulomb}, 
\label{intH}
\end{equation}
where $T_{\rm rel} $ is the relative kinetic energy $T_{\rm rel} =\frac{1}{A}\sum_{i<j}\frac{(\vec p_i - \vec p_j)^2}{2m}$ ($m$ is the nucleon mass), the $V_{NN}$ is the nucleon-nucleon interaction, $V_{NN}=\sum_{i<j}^A (V_{NN})_{ij}$  (and possibly, $V_{3N}=\sum_{i<j<k}^A (V_{NNN})_{ijk}$, $V_{4N}$, ... interactions) included along with the Coulomb interaction between the protons.  The Hamiltonian may include additional terms, e.g., higher-order electromagnetic interactions such as magnetic dipole-dipole terms.  

\item[{\rm \em Basis configurations} ] A complete orthonormal basis $\psi_i$ is adopted, such that the expansion  $\Psi(\vec r_1, \vec r_2, \ldots, \vec r_A)$ in terms of  unknown coefficients $c_k$,
$\Psi(\vec r_1, \vec r_2, \ldots, \vec r_A) = \sum_{k} c_k \psi_k(\vec r_1, \vec r_2, \ldots, \vec r_A)$,
 renders Eq. (\ref{ShrEqn}) into a matrix eigenvalue equation,
\begin{equation}
\sum_{k'} H_{k k'} c_{k'} = E c_k,
\end{equation}
where the many-particle Hamiltonian matrix elements are
$H_{k k'} = \langle \psi_k | H | \psi_{k'} \rangle$ and are calculated for the given interaction (\ref{intH}). Typically, the basis is a finite set of antisymmetrized products of  single-particle states (Slater determinants), referred to as a ``model space", where the single-particle states  of a three-dimensional spherical harmonic oscillator (HO) are used, $\phi_{\eta ljmt_z}(\vec r; b)$, where $\eta=2n_r+l$, $l$ is coupled to spin-\half~ to $j$, $t_z$ distinguishes between protons and neutrons, and  the oscillator length $b=\sqrt{ \frac{\hbar }{m\Omega} }$ with oscillator frequency $\Omega$.  Such a basis allows for preservation of translational invariance of the nuclear self-bound system and provides solutions in terms of single-particle wave functions that are analytically known. With larger model spaces utilized in the  shell-model theory, the  eigensolutions converge to the exact  values.
\end{description}

Depending on the interaction used and the model space adopted, there are various nuclear shell-model approaches. In particular,   the {\it ab initio} shell model uses  high-precision $NN$ ($NNN$) potentials fitted to two-body (three-body) data, in particular, to scattering phase shifts and properties of the deuteron (and  triton) (see, e.g., \cite{Machleidt01,GazitDQN09}).  The $NN$ potentials include  AV18 \cite{WiringaSS95}, CD-Bonn \cite{Machleidt01}, and N$^3$LO  \cite{EntemM03, Epelbaum06}.  Another high-precision $NN$ interaction is JISP16 \cite{ShirokovMZVW07} based on $J$-matrix version of inverse scattering
theory that is adjusted, in addition, to binding energies up to $A=16$ and  typically leads to rapid convergence in large-scale shell-model evaluations, describes $NN$ data to high accuracy and minimizes the contribution of the $NNN$ forces.
Various renormalization techniques, such as Okubo-Lee-Suzuki (OLS) \cite{LeeSuzuki80}, Similarity Renormalization Group (SRG) \cite{BognerFP07}, and Unitary Correlation Operator Method (UCOM) \cite{RothRH08}, 
aim to achieve a softer (renormalized or effective) interaction that enables the use of smaller manageable model spaces.
Phenomenological interactions, including schematic interactions (adopting a simple spatial form, such as the $\delta$ interaction) and empirical interactions (adjusting matrix elements of the residual interaction to nuclear data),  are fitted to many-body  nuclear properties, e.g., binding energies, excitation spectra, and possibly other spectral properties. Furthermore, the no-core shell model (NCSM) \cite{NavratilVB00,BarrettNV13} treats all $A$ particles active, while  the valence shell model \cite{Brown01,CaurierMNPZ05} assumes a core of inactive particles and a subset of valence particles within the valence partially-filled shell, which takes all the burden to account for particle correlations.

In the nuclear shell model, two principal limitations are encountered: (i)
the number of configurations necessary to describe a nuclear state  is typically huge and grows combinatorially with the number of particles and the size of the space they occupy, and (ii) phenomenological interactions typically yield predictions of nuclear properties that highly diverge outside the nuclear region they were fitted; present high-precision potentials hold predictive power, however, they generate strong -- to some degree or another -- short-range correlations  (coupling to high momenta), and these together with long-range correlations responsible for enhanced collectivity, large spatial deformation and $\alpha$-cluster substructures (wave function tail spreading to large distances) require ultra-large shell-model spaces,
often inaccessible on the best of modern-day supercomputers.
 An approach that addresses these limitations
invokes symmetries and is related to the fact that the wave functions of a quantum
mechanical system can be characterized by their invariance properties under certain group
transformations. In addition, if one can recognize near symmetries that survive within the nuclear dynamics, they
can be used to help reduce the dimensionality of a model space to tractable sizes.
This approach constitutes a major class of group-theoretical fermion models 
\cite{Draayer92}.

\section{Symmetries of strongly interacting particles -- organization of the shell-model space}

\subsection{Conventional coupling schemes}
%\begin{itemize}
%\item $jj$-coupling scheme ($M$ scheme and $J$ scheme  NCSM for a fixed $J$)
%\item $LS$-coupling scheme 
%\end{itemize}
We briefly review the $jj$-coupling scheme, together with $M$ scheme and $J$ scheme used in NCSM calculations \cite{NavratilVB00,BarrettNV13}, as well as the $LS$-coupling scheme, which underpins the SA-NCSM.

In the 1950s, two simple models  of nuclear structure, complementary in nature,  were advanced, and eventually merit a Nobel prize. These are the independent-particle model  of Mayer and Jensen \cite{Mayer49,HaxelJS49,Mayer50}, and the collective model of Bohr and  Mottelson \cite{BohrMottelson69,Mottelson_NP}. The first of these, which is microscopic in nature, recognizes that nuclei can be described by particles independently moving in a mean field, with the harmonic oscillator (HO) potential being a very good first approximation to the average potential experienced by each nucleon in a nucleus. This is augmented by a spin-orbit  $\mathbf{l}\cdot \mathbf{s}$ term and an orbit-orbit $\mathbf{l}^2$ force (that
shifts higher-$l$ levels downward) that lead to a successful reproduction of  the ``magic numbers" pattern. 
For a strong spin-orbit splitting, as the one observed for heavy nuclei, $\mathbf{l}\cdot \mathbf{s}$ energetically separates orbits with the same $l$ but different $j$, yielding the $jj$-coupling scheme with single-particle states labeled by $\eta (ls)  jm t_z$, or simply $\eta l jm t_z$ for $s=\half$. 

The second of these models, the collective model of Bohr and  Mottelson recognizes that deformed shapes dominate the nuclear dynamics. While enhanced deformation has been evident in heavy nuclei and those away from closed shells, deformed configurations are found to be important even in a nucleus such as $^{16}$O, which is commonly treated as spherical in its ground state, but about 40\% of the latter is governed by deformed shapes \cite{DytrychSBDV_PRCa07}; in addition, the lowest-lying excited $0^+$ states in $^{16}$O and their rotational bands are dominated by large deformation (see, e.g., \cite{RoweTW06}).  Bohr \& Mottelson offered a simple but important description of nuclei  in terms of the deformation of the nuclear surface and associated vibrations and rotations. While this model was not microscopic, it discussed spatial degrees of the  combined many-particle system (spatial deformation and rotations of ``shapes"), which suggested a relevant $LS$-coupling scheme, with single-particle states labeled by $\eta l m_l sm_s t_z$,  for which, e.g., a two-particle basis state looks like,  $\{a^\dagger_{\eta l s t_z}\times a^\dagger_{\eta' l' s' t_z'}\}^{(LS)JM}\ket{0}$, where $a^\dagger$ is the usual particle creation operator. Indeed, the microscopic Elliott model \cite{Elliott58, Elliott58b, ElliottH62} and its multi-shell expansion, the symplectic shell model  \cite{RosensteelR77,Rowe85}  that provides a microscopic formulation of the Bohr-Mottelson collective model, have soon after confirmed the relevance of the $LS$-coupling scheme, while providing a unique and physically relevant organization of the shell-model space, as discussed in Secs.  \ref{su3} and \ref{sp3r}.

Following the success of the independent-particle model  of Mayer and Jensen, many valence shell models and the no-core shell model  utilize the  $jj$-coupling scheme, together with  an important exact symmetry of the nuclear Hamiltonian, that is, it is invariant under rotation or a scalar with respect to the \SO{3} group. This implies that  $J$ and $M$ are good quantum numbers in nuclear states  and those can be used to enumerate shell-model spaces (parity and, sometimes, isospin are also adopted as good quantum numbers for the basis states, but they are mixed by comparatively weaker parity and isospin nonconserving inter-nucleon interactions and, in the case of isospin, the Coulomb potential). In particular, individual particle $j$'s can be coupled to a good total angular momentum $J$, leading to basis states with good $J$, which is the $J$ scheme. A basis state for two particles looks like, $\{a^\dagger_{\eta l j t_z}\times a^\dagger_{\eta' l' j' t_z'}\}^{JM}\ket{0}$. Alternatively, one can simply construct Slater determinants with fixed total $z$-component $M$, called an $M$ scheme --  e.g., for $A=2$, $a^\dagger_{\eta l j m t_z} a^\dagger_{\eta' l' j' m' t_z'}\ket{0}$, with $m+m'=M$.  The $M$-scheme basis states are easy to work with, but each of them is an admixture of states of different total angular momentum $J$.   Compared to the $M$ scheme, the $J$-scheme dimensionality is typically an order of magnitude smaller (see Fig. \ref{fig:dims} of Sec. \ref{appSANCSM}), however, the Hamiltonian matrices are denser and computing matrix elements is more time-consuming.

\subsection{\SU{3} scheme \label{su3}}
%SU(3): 3D harmonic oscillator\\
%Work by  Elliott, Harvey, Draayer, Rosensteel, Hirsch
%
In place of the spherical quantum numbers $\ket{\eta l m_l}$, the single-particle HO basis can be specified by $\ket{\eta_z \eta_x \eta_y}$, the HO quanta in the three Cartesian directions, $z$, $x$, and $y$, with $\eta_x + \eta_y + \eta_z=\eta$ ($\eta=0, 1, 2, \dots$ for $s$, $p$, $sd$, ... shells ). 
For a given  HO major shell,  the complete shell-model space is then specified by all distinguishable distributions of $\eta_z, \eta_x$ and  $\eta_y$. E.g., for $\eta=2$, there are 6 different distributions, $(\eta_z,\eta_x,\eta_y)=(2,0,0),(1,1,0),(1,0,1),(0,2,0),(0,1,1)$ and $(0,0,2)$. The number of these configurations is $\Omega_{\eta}=(\eta+1)(\eta+2)/2$ (spatial degeneracy) and the associated symmetry is  described by the $U(\Omega_\eta)$ unitary group. Each of these $(\eta_z,\eta_x,\eta_y)$ configurations  can be either unoccupied or has  maximum of two particles with spins $\uparrow \downarrow$.  

As a simple example for an \SU{3}-scheme basis state, consider  $A=2$ protons in the $sd$ shell ($\eta=2$) with a particle in the $(2,0,0)$ level with spin $\uparrow $ and another in the $(1,1,0)$ level with spins $\uparrow $. The total  number of quanta in each direction is $(\eta_z^{\rm tot},\eta_x^{\rm tot},\eta_y^{\rm tot},)=(3,1,0)$, or equivalently,  $\eta^{\rm tot} (\lambda\,\mu)=4 (2\,1)$, where  $\eta^{\rm tot}=\eta_x^{\rm tot}+\eta_y^{\rm tot}+\eta_z^{\rm tot}$, together with $\lambda= \eta_z^{\rm tot}-\eta_x^{\rm tot}$ and $\mu= \eta_x^{\rm tot}-\eta_y^{\rm tot}$ labeling an \SU{3} irrep, in addition to  the total intrinsic spin and its projection $S M_S$. For given $(\lambda \,\mu)$, the quantum numbers $\kappa$, $L$ and $M_L$ are given by Elliott \cite{Elliott58, Elliott58b}, according to the $\SU{3}  \stackrel{ \kappa}{\supset}\SO{3}_L{\supset}\SO{2}_{M_L}$, where the label $\kappa$ distinguishes multiple occurrences of the same orbital momentum $L$ in the parent irrep $(\lambda\,\mu)$. 
For our example, $(\lambda \,\mu)=(2\,1)$ with $\kappa=1$, $L=1,2,3$, and $M_L=-L,-L+1,\dots,L$. Hence, the set $\{{\eta}^A  (\lambda\,\mu) \kappa (LS) JM\}$ completely labels a 2-proton \SU{3}-scheme basis state (with $\eta^{\rm tot}=A\eta$). A basis state in this scheme for a 2-particle system looks like, $\{a^\dagger_{(\eta\, 0) s t_z}\times a^\dagger_{(\eta' 0) s' t_z'}\}^{(\lambda\,\mu)\kappa (LS)JM}\ket{0}$, which is an \SU{3}-coupled product,
provided that $a^\dagger$ is a proper  \SU{3} tensor; incidentally, the  \SU{3} tensor $a^\dagger$  of rank $(\lambda\,\mu)=(\eta\, 0)$ coincides with the familiar particle creation operator, $a^\dagger_{(\eta\, 0)lm s \sigma t_z} \equiv a^\dagger_{\eta lm s \sigma t_z}$, while the particle annihilation  \SU{3} tensor of rank $(\lambda\,\mu)=(0\,\eta)$ is given as $\tilde a_{(0\,\eta)l -m s -\sigma t_z} = (-1)^{\eta+l-m+s-\sigma} a_{\eta l m s \sigma t_z}$. 
Note that for $\eta=\eta'=2$, e.g., there are only a few 2-proton configurations $(\lambda\,\mu)L = (4\,0)L=0,2,4, (2\,1)L=1,2,3,$ and $(0\,2)L=0,2$. Furthermore, 
these basis states are related to $LS$-coupled basis states (similarly, to $jj$-coupled basis states) via a simple unitary transformation, $\{a^\dagger_{(\eta\, 0) s t_z}\times a^\dagger_{(\eta' 0) s' t_z'}\}^{(\lambda\,\mu)\kappa (LS)JM}\ket{0}=
\sum_{l,l'} \RedCG{(\eta\, 0)l}{(\eta' 0)l'}{(\lambda\,\mu)\kappa L}
\{a^\dagger_{\eta l s t_z}\times a^\dagger_{\eta' l' s' t_z'}\}^{(LS)JM}\ket{0}$, where $\RedCG{\dots}{\dots}{\dots}$ is the \SU{3} analog of the  familiar reduced Clebsch-Gordan coefficient (note that there is no dependence on the particle orbital momenta, $l$ and $l'$, in the \SU{3}-scheme basis states). 
\begin{table}[th]
\begin{center}
    \begin{tabular}{ll|l}
      \hline
      \hline
      \multicolumn{2}{c|}{Spatial d.o.f.} & Spin d.o.f. \\
      \hline 
        $\Un{10}$  \hspace{23pt} $\supset$   &  \SU{3} & $\SU{2}$\\
      $[f_1f_2\dots f_{10}]$ & $(\lambda\,\mu)$ &  $S$  \\
      \hline
&& \\
$\yng(2,2)$
&    
$(8\,2),(7\,1),(4\,4)^2,(5\,2),(0\,6),(6\,0), (3\,3)$
&
$\yng(2,2)$
\\
$[2^2]$& 
$(1\,4),(4\,1),(2\,2)^2,(1\,1)$ 
&
$S=0$ \\
&& \\
$\yng(2,1,1)$
&  
$(9\,0),(6\,3), (7\,1), (4\,4), (2\,5), (5\,2)^2, (3\,3)^2$
&
$\yng(3,1)$   
\\
$[21^2]$& 
$(1\,4)^2,(4\,1)^2,(2\,2),(0\,3),(3\,0)^2,(1\,1)$ 
&
$S=1$\\
&& \\
$\yng(1,1,1,1)$
&
$(5\,2),(0\,6),(3\,3),(2\,2),(3\,0)$    
&
$\yng(4)$    
\\
$[1^4]$& & $S=2$ 
 \\
      \hline
    \end{tabular}
\caption{\SU{3}$\times$\SU{2}$_S$ configurations for 4 protons (neutrons) in the $pf$ shell ($\eta=3$ with $\Omega_\eta=10$). Note that a spatial symmetry represented by a Young tableau $\left[f_{1},\dots, f_{\Omega_{\eta}}\right]$ is uniquely determined by its complementary spin symmetry  of a given intrinsic spin $S_{\eta}$ (conjugate Young tableaux) ensuring the overall antisymmetrization of each $U(\Omega_\eta)$$\times$\SU{2}$_{S_\eta}$ configuration with respect to spatial and spin degrees of freedom (d.o.f.) \cite{DraayerLPL89}.
 } 
\label{UNtoSU3}
\end{center}
\end{table}

An important feature of the \SU{3} scheme is that  all possible 
configurations  within a major HO shell $\eta$ (for protons or neutrons) are not constructed using the tedious procedure of coupling of creation operators referenced above, but are readily available based on  the $U(\Omega_\eta)$ unitary group of the many-body three-dimensional HO. In particular, the basis construction is implemented according to the reduction \cite{DraayerLPL89}
\begin{equation}
\begin{array}{cccccc}
   \Un{\Omega_\eta}        &     & & \times  &  \SU{2}       \\
   
          \left[f_1,f_2,\dots f_{\Omega_\eta}\right] &     & & & S_\eta       \\
  \cup   &  \alpha_\eta   & &    &     \\
\SU{3}   & &   & &     \\
(\lambda_{\eta}\,\mu_{\eta})   & &   & &     \\
\end{array},
\label{UtoSU3}
\end{equation}
with $\SU{3}_{(\lambda_{\eta}\,\mu_{\eta})}  \stackrel{ \kappa_\eta}{\supset}\SO{3}_{L_\eta}{\supset}\SO{2}_{M_{L_\eta}}$ \cite{Elliott58, Elliott58b}, where a multiplicity index $\alpha_{\eta}$ distinguishes multiple occurrences of an
\SU{3} irrep $(\lambda_{\eta}\,\mu_{\eta})$ in a given $\Un{\Omega_{\eta}}$ irrep labeled by Young tableaux, $[\mathbf{f}]=[f_1,f_2,\dots,f_{\Omega_\eta}]$,  with $f_1 \ge f_2 \ge \dots \ge f_{\Omega_\eta}$ and $f_i=0$ (unoccupied), $1$ (occupied by a particle), or $2$ (occupied by 2 particles of spins $\uparrow \downarrow$). An illustrative example for 4 particles in the $pf$ shell ($\eta=3$) is shown in Table \ref{UNtoSU3}. 

\vspace{12pt}
We note that the \SU{3} scheme provides a classification of the  complete shell-model space (in a single shell as illustrated above and in multiple shells as described in Sec. \ref{multiSU3}) and is related to the $LS$-coupling and $jj$-coupling schemes via a unitary transformation. It divides the space into basis states of definite $(\lambda\,\mu)$ quantum numbers of \SU{3} that are linked to the intrinsic quadrupole deformation according to the established  mapping \cite{RosensteelR77b,LeschberD87,CastanosDL88}. For example, the simplest cases, $(0\, 0)$, $(\lambda\, 0)$, and $(0\,\mu)$,  describe spherical, prolate, and oblate deformation, respectively\footnote{
Following this mapping, quadrupole moments of $(0\, 0)$, $(\lambda\, 0)$, and $(0\,\mu)$ configurations -- in a simple classical analogy to rotating spherical, prolate, and oblate spheroids in the lab frame \cite{RoweW2010book} --  are zero, negative, and positive, respectively.
},
 while a general nuclear state is typically a superposition of several hundred various triaxially deformed configurations.  Note  that, in this respect, basis states can have little to no deformation,  and, e.g.,  about 60\% of the ground state of the closed-shell $^{16}$O is described by a single \SU{3} basis state, the spherical $(0\, 0)$ (see Table \ref{tab:dominantConfig} of Sec. \ref{spin}).  

%%%%%%%%%%%%%%%%%%%%%%%%%%%%%%%%%%
%%%%%%%%%%%%%%%%%%%%%%%%%%%%%%%%%%
\subsubsection{Elliott model}

\begin{figure}[th]
\centerline{
\includegraphics[height=3.5in]{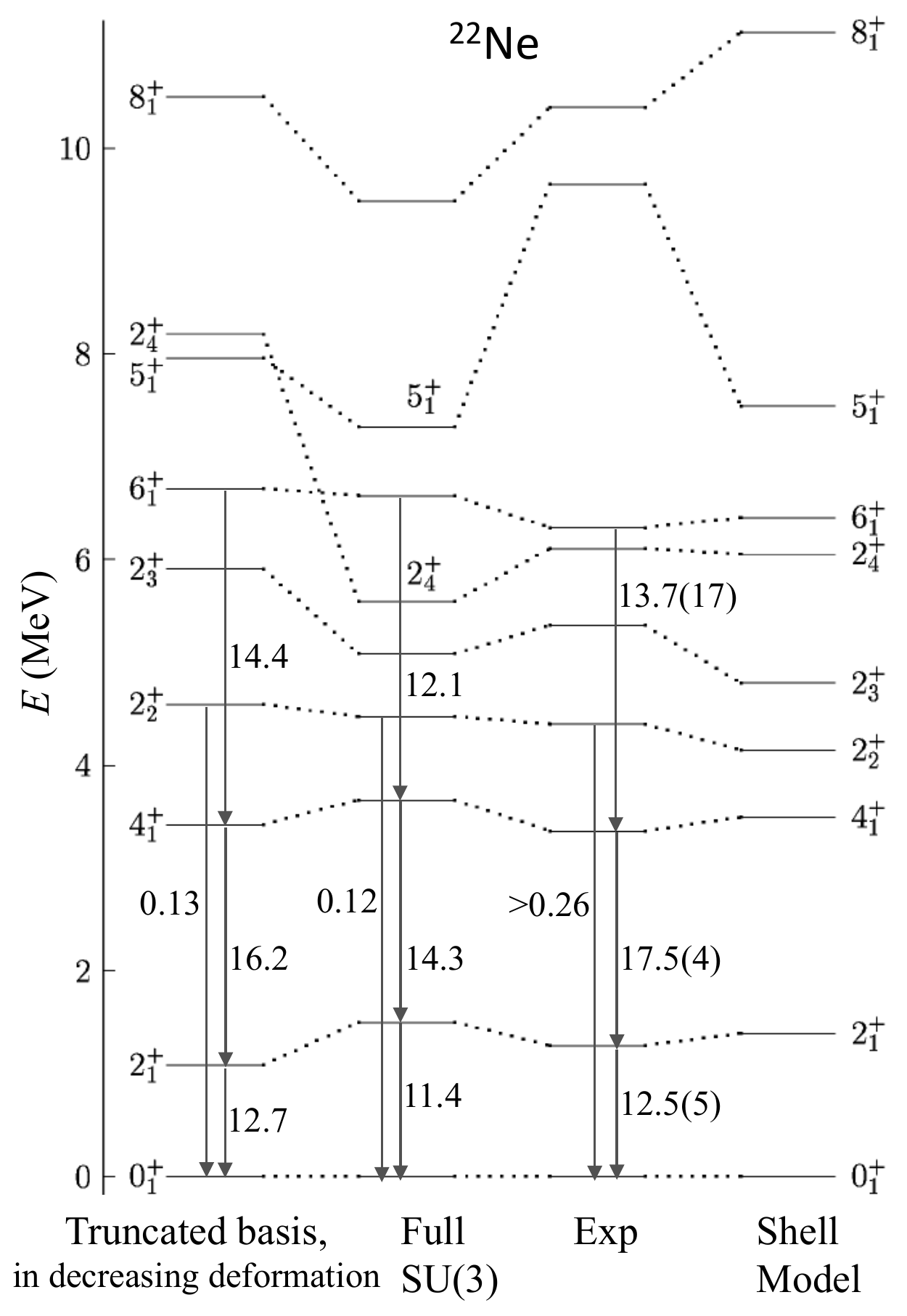}
}
\caption{ 
Elliott's model with an \SU{3}-preserving interaction $+$ pairing in the $sd$ valence shell for $^{22}$Ne. $B(E2)$ transitions strengths (W.u.) are calculated for  proton $e_{\rm eff}=1.3$ effective charge. Figure adapted from \cite{Vargas01}.
}
\label{Vargas01_Ne22}      
\end{figure}

The seminal work of Elliott \cite{Elliott58, Elliott58b, ElliottH62} focused on the key role of \SU{3}, the exact symmetry  of the three-dimensional spherical HO (see also Refs.\cite{Moshinsky62, MoshinskyPSW75,Millener78}).  Within a shell-model framework, Elliott's model utilizes  an \SU{3}-scheme basis  that is  related via a unitary transformation to  the basis used in the conventional shell model. For \SU{3}-symmetric interactions, the model can be solved analytically. But regardless whether a simple \SU{3}-preserving interaction is used (see Figs. 5.1-5.6 of Ref. \cite{Harvey68}), or an \SU{3}-symmetry breaking interaction (see  Fig. 1 of Ref. \cite{ElliottH62}), the results have revealed  a striking  feature, namely, the dominance of a few most deformed configurations. This has  been shown for $sd$-shell nuclei, such as $^{18}$Ne, $^{20}$Ne,  $^{22}$Ne, $^{22}$Mg, $^{24}$Mg, and $^{28}$Si, that have been known to possess a clear collective rotational structure in their low-lying states \cite{Elliott58, Elliott58b, ElliottH62}. It has been also observed in heavier nuclei, where pseudo-spin symmetry \cite{HechtA69,ArimaHS69} and its
pseudo-\SU{3}  complement \cite{RatnaRajuDH73} have been shown to play a similar role in accounting for deformation in the upper $pf$ and lower $sdg$ shells, and in particular, in strongly deformed nuclei of the rare-earth and actinide regions \cite{Draayer91,BahriDM92} (see Sec. \ref{pseudoSU3}). In this mass region, an approximate ``quasi-\SU{3}" symmetry has been
also suggested \cite{ZukerRPC95}. Furthermore, the pairing interaction has been microscopically incorporated into the Elliott model where it breaks the \SU{3} symmetry and mixes different  $(\lambda\,\mu)$ configurations. It has been shown in Ref. \cite{Vargas01}  that using an \SU{3}-symmetric  interaction-plus-pairing yields results close to experiment and to the energies obtained using shell-model calculations in the full $sd$ shell \cite{PreedomW72} (Fig. \ref{Vargas01_Ne22}). It is remarkable that, even in the presence of pairing, comparable results have been obtained in a truncated model space that includes only about 10 most deformed configurations.

Two other complementary models, that are not based on the fermion shell models  but have figured prominently in informing the importance of deformation and pairing, are the Geometric Collective Model  \cite{EisenbergG87,GneussMG69} advanced by Greiner and collaborators, and the Interacting Boson Model (IBM) of Iachello and associates \cite{IachelloA87,ArimaI75}. The latter has
offered a bosonic realization of these phenomena, suitable for systematic classifications of the large nuclear data, in terms of 
a common overarching \Un{6} algebraic structure and its physical subgroups, \Un{5} for pairing modes, \SU{3}$\supset$\SO{3} for rotations and \On{6}$\supset$\SO{3} for triaxial systems. 

%%%%%%%%%%%%%%%%%%%%%%%%%%%%%%%%%%
%%%%%%%%%%%%%%%%%%%%%%%%%%%%%%%%%%
\subsubsection{\textit{Ab initio} symmetry-adapted no-core shell model (SA-NCSM) \label{multiSU3}}

The symmetry-adapted no-core shell model (SA-NCSM) \cite{DytrychLMCDVL_PRL12} is a multi-shell generalization of the \SU{3} scheme used in the Elliott model. It adopts the first-principle concept and is a no-core shell model (NCSM) carried forward in an \SU{3} scheme. The  many-nucleon basis states of the SA-NCSM are constructed using efficient group-theoretical algorithms based on \SU{3}$\times$\SU{2}$_S$ configurations (irreps) labeled by $(\lambda\,\mu)$ quantum numbers and the   intrinsic spin $S$ \cite{DytrychLMCDVL_PRL12,DytrychHLDMVLO14}. 

\begin{figure}[th]
\centerline{
\includegraphics[width=0.75\textwidth]{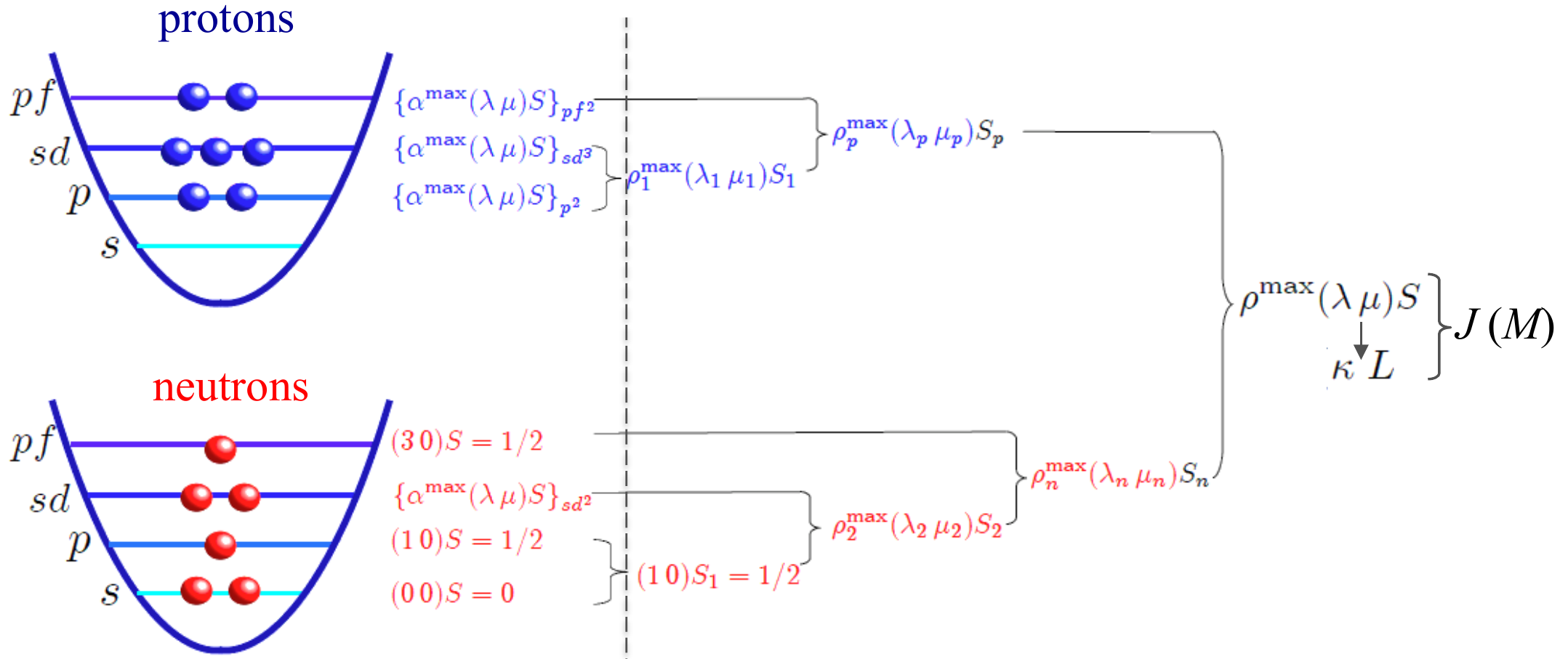}
}
\caption{ 
Example for an \SU{3}-scheme basis state of good $(\lambda\,\mu)S$ and $J$. All additional quantum numbers needed to specify the basis state are shown in the illustration. The dashed line divides single-shell  (left) and inter-shell (right) quantum numbers.
}
\label{su3scheme}      
\end{figure}

In particular, the many-particle basis states of the SA-NCSM  are nuclear configurations of fixed parity, consistent with the Pauli principle, and truncated by a cutoff $N_{\max}$. The $N_{\max}$ cutoff is defined as the maximum number of HO quanta
allowed in a many-particle state above the minimum for a given nucleus.
For a given $N_{\max} $, the SA-NCSM many-particle basis states (Fig. \ref{su3scheme}) are constructed in the proton-neutron
formalism, that is, we treat neutron and proton orbitals independently so total isospin is not conserved.  
For all possible distributions  of protons $\{Z_{0}, Z_{1}, Z_{2}, \dots\}$ and  neutrons $\{N_{0}, N_{1}, N_{2}, \dots\}$ over the major HO shells $\eta$ ($\eta=0,1,2,\dots$ for the $s$, $p$, $ds$,\dots HO shell), limited by the number of HO quantum excitations up through $N_{\max}$, the \SU{3}$_{(\lambda_{\eta}\,\mu_{\eta})}\times$\SU{2}$_{S_\eta}$ configurations are first enumerated for every major HO shell, following the $U(\Omega_\eta)\supset \SU{3}_{(\lambda_{\eta}\,\mu_{\eta})}$ reduction (\ref{UtoSU3}). This is followed by an inter-shell \SU{3}$\times$\SU{2}$_{S}$ coupling of the in-shell configurations. 
Finally, the resulting proton and neutron configurations are coupled to good
quantum numbers $(\lambda\,\mu)\kappa L$ of the \SU{3}$_{(\lambda\,\mu)}\underset{\kappa}{\supset}$\SO{3}$_L$ group chain, together with proton, neutron, and total
intrinsic spins $S_{p}$, $S_{n}$, and $S$ of the complementary \SU{2} spin group. The orbital angular momentum $L$
is coupled with $S$ to the total angular momentum $J$ with a projection
$M$. Each basis state in this scheme is labeled schematically as
$ |\vec{\gamma}\, N(\lambda\,\mu)\kappa L;(S_{p}S_{n})S;J M\rangle$, where $N$ is the total number of HO excitation quanta and 
 $\vec{\gamma}$ denotes additional quantum numbers needed to
distinguish among configurations carrying the same $N(\lambda\,\mu)$ and
($S_{p}S_{n})S$ labels. In this way, a complete  shell-model basis is classified. 

There are two major advantages that follow from the use of  an \SU{3}-scheme basis that empower the SA-NCSM with two unique and important features:
\begin{enumerate}
\item The organization of the model space allows the complete $N_{\rm max}$ space to be down-selected to the physically relevant subspace. 
\item Within the  space down-selected to a subset of  $(\lambda\,\mu)$ irreps and intrinsic spins $(S_{p}S_{n}S)$, the spurious center-of-mass (CM) motion can be factored out exactly~\cite{Verhaar60,Hecht71,Millener92}. This ensures the translational invariance of the SA-NCSM wave functions.
\end{enumerate}

The underlying principle behind the SA-NCSM kernel  is an \SU{3}-type
Wigner-Eckhart theorem, which factorizes Hamiltonian matrix elements into the
product of \SU{3} reduced matrix elements  and the associated \SU{3}
coupling coefficients.  
To compute the Hamiltonian matrix elements, the computational  realization of the SA-NCSM, dubbed \texttt{LSU3shell} \cite{LSU3shell13}, 
 adopts state-of-the-art group-theoretical methods~\cite{DraayerLPL89} and
optimized numerical subroutines~\cite{AkiyamaD73} for computing  \SU{3} coupling/recoupling coefficients.  
Recent developments and applications in the framework of the {\it ab initio} SA-NCSM are reviewed in Sec. \ref{appSANCSM}.

\subsection{Symplectic \SpR{3} scheme \label{sp3r}}
%Sp(3,R)$\subset$Sp(3(A-1),R): $A$-particles in phase space\\
%Work by  Rowe, Rosensteel, Hecht, Draayer \\
%
For $A$ particles in three-dimensional space, the complete basis for the shell model is described by \SpR{3A}$\times$\Un{4} \cite{Rowe96}, where \SpR{3A} is the group of all linear canonical transformations of the $3A$-particle phase space and Wigner's supermultiplet group $\Un{4}$ describes the complementary spin-isospin space (see Sec. \ref{su4}). A complete translationally invariant shell-model basis is classified according to (see, e.g., \cite{Rowe85,Rowe96}),
\begin{equation}
\begin{array}{cccccc}
   \SpR{3(A-1)}        &     & & \times  &  \Un{4}       \\
  \cup      &     & &    & \cup     \\
\SpR{3} \times  \On{A-1}    & &   & & \SU{2}_S  \times  \SU{2}_T      \\
\label{phasespace}
\end{array}.
\end{equation}
The \SpR{3} scheme utilizes the symplectic group \SpR{3}, which consists of all particle-independent linear canonical transformations of the single-particle phase-space observables, $x_{i \alpha} \rightarrow \sum_{\beta} a_{\alpha \beta} x_{i \beta} +b_{\alpha \beta} p_{i \beta}$ and $p_{i \alpha} \rightarrow \sum_{\beta} c_{\alpha \beta} x_{i \beta} +d_{\alpha \beta} p_{i \beta}$ ($i=1,\dots,A$ and $\alpha,\beta=x,y,z$), that preserve the commutation relations $[x_{i\alpha},p_{j \beta}]=i\hbar \delta_{ij}\delta_{\alpha \beta}$ \cite{Rowe13}.
The \SpR{3} scheme further utilizes the group reduction to classify many-particle basis states $|\sigma n\rho\omega \kappa L M\rangle $ of a symplectic irrep,
\begin{equation}
\begin{array}{cccccccc}
\SpR{3}       & \supset & U(3)  & \supset & \SO{3} & \supset & SO(2)   \\
\sigma         & n\rho     & \omega & \kappa  & L    & & M
\end{array},
\end{equation}
where $\sigma$ $\equiv $ $N_\sigma\left(\lambda_{\sigma}\, \mu_{\sigma}\right)$
labels the \SpR{3} irrep, $n\equiv N_n\left(\lambda_{n}\,
\mu_{n}\right)$, $\omega\equiv N \left(\lambda_{\omega}\,
\mu_{\omega}\right)$, and  $N=N_{\sigma}+N_n$ is
the total number of HO quanta  ($\rho$ and $\kappa$ are multiplicity labels) \cite{Rowe85}. The relation of these symplectic basis states to $M$-scheme states of the NCSM is provided in Ref.  \cite{DytrychSDBV08_review}.
The classification of basis states based on the dual  \On{A-1} of the reduction (\ref{phasespace}) is briefly discussed in Sec. \ref{oA}. 

The key importance of  the  symplectic \SpR{3} group for a microscopic description of a quantum many-body system of interacting particles emerges from the physical relevance of its 21
generators, which are directly related to the particle momentum  and position coordinate  and realize 
important observables, as shown below. Namely, the many-particle kinetic energy,  the 
HO potential (or equivalently, the monopole operator), the mass  quadrupole moment, and angular momentum operators are all generators of \SpR{3} and preserve the symplectic symmetry. In addition, the model includes multi-shell collective vibrations and vorticity degrees
of freedom for a description from irrotational to rigid rotor flows.
 Briefly, the translationally invariant (intrinsic) symplectic \SpR{3} generators  can be written as \SU{3} tensor operators in terms of the harmonic oscillator raising, $b_{i \alpha}^{\dagger(1\,0)}=\frac{1}{\sqrt{2}}(X_{i \alpha}-iP_{i \alpha})$, and lowering $b^{(0\,1)}$ dimensionless operators (with $\mathbf X$ and $\mathbf P$ the lab-frame position and momentum coordinates and $\alpha=1,2,3$ for the three spatial directions), 
\begin{eqnarray}
A^{(2\,0)}_{\mathfrak{L}M}\!\!&=&\!\! 
\frac{1}{\sqrt{2}} \sum_{i=1}^A
\{b_{i}^{\dagger}\times b_{i}^{\dagger}\}^{(2\,0)}_{\mathfrak{L}M}
- \frac{1}{\sqrt{2}A} \sum_{s,t=1}^A
\{b^{\dagger}_{s}\times b^{\dagger}_{t}\}^{(2\,0)}_{\mathfrak{L}M}
 \label{sp3RgenA}\\
C^{(1\,1)}_{\mathfrak{L}M}\!\!&=&\!\!
\sqrt{2} \sum_{i=1}^A
\{b_{i}^{\dagger}\times b_{i}\}^{(1\,1)}_{\mathfrak{L}M}
\!- \frac{\sqrt{2}}{A} \sum_{s,t=1}^A
\{b^{\dagger}_{s}\times b_{t}\}^{(1\,1)}_{\mathfrak{L}M},
\label{sp3RgenC}
\end{eqnarray}
together with $B^{(0\,2)}_{\mathcal{L}M}=(-)^{\mathcal{L}-M}(A^{(2\,0)}_{\mathcal{L}-M})^{\dagger}$ ($\mathcal{L}=0,2$) and
$H_{00}^{(00)}= \sqrt{3} \sum_{i} \{b_{i}^{\dagger}\times b_{i}\}^{(00)}_{00}
-\frac{\sqrt{3}}{A} \sum_{s,t}
\{b^{\dagger}_{s}\times b_{t}\}^{(0\,0)}_{00}
+\frac{3}{2}(A-1)$, 
where the sums run over all $A$ particles of the system. 

Equivalently, the symplectic generators, being one-body-plus-two-body operators can be  expressed in terms of the creation operator $a^\dagger_{(\eta\, 0)}=a^\dagger_\eta$ and its \SU{3}-conjugate annihilation operator, $\tilde a_{(0\, \eta)}$. This is achieved by using the known matrix elements of the position and momentum  operators in a HO basis, and hence, e.g., the first sum of $A^{(2\,0)}_{\mathfrak{L}M}$ in Eq. (\ref{sp3RgenA}) becomes, 
$ \sum_{\eta}\sqrt{\frac{(\eta +1)(\eta +2)(\eta +3)(\eta +4)}{12}}
\left\{ a_{(\eta+2\, 0)}^{\dagger}\times \tilde a_{(0\,\eta)}\right\}^{(2\,0)}_{\mathfrak{L}M}$ \cite{EscherD98}. Note that this operator describes excitations of a nucleon from the $\eta$ shell to the $\eta+2$ shell, which corresponds to creating two single-particle HO excitation quanta, as manifested in the first term of Eq. (\ref{sp3RgenA}).

The eight  0\ho~operators $C^{(1\,1)}_{\mathcal{L},M}$ ($\mathcal{L}=1,2$) generate the \SU{3} subgroup of
\SpR{3}. They realize the angular momentum operator (dimensionless):
\begin{equation}
L_{1M}=C^{(1\,1)}_{1M},\, M=0,\pm1,
\label{Lgen}
\end{equation}
 and the Elliott ``algebraic" quadrupole moment tensor $\mathcal{Q}^{a}_{2M}=\sqrt{3}C^{(1\,1)}_{2M},\, M=0,\pm1,\pm2$. 

It is important to note that, in addition to the orbital  angular momentum $L$, operators of a physical significance are \SpR{3}-preserving and can be constructed in terms of the symplectic generators:
\begin{enumerate}
\item Mass quadrupole moment (dimensionless):
\begin{equation}
Q_{2M}=\sqrt{3}(A^{(2\,0)}_{2M}+C^{(1\,1)}_{2M}+B^{(0\,2)}_{2M});
\label{Qgen}
\end{equation}
\item Many-particle kinetic energy:
\begin{equation}
\frac{T}{\ho}=\frac{1}{\ho}\sum_i\frac{{\mathbf p}_{i}^2}{2m}=\frac{1}{2}H_{00}^{(00)} -\sqrt{\frac{3}{8}}(A^{(2\,0)}_{00}+B^{(0\,2)}_{00});
\label{KE}
\end{equation}
\item HO potential (monopole operator):
\begin{equation}
\frac{V_{HO}}{\ho}=\frac{1}{\ho}\sum_i\frac{m\Omega^2 {\mathbf r}_{i}^2}{2} =\frac{1}{2}H_{00}^{(00)} +\sqrt{\frac{3}{8}}(A^{(2\,0)}_{00}+B^{(0\,2)}_{00}).
\label{r2}
\end{equation}
\end{enumerate}
Therefore, none of these operators mixes symplectic irreps.

\begin{figure}[th]
\centerline{
\includegraphics[width=0.65\textwidth]{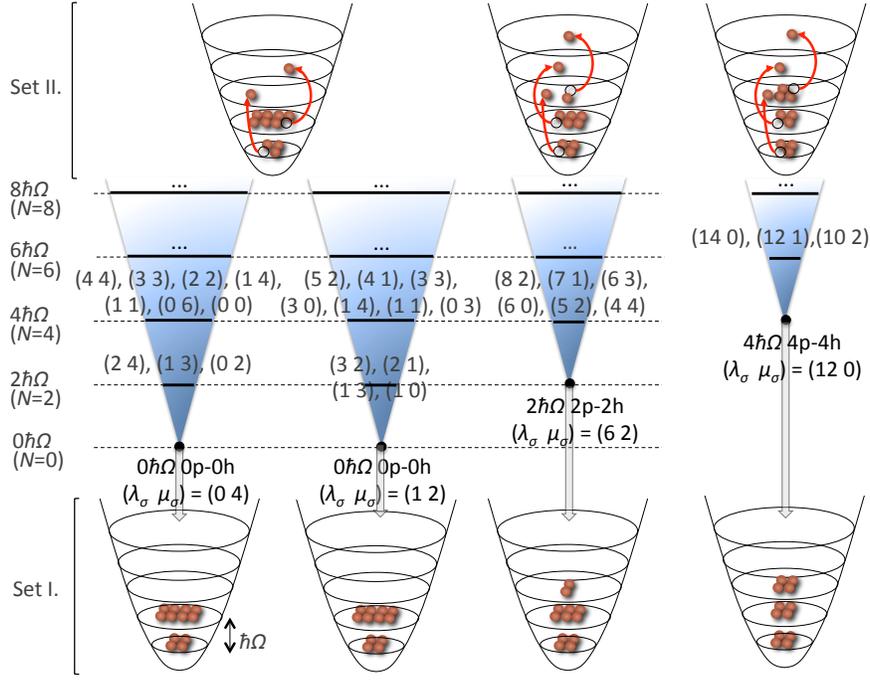}
}
\caption{ 
Four \SpR{3} irreps (vertical cones) that dominate low-lying states in $^{12}$C. Basis states of an irrep that have good $(\lambda\,\mu)$  are built by 2\ho~\ph{1} monopole or quadrupole excitation (Set II) over a bandhead. The symplectic  bandhead (Set I) is an \SU{3}-coupled many-body state with a given nucleon distribution over the HO shells. 
}
\label{sp3Rpicture}      
\end{figure}

The symplectic  structure accommodates relevant particle-hole (p-h) configurations:  2\ho~\ph{1}  monopole excitations (one particle raised by two shells) are driven by the monopole operator (\ref{r2}), while  2\ho~\ph{1}  quadrupole excitations are driven by the $Q$ operator (\ref{Qgen}), or equally,  by $A^{(2\,0)}_{\mathfrak{L}M}$ with $\mathfrak{L}=0$ ($\mathfrak{L}=2$) for the monopole (quadrupole) excitations. Hence, the basis states of an \SpR{3} irrep (vertical cone) are built over a bandhead $\ket{\sigma}$  (Fig. \ref{sp3Rpicture}, Set. I) by 2\ho~ \ph{1} 
monopole or quadrupole excitations (Fig. \ref{sp3Rpicture}, Set. II), realized by the first term in $A^{(2\,0)}_{\mathfrak{L}M}$ of Eq. (\ref{sp3RgenA}),  together with a smaller 2\ho~\ph{2} correction for eliminating the spurious center-of-mass (CM) motion, realized by the second term in $A^{(2\,0)}_{\mathfrak{L}M}$:
\begin{equation}
|\sigma n\rho\omega \kappa L M_L\rangle 
= \{ \{A^{(2\,0)}\times A^{(2\,0)} \dots \times A^{(2\,0)}\}^{n}\times\ket{\sigma}\}^{\rho\omega}_{\kappa L M_L}.
\label{basis}
\end{equation}
Remarkably, these \SpR{3} basis states are in one-to-one correspondence with a coupled product of the states of the Bohr vibrational model (realized in terms of giant monopole-quadrupole resonance states with irrotational flows), $\{ \{A^{(2\,0)}\times A^{(2\,0)} \dots \times A^{(2\,0)}\}^{n}\times\ket{N_\sigma(0\,0)}\}^{(\lambda_n\, \mu_n)}$, and $(\lambda_\sigma\,\mu_\sigma)$ deformed states of an \SU{3} model \cite{Rowe13}.

 Including spin  degrees of freedom, $\SpR{3} \times\SU{2}_S$, the many-particle basis states become, 
\begin{equation}
|\sigma n\rho\omega \kappa (L S_{\sigma}) JM \rangle = \sum_{M_LM_S} \CG{L M_L}{S_{\sigma} M_S}{J M} |\sigma n\rho\omega \kappa L M_L S_{\sigma} M_S \rangle.
\end{equation}
States within a symplectic irrep have the same spin value, which are given by the spin $S_\sigma$ of the bandhead  $\ket{\sigma; S_{\sigma}}$. 
Symplectic basis states span the entire shell-mode space. A complete set of labels includes additional quantum numbers $\ket{\left\{\alpha \right\}\sigma}$ that distinguish different bandheads with the same $N_\sigma\left(\lambda_{\sigma}\, \mu_{\sigma}\right)$.  

The symplectic  bandhead $\ket{\sigma}$ is defined by the usual requirement that the symplectic lowering operators $B^{(0\,2)}_{\mathcal{L}M}$ annihilate it.  The  bandhead, $\ket{\sigma; \kappa_\sigma L_\sigma M_\sigma}$, is an \SU{3}-coupled many-particle state with a given nucleon distribution over the HO shells and while not utilized in the \SpR{3} scheme, can be obtained in terms of the particle creation operators. For example, for a given 0\ho~$(\lambda_{\sigma}\, \mu_{\sigma})$ bandhead, the nucleon distribution is a single configuration,
\begin{equation}
\left\{ a^\dagger_{(\eta_1\, 0)} \times a^\dagger_{(\eta_2\, 0)} \times  \dots \times  a^\dagger_{(\eta_A\, 0)} \right\}^{(\lambda_{\sigma}\, \mu_{\sigma})}_{\kappa_\sigma L_\sigma M_\sigma}\ket{0}
\end{equation}
with $N_ \sigma = \eta_1+ \eta_2 + \dots + \eta_A+\frac{3}{2}(A-1)$, such that $N_\sigma \ho$ includes the HO zero-point energy and $3/2$ is subtracted to ensure a proper treatment of the center-of-mass.

An example for the symplectic basis states follows for $^{24}$Mg. Its lowest HO-energy configuration is given by $N_\sigma= 62.5$ or 0\ho, while the 4\ho~$(20\,0)$ symplectic irrep includes:
\begin{enumerate}
\item A  bandhead  ($N_n=0$) with  $N_\sigma= 66.5$ (or 4\ho) and $(\lambda_ \sigma \,\mu_ \sigma)=(20\,0)$;
\item  $N_n=2$  states with $N $=68.5 and $(\lambda_ \omega \,\mu_ \omega)= (22\, 0)$, $(20\, 1)$, and $(18\, 2)$;
\item $N_n=4$  states with $N $=70.5 and $(\lambda_ \omega \,\mu_ \omega)= (24\, 0)$, $(22\, 1)$, $(20\, 2)^2$, $(19\, 1)$, $(18\, 3)$, $(18\, 0)$ and $(16\, 4)$; there are two occurrences of $(\lambda_ \omega \,\mu_ \omega)=(20\, 2)$, one of which results from the coupling of $(\lambda_ \sigma \,\mu_ \sigma)=(20\,0)$ to $(\lambda_ n \,\mu_ n)=(4\, 0)$ and the other from the coupling of  $(20\,0)$ to $(\lambda_ n \,\mu_ n)=(0\, 2)$.
\item and so forth for higher $N_n$.
\end{enumerate}
For each $(\lambda_ \omega \,\mu_ \omega)$, the quantum numbers $\kappa$, $L$ and $M$ are given by Elliott \cite{Elliott58, Elliott58b}, as discussed in Sec. \ref{su3} and in Ref. \cite{Rowe13}. For example, for $(20\,0)$, $\kappa=0$, $L=0,2,4,\dots,20$, and $M=-L,-L+1,\dots,L$.

\subsubsection{Symplectic model}
A significant breakthrough for the nuclear modeling is the microscopic symplectic model, developed by Rosensteel  and Rowe ~\cite{RosensteelR77,Rowe85}. It provides a microscopic framework for understanding deformation-dominated collective phenomena in atomic nuclei~\cite{Rowe85} and accommodates particle-hole excitations across multiple shells.  
Indeed, the symplectic \SpR{3} symmetry underpins the symplectic shell model that provides a microscopic formulation of the Bohr-Mottelson collective model and is a multiple-shell generalization of the successful Elliott \SU{3} model. The classical realization of this symmetry underpins the dynamics of
rotating bodies and has been used, for example, to describe the rotation of deformed
stars and galaxies \cite{Rosensteel93}.

In its simplest depiction \cite{RoweTW06}, the symplectic shell model is based on nucleons occupying HO shells with important correlations within each shell and between shells differing by $\pm2\ho$. The in-shell correlations are dominated by interactions of the quadrupole-quadrupole type, as first introduced by Elliott \cite{Elliott58, Elliott58b}, while the inter-shell correlations are of the giant monopole and giant quadrupole type. The inter-shell correlations enhance the electric quadrupole collectivity in such a way as to eliminate the need for effective charges. It is found that in many strongly deformed heavy nuclei the \SU{3} quantum numbers, $\lambda$ and $\mu$, possess very large values, e.g., $\lambda \sim 100$ and $\mu \sim 10$ \cite{JarrioWR91}.
This naturally leads to contraction of the \SU{3} model to a rotor model \cite{Rowe96,Rowe13}.

The symplectic model with \SpR{3}-preserving interactions have achieved a remarkable reproduction of rotational bands and transition rates without the need for introducing effective charges, while only a single \SpR{3} irrep is used \cite{Rowe85,BahriR00}. The model of Ref. \cite{BahriR00} adopts a Davidson potential, $V(Q)=\chi (Q\cdot Q+\varepsilon / Q \cdot Q)$ \cite{Davidson32}, used to describe diatomic molecules \cite{RoweB98}. The symplectic model is used to construct rotational states for a rare-earth nucleus with microscopic wave functions. Analysis of the states in terms of their \SU{3} content shows that \SU{3} is a very poor dynamical symmetry (mixing of many irreps) but an excellent {\it quasi-dynamical} symmetry for the
model (the same \SU{3} irreps and their contribution propagates for different $L$'s). 

\begin{figure}[tH]
\centerline{
\includegraphics[width=0.52\textwidth]{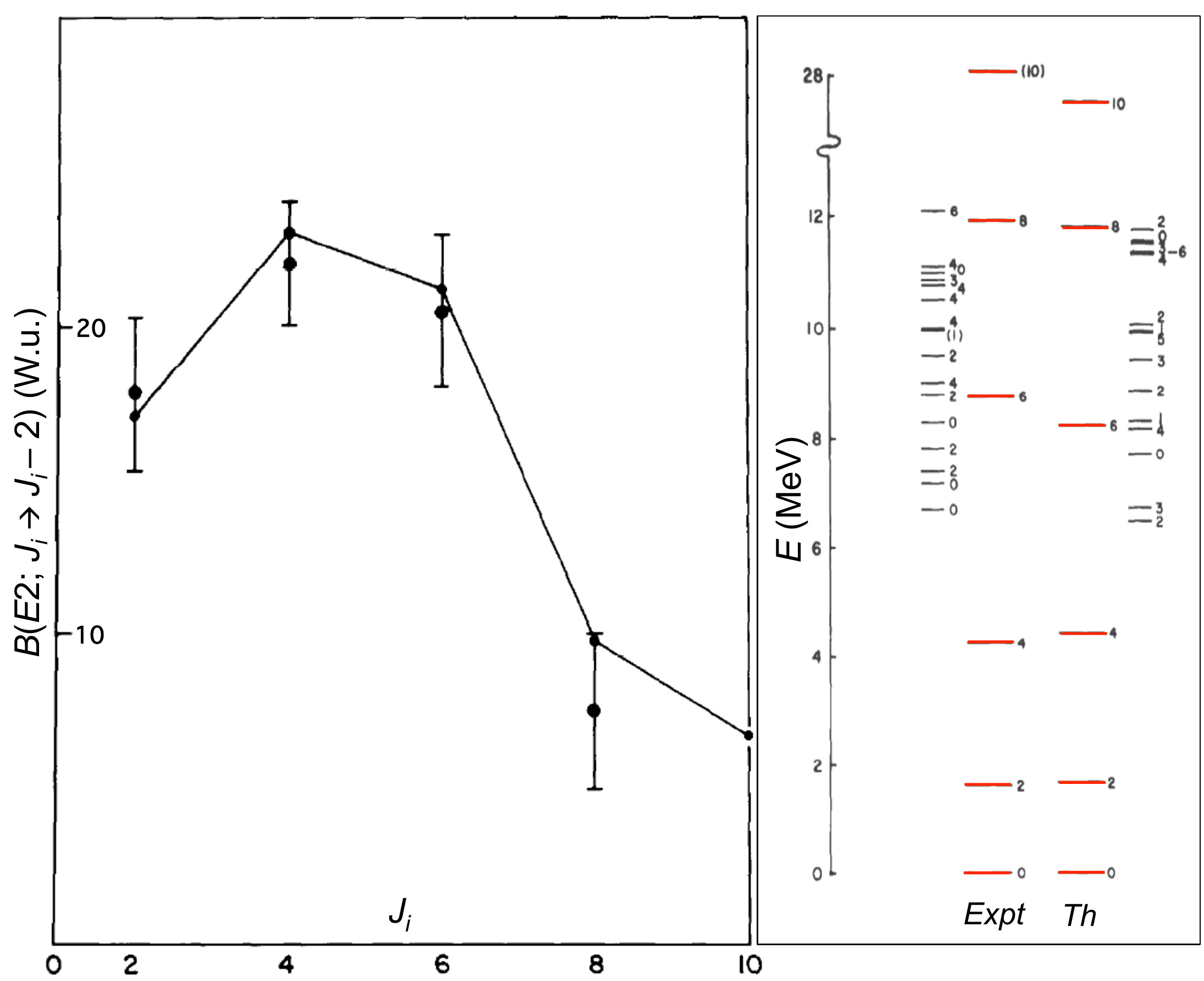}
\hspace{0.1in}
\includegraphics[width=0.45\textwidth]{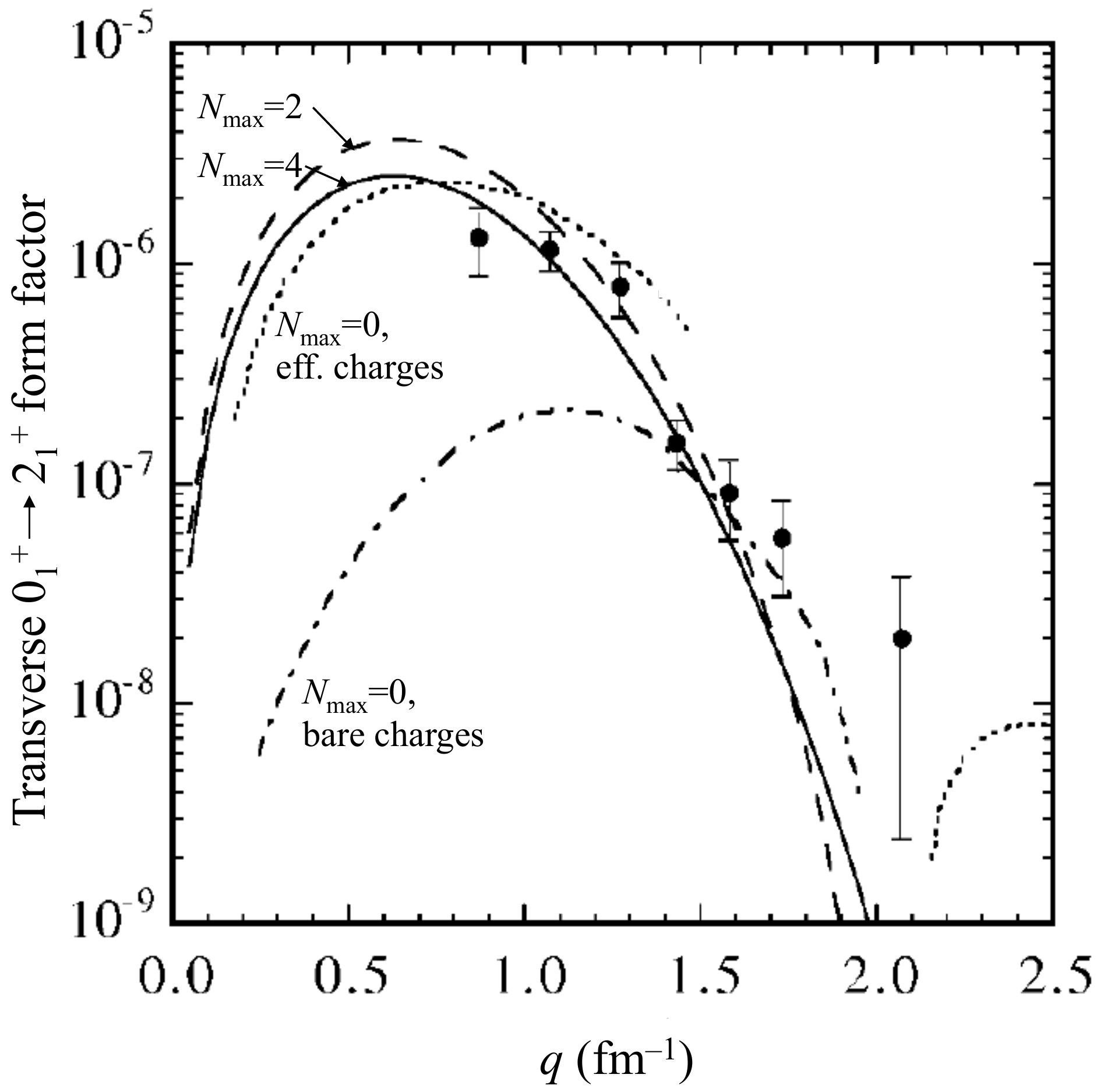}
}
\hspace{2in}(a) \hspace{3.5in} (b)
\caption{ 
(a) Microscopic symplectic model with a set of effective single-particle energies, a $Q\cdot Q$-type interaction$+$pairing  for $^{20}$Ne. Left: Calculated $B(E2\downarrow)$ transition strengths {\it without effective charges} fall within the uncertainties of the corresponding experimental measurements. Right: Calculated  energy spectrum for $^{20}$Ne  as compared to the experiment. Figure adapted from Ref. \cite{DraayerWR84}.
(b) Transverse $0_1^+ \rightarrow 2_1^+$ electron scattering form factors for $^{24}$Mg in the symplectic model for $N_{\rm max}=2$ and $4$ {\it with no effective charges}, as compared to valence-shell calculations with bare and effective charges \cite{Hotta87}
and experiment \cite{Hotta87,LiYS74}.
Form factors are corrected for the center-of-mass motion and the finite-size effect of the
nucleon. Figure adapted from Ref. \cite{EscherD99}.
}
\label{sd_Sp}      
\end{figure}

Another successful extension to  multiple shells has been achieved and applied to the $^{24}$Mg ground-state rotational band \cite{PetersonH80}, where an interaction given as a polynomial in $Q$ up through $(Q\cdot Q)^2$ was employed.  Furthermore, a shell-model study in a symplectic basis that allows for  mixing of \SpR{3} irreps due to pairing  and non-degenerate single-particle energies above a $^{16}$O core  \cite{DraayerWR84}  has found that using only seven \SpR{3} irreps, which extend up through 15 HO shells,  is sufficient to achieve a remarkable reproduction of the $^{20}$Ne energy spectrum,  as well as of $E2$ transition rates without effective charges  (Fig. \ref{sd_Sp}a). Recently, an \SpR{3}-based study using self-consistent arguments has been successful to give further insight into  the  cluster states of $^{16}$O and shape-coexistence \cite{RoweTW06}.  

Electron scattering form factors for transitions between low-lying states of $^{24}$Mg have been calculated in
the symplectic shell model using up to 4-body $Q$- and $L$-dependent  interactions  \cite{EscherD99}. Parameters of the Hamiltonian
have been fitted to reproduce measured energies and reduced
transition probabilities in up to $N_{\rm max}=20$ model spaces, and no further adjustments have been made
in obtaining the predicted form factors. The symplectic form factors
demonstrate excellent agreement with the available data (Fig. \ref{sd_Sp}b), indicating that larger-$N_{\rm max}$ spaces and associated correlations play a substantial role
in describing nuclear current and charge densities.

\subsubsection{No-core symplectic shell model (NCSpM) and the elusive Hoyle state \label{NCSpM}}
Using the \SpR{3} scheme,  the no-core symplectic shell model (NCSpM) \cite{DreyfussLTDB13}  offers $N_{\rm max}=12-24$ shell-model descriptions of low-lying states in deformed $sd$-shell nuclei ($^{20}$O, $^{20,22}$Ne, and $^{20,22,24}$Mg)  \cite{TobinFLDDB14} and of  phenomena tied to giant monopole and quadrupole resonances \cite{DreyfussLTDBDB16}, as well as to collectivity and alpha-clustering in $^{12}$Be  \cite{LauneyDDTFLDMVB13} and $^{12}$C, and in particular, the challenging Hoyle state and its first $2^+$ and $4^+$ excitations \cite{DreyfussLTDB13,LauneyDDDB14}. While such ultra-large model spaces remain inaccessible by {\it ab initio} shell models, the NCSpM  addresses a long-standing challenge \cite{EllisE70,EngelandE72,SuzukiH86}, namely, understanding highly-deformed spatial configurations from a shell-model perspective. Our present-day knowledge of various phenomena of astrophysical significance, such as nucleosynthesis, the evolution of primordial stars in the Universe, and X-ray bursts  depends on reaction rates for the stellar triple-$\alpha$ process, which can considerably affect, e.g., results of core-collapse supernovae simulations and stellar evolution models, predictions regarding X-ray bursts,  as well as estimates of carbon production in asymptotic giant branch (AGB) stars \cite{Fynbo05}. These rates, in turn, are greatly influenced by accurate measurements and theoretical predictions of several important low-lying states in $^{12}$C, including the second $0_2^+$ (Hoyle) state and its $2^+$ excitation that has fostered long-lasting debate in experimental studies \cite{Freer0709,Hyldegaard10,Itoh11,ZimmermanDFGS11,Raduta11,Zimmerman13, Marin14}. Further challenges relate to the  $\alpha$-cluster substructure of these states that has been  explored within cluster-tailored  \cite{KanadaEnyo98,FunakiTHSR03,YamadaS05,ChernykhFNNR07,KhoaCK11,NeffF14} or self-consistent \cite{ZamickZLCG91,UmarMIO10} framework, but has hitherto precluded a fully microscopic {\it ab initio} no-core shell-model  description  \cite{RothLCBN11} (see, e.g., detailed reviews on the topic \cite{HoriuchiIK12,FunakiHT15}). Only recently, first {\it ab initio} state-of-the-art calculations have been attempted  using lattice  effective field theory (EFT) \cite{EpelbaumKLM11,EpelbaumKLLM12}.
\begin{figure}[t]
\begin{center}
\begin{minipage}{0.57\textwidth}
\includegraphics[width=1\textwidth]{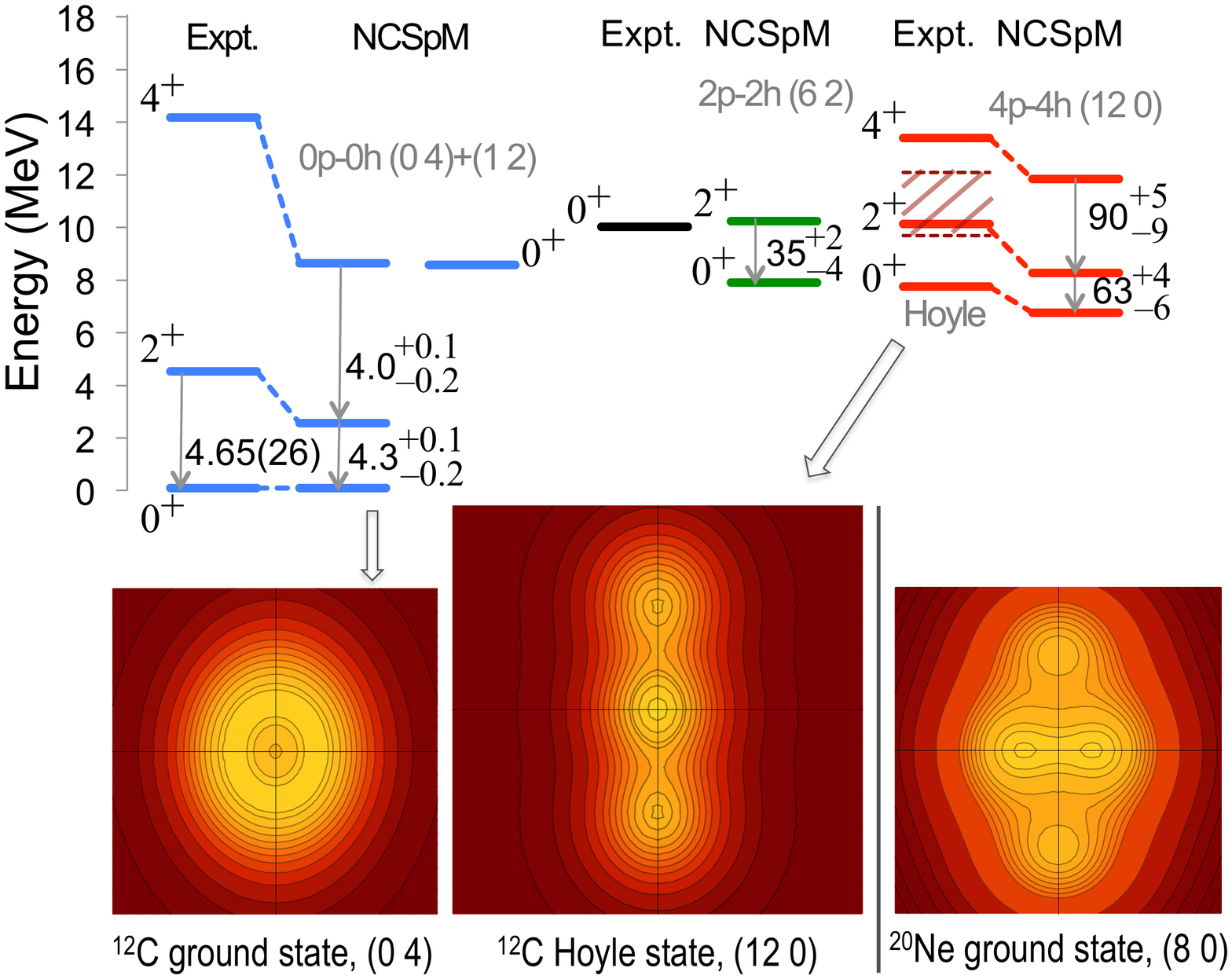}\\
\end{minipage}
\hspace{0.03in}
\begin{minipage}{0.41\textwidth}
\includegraphics[width=0.95\textwidth]{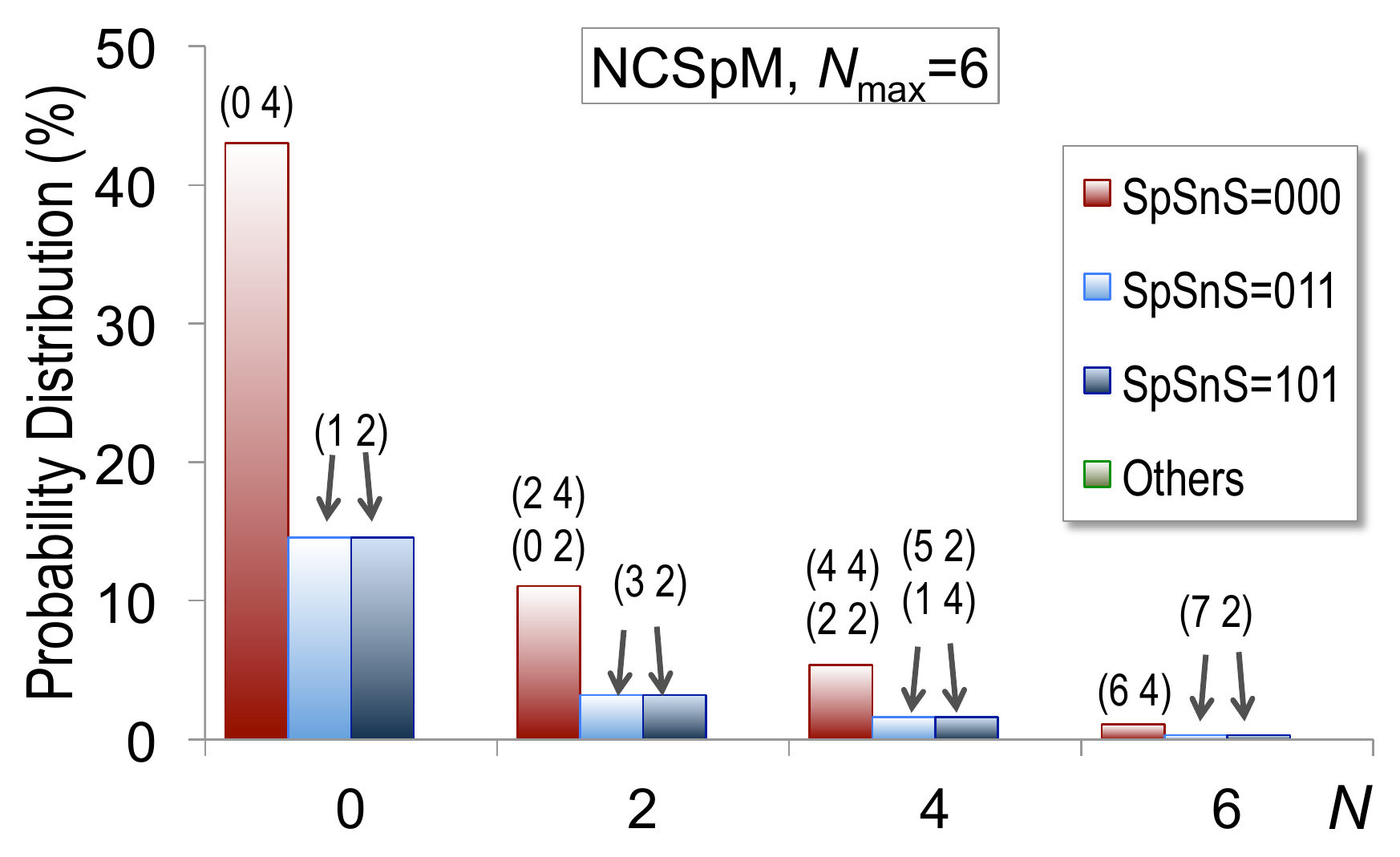}\\
\includegraphics[width =0.95\textwidth]{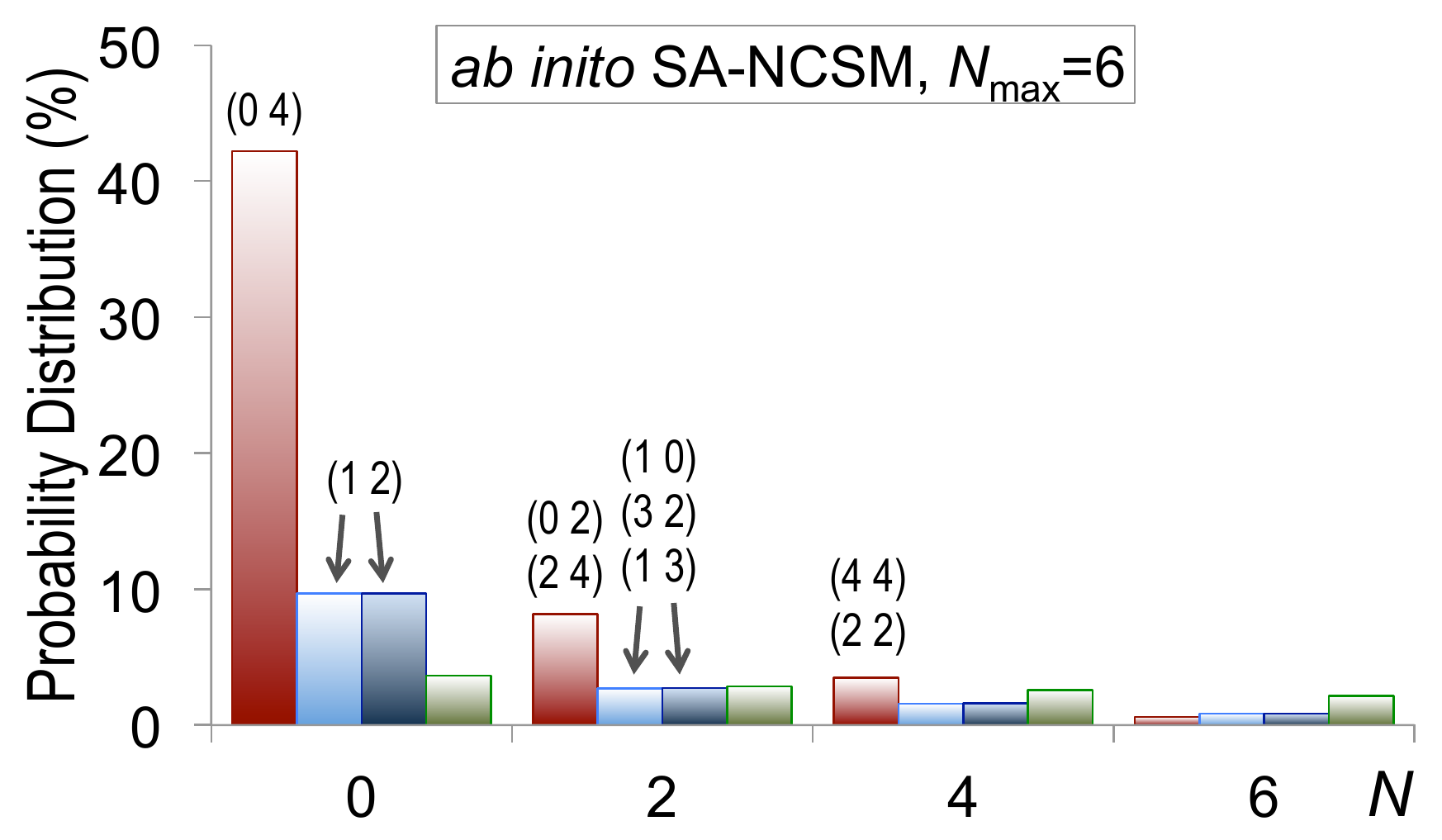}
\end{minipage}\\
\hspace{0.7in} (a)\hspace{3.5in}(b)
\end{center}
\caption{
(a) $N_{\rm max}=20$ NCSpM energy spectrum of $^{12}$C. 
Experimental data  is from \cite{ASelove90}, except the latest results for  $0_3^+$ \cite{Itoh11} and the states above the Hoyle state, $4^+$  \cite{Freer11} and $2^+$  \cite{Zimmerman13} (with a shaded area showing the  energy range from \cite{Freer0709,Hyldegaard10,Itoh11,ZimmermanDFGS11,Raduta11}). 
$B(E2)$  transition rates are in W.u.; theoretical uncertainties are estimated for a $\pm 60\%$ deviation of the Hoyle state energy. One-body density profiles {\it in the intrinsic frame} are shown for the $0^+_{gs}$ (for the most dominant symplectic irrep) and the Hoyle state  in $^{12}$C, and the ground state in $^{20}$Ne. Note the torus-like shape for the $^{12}$C $0^+_{gs}$ (a dip in the middle), and the overlapping clusters in the $^{12}$C Hoyle state.
Figure adapted from Ref. \cite{DreyfussLTDB13}. 
(b)
Probability distribution for the $^{12}$C ground state  vs. the $N$ total excitations, as calculated by NCSpM (top) and SA-NCSM (bottom) for $N_{\rm max}=6$ and $\ho=18$ MeV. The dominant  \SU{3} modes (with probability amplitude $ \ge 1\%$), specified by $(\lambda\,\mu)$, are also shown. Very similar results are obtained for $2_1^+$ and $4_1^+$ \cite{DreyfussLTDBDB16}.
}
\label{enSpectrumC12}
\end{figure}
\begin{table}[th]
\caption{ NCSpM point-particle rms matter
radii and electric quadrupole moments for $^{12}$C (for $N_{\rm max}=20$) and $sd$-shell nuclei (for $N_{\rm max}=12$) \cite{DreyfussLTDB13 ,TobinFLDDB14} compared to experimental (experimentally deduced) data. See text for a comparison to $r_{\rm rms}$ predictions  of other models. Experimentally deduced matter radii  are summarized in Ref. \cite{OzawaST01} and each of the original references is provided in the table; measured $Q$ moments are taken from Refs. \cite{Tilley98A20,Firestone05A22} for $A=20$ and $22$, respectively.
}
{\footnotesize
\begin{tabular}{llllll}
\hline
&&	 \multicolumn{2}{l}{matter   $r_{\rm rms}$, fm} 					&	 \multicolumn{2}{l}{$Q$, $e\,$fm$^2$} \\			
&	&	  Expt. 	&	 NCSpM 	 			&	  Expt. & NCSpM \\ 
			\hline									
$^{12}$C		&	$0^+_{gs}$  	&	 $2.43(2)^{a}$ 	&	 $2.43(1)$ 	&		&	 \\	
	&	$2_1^+$  	&	 $2.36(4)^{b*}$ 	&	 $2.42(1)$  	&	 $+6(3)^{d}$ 	&	 $+5.9(1)$\\	
	&	$4_1^+$  	&	--	&	 $2.41(1)$  	&	 -- 	&	 $+8.0(3)$\\	
	&	$0^+_{2}$  {\scriptsize (Hoyle)}  	&	 $2.89(4)^{b*}$  	&	 $2.93(5)$ 	&		&	 \\	
	&	 $2^+$ above $0^+_{2}$  	&	 $3.07(13)^{c*}$ 	&	 $2.93(5)$  	&	  --  	&	 $-21(1)$\\	
	&	$4^+$ above $0^+_{2}$  	&	--	&	 $2.93(5)$  	&	  --  	&	 $-26(1)$\\	
	&	$0^+_{3}$ 	&	--	&	 $2.78(4)$  	&		&	 \\	
$^{20}$O		&	$0^+_{gs}$  	&	$	2.69(3)^e	$	&	$	2.73	$	&		&	\\	
	&	$2_1^+$ [$4_1^+$]	&		&		&	 -- 	&	$	-8.45	$ [$ -11.11 $]	\\
$^{20}$Ne		&	$0^+_{gs}$  	&	$	2.87(3)^e	$	&	$	2.79	$	&		&		\\
	&	$2_1^+$ [$4_1^+$, $6_1^+$    ]	&		&		&	$	 -23(3)	$	&	$	-15.69	$ [$ -19.69 $, $ -21.05 $]	\\
$^{22}$Ne		&	$0^+_{gs}$  	&		--		&	$	2.82	$	&		&		\\
	&	$2_1^+$ [ $4_1^+$, $6_1^+$    ]	&		&		&	$	-17(3)	$	&	$	-14.90	$ [$ -19.22 $, $ -21.61 $]	\\
$^{20}$Mg		&	$0^+_{gs}$  	&	$	2.88(4)^e	$	&	$	2.73	$	&		&		\\
	&	$2_1^+$ [$4_1^+$  ]	&		&		&		--		&	$	-12.67	$ [$ -16.67 $]	\\
$^{22}$Mg		&	$0^+_{gs}$  	&	$	2.89(6)^f 	$	&	$	2.82	$	&		&		\\
	&	$2_1^+$ [$4_1^+$, $6_1^+$  ]	&		&		&	--	&	$ -17.88 $ [$ -23.07 $, $ -25.93 $]	\\
$^{24}$Mg		&	$0^+_{gs}$  	&	$	2.97(12)^g 	$	&	$	3.03	$	&		&		\\
	&	$2_1^+$	&		&		&	$-16.6(6)^d$	&	$ -22.7 $ 	\\
 \hline 
 \end{tabular}
}\\
{\scriptsize $^a$Ref. \cite{Tanihata85};  $^b$Ref. \cite{DanilovBDGO09}; $^c$Ref. \cite{Ogloblin13}; $^d$Ref. \cite{ASelove90}, $^e$Ref. \cite{OzawaST01ref28}; and $^f$Ref. \cite{OzawaST01ref19}; and $^g$Ref. \cite{VernotteBKT82}.}\\
{\scriptsize *Experimentally deduced, based on model-dependent analyses of diffraction scattering; $0^+_{gs}$ $r_{\rm rms}= 2.34$ fm.}
\label{Observables}
\end{table}
%%%%%%%%%

With the goal to inform key features of nuclear structure and the interaction, enhanced collectivity and cluster substructures are studied in the NCSpM by down-selecting, first, to the most physically relevant nuclear configurations and, second, to pieces of the nucleon-nucleon ($NN$) interaction that enter in commonly used nuclear potentials \cite{Elliott58, Elliott58b,Harvey68,BohrMottelson69,KanadaEnyo98}. Specifically, the physically relevant symplectic irreps are chosen among all possible symplectic \SpR{3} irreps within an $N_{\rm max}$ model space (e.g., 4 irreps for $^{12}$C extended up to $N_{\rm max}=20$ of dimensionality of $6.6\times10^3$, shown in Fig. \ref{sp3Rpicture}). For the interaction, the long-range part of the central $NN$ force and a spin-orbit term  are considered (better accounting of the symplectic symmetry mixing has been explored in Ref. \cite{LauneyDDDB14} with preliminary results that use the bare JISP16 $NN$ instead of the spin-orbit term).  The interaction is augmented by $e^{-\gamma Q\cdot Q}$, with a single adjustable parameter $\gamma$ that controls the contribution of the many-nucleon forces \cite{LeBlancCVR86}.  While short-range and tensor forces are indispensable for accurate descriptions, they appear to be of secondary importance to the primary physics responsible for the formation of clusters and the emergence of collectivity, as suggested by  the reasonably close agreement of  the model outcome with experiment and {\it ab initio} results in smaller spaces (Fig. \ref{enSpectrumC12}). 
\begin{figure}[t]
\centering
\includegraphics[width =0.47\textwidth]{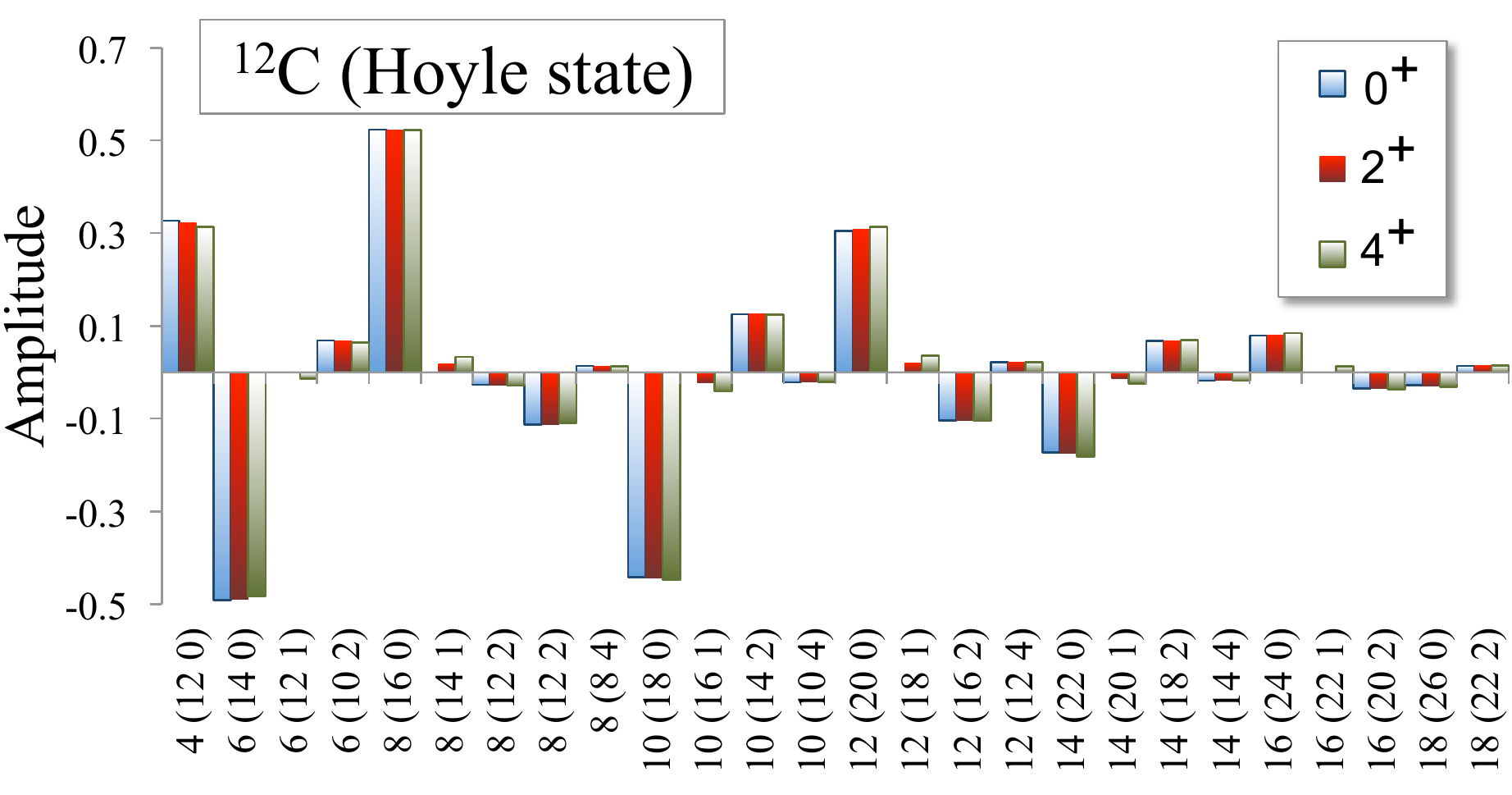} 
\includegraphics[width=.5 \textwidth]{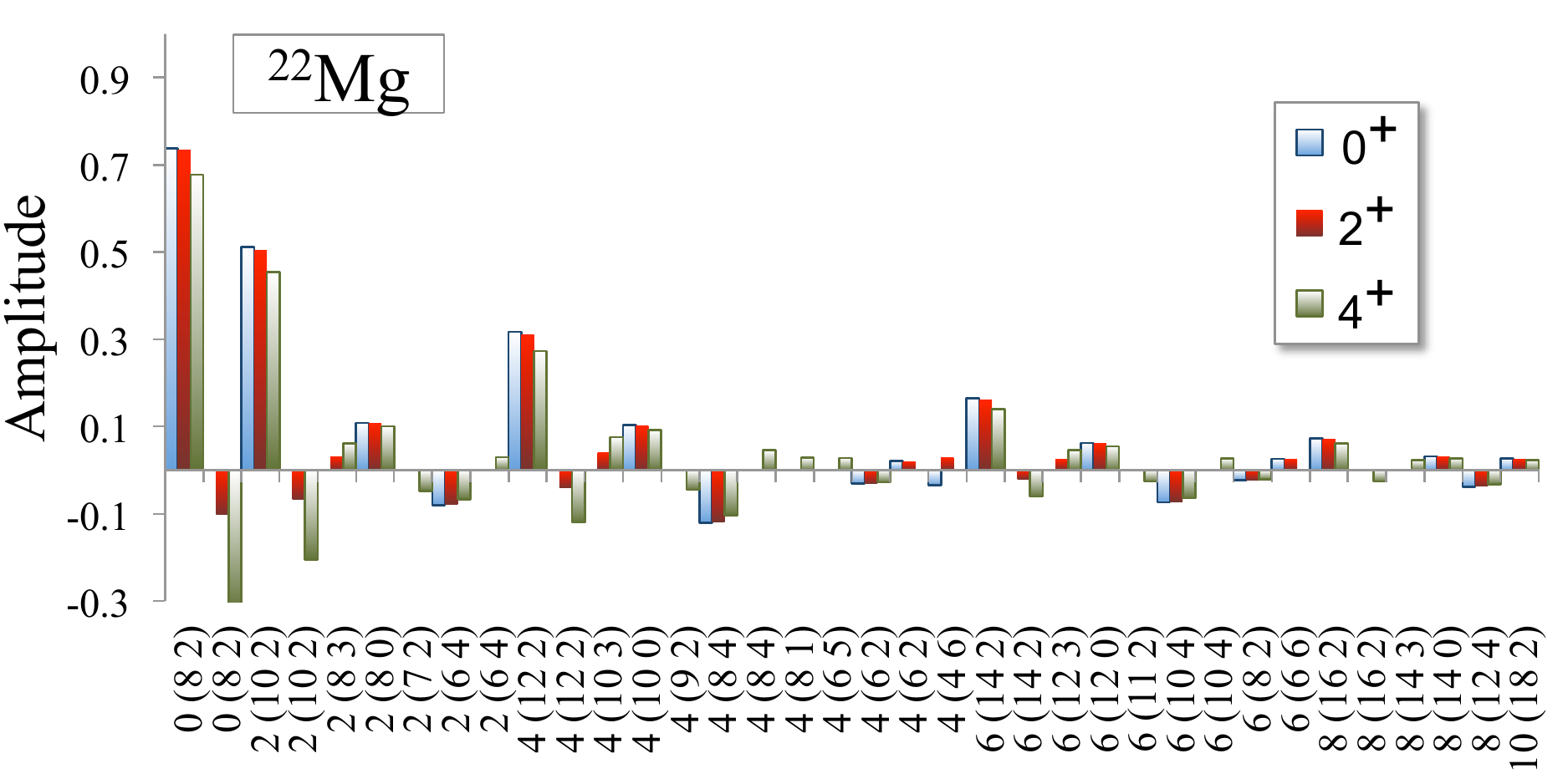} \\ 
(a) \hspace{3.5in}(b)
\caption{
NCSpM amplitudes  of relevant symplectic basis states (horizontal axis) labeled by $N(\lambda_\omega\,\mu_ \omega)$  for the $0^+$ (blue) and the excited $2^+$ (red) and $4^+$ (green) states in the rotational  band for  (a) the Hoyle state in $^{12}$C and (b) the ground state of $^{22}$Mg \cite{TobinFLDDB14, LauneyDDTFLDMVB13}. 
Multiplicities to distinguish repeated $N(\lambda_\omega\,\mu_ \omega)$ labels are not shown. Note that the Hoyle-state probability distribution peaks at 8\ho.
}
\label{wvfnProbabilities}
\end{figure}

Once the $\gamma$ parameter is fixed by the order  of the three low-lying $0^+$ states in $^{12}$C,  excitation energies and other observables such as matter rms radii, electric quadrupole moments and $E2$ transition rates for various $p$- and $sd$-shell nuclei are found,  with no parameter adjustment,  to be in a remarkable agreement with the experiment  \cite{DreyfussLTDB13,TobinFLDDB14, LauneyDDTFLDMVB13} (see also Fig. \ref{enSpectrumC12}  and Table \ref{Observables}).   For $^{12}$C, the NCSpM realizes a very reasonable $r_{\rm rms}$ 
for the ground state, and  the Hoyle-state  $r_{\rm rms}$ is found to lie close to a recent value deduced from experiment \cite{DanilovBDGO09}, and as well tracks with the {\it ab initio} lattice EFT results at a leading order \cite{EpelbaumKLM11}, but differs considerably from predictions of cluster models, e.g., 3.4-4.3 fm  \cite{ChernykhFNNR07, FunakiTHSR03,YamadaS05}.
Furthermore, the model yields a positive $Q_{2_1^+}$ very close to the experimental value, and a large negative one for the $2^+$ and $4^+$ states above the Hoyle state (Table \ref{Observables}), which as mentioned above, indicates oblate (prolate) deformation for the ground-state (Hoyle-state) rotational band. This, together with the dominance of prolate $(\lambda\,\mu)$ configurations in the Hoyle state, $(12\,0)$, $(14\,0)$, $(16\,0)$, $(18\,0)$, and $(20\,0)$ (Fig. \ref{wvfnProbabilities}a) and the $2^+$ and $4^+$ states above it,  indicates a substantial prolate deformation for these states. Such a deformation, albeit not  so pronounced, has been also suggested  by the {\it ab initio} lattice EFT \cite{EpelbaumKLM11}.  Indeed, the one-body density profile for the Hoyle state  very clearly supports an underlying $\alpha$-particle, cluster-like picture for its structure (Fig. \ref{enSpectrumC12}a). We emphasize that this cluster structure is essentially very different from a simple $\alpha$ chain suggested by cluster models, as the clusters partially overlap. 

For intermediate-mass nuclei, we find that it is imperative that model spaces be expanded well beyond the current limits up through 15 major shells to accommodate particle excitations, which
appear critical to highly deformed spatial structures (Figs. \ref{enSpectrumC12}a and \ref{wvfnProbabilities}b) and the convergence of associated observables, as detailed in Ref.  \cite{TobinFLDDB14}.

\subsection{Pseudo-spin symmetry for heavy nuclei \label{pseudoSU3}}
%Pseudo-SU(3) and Pseudo-Sp(3,R)\\
%Work by Hecht, Arima, Draayer, Blokhin

For heavy nuclei ($A\gtrsim 100$), the discovery of the pseudo-spin
symmetry\cite{HechtA69,ArimaHS69} together with its fundamental nature\cite{Draayer91,BahriDM92,BlokhinBD95} has
established the pseudo-\SU{3} model\cite{RatnaRajuDH73}. In particular, the microscopic origin of the pseudo-spin symmetry has been unveiled in Ref. \cite{BlokhinBD95}, which has identified that the  many-particle $p$-helicity operator generates a transformation to the pseudo-spin basis in heavy nuclei, while satisfying all other global symmetry requirements. Both mean-field and many-particle estimates demonstrate that in the helicity-transformed representation, the nucleons move in a finite-depth non-local potential with a reduced spin-orbit strength.  
Furthermore, the approximate independence of the single-nucleon spectrum in an infinite medium on the helicity transformation and the consistency of the microscopic estimates for the single-particle nuclear potentials with the Dirac-Brueckner calculations, allow one to connect the pseudo-spin symmetry to the one boson-exchange (OBEP) nature of the nucleon-nucleon $NN$ interaction. Based on this link and because of the close relation (coincidence in the chiral symmetry limit) of the helicity and chirality operations, the goodness of pseudo-spin symmetry may be expected to increase with rising densities (or energy per particle) in hadronic systems, and actually yield to the chiral symmetry of massless hadrons in the high energy region  \cite{BlokhinBD95}.

\begin{figure}[th]
\centerline{
\includegraphics[width=0.75\textwidth]{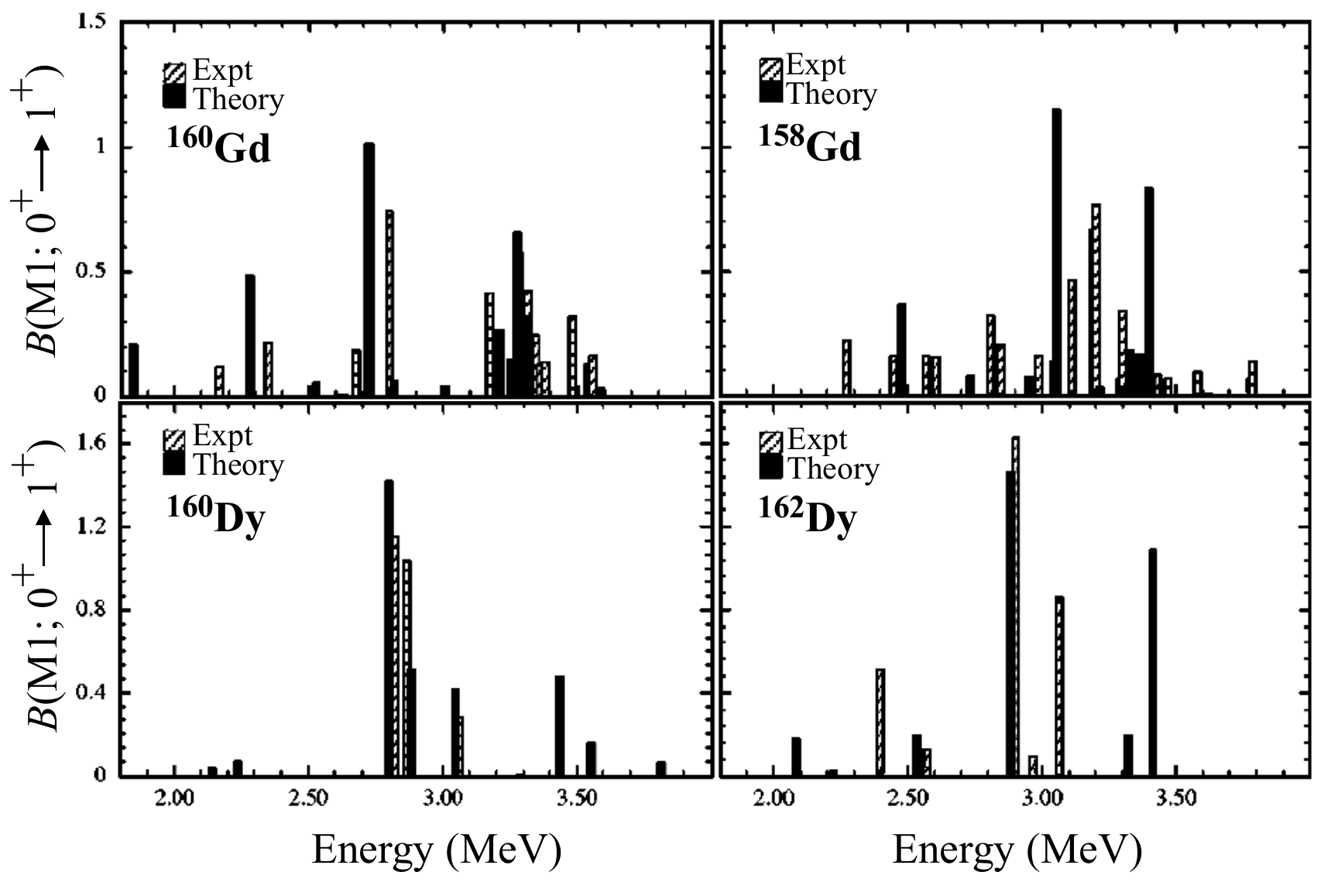}
}
\caption{
Comparison of experimental (crosshatched bars) \cite{KneisslPZ96} and theoretical (solid bars) $M1$ strength distributions. In each case the
eigenstates were determined by fitting parameters in the Hamiltonian to the experimental energy spectrum and associated $B(E2)$
transitions. 
Figure adapted from Ref. \cite{BeuschelDRH98}.
}
\label{M1strength_pseudoSU3}
\end{figure}

Pseudo-spin scheme is an excellent starting point for a many-particle description of heavy nuclei,
whether or not they are deformed. As for the \SU{3} shell model, in many cases leading-irrep
calculations (e.g., see \cite{DraayerW83}) or mixed-irrep calculations (e.g., see \cite{PopaHD00})
achieve good agreement with experimental data. The pseudo-\SU{3} shell model provides a further
understanding of the $M1$ transitions in nuclei such as the even-even
$^{160-164}$Dy and $^{156-160}$Gd isotopes, specifically it reflects on the scissors and twist
modes as well as the observed fragmentation (Fig. \ref{M1strength_pseudoSU3}), that is, the break-up of the $M1$ strength among several
levels closely clustered around a few strong transition peaks in the 2-4 MeV energy
region \cite{BeuschelDRH98} (and references therein; for a detailed review on magnetic dipole excitations in nuclei, see Ref. \cite{HeydeNR10}).
In medium-mass and heavy nuclei, where the pseudo-spin valence space is intruded by the highest-$j$ orbit from the shell
above, a major step towards understanding the significance of the intruder level is achieved by the
pseudo-\SU{3} plus intruder level shell model\cite{WeeksHD81}.

Furthermore, the advantages of the symplectic \SpR{3} extension of the \SU{3}
model can be employed beyond the light nuclei domain towards a description of heavy nuclei in the
framework of the pseudo-\SpR{3} shell model. For example, this model with a symmetry-preserving (in terms of the pseudo-$Q$ and pseudo-$L$) two-, three-, and four-body interaction  has been applied to $^{238}$U with a focus on the energy spectrum and $B(E2)$ transition strengths from $J_i=2,4,6,8,10,$ and $12$ to $J_i-2$ within the ground-state rotational band  \cite{CastanosHDR91}. In this study, the experimental $B(E2)$ values are remarkably  well reproduced without the need of an effective charge. The results are comparable to a valence-shell pseudo-\SU{3} model with effective charges, however, larger discrepancies have been observed for  the higher-lying states \cite{CastanosHDR91}. While a pioneer work has revealed the power of this model for heavy nuclear systems, it still remains not fully explored.

\subsection{O($A-1$) scheme and \textit{ab initio} hyperspherical harmonics method \label{oA}}

As shown in Eq. (\ref{phasespace}) of  Sec. \ref{sp3r}, the complete translationally invariant shell-model basis is classified according to a reduction chain of \SpR{3(A-1)}$\times$\Un{4}, which for the spatial degrees of freedom, invokes $\SpR{3} \times  \On{A-1}$.    While the \SpR{3} scheme utilizes the \SpR{3} group (transformations acting on $x$, $y$, and $z$ components), one can organize the $A$-particle model space according to the complementary group   $\On{A-1}$, with $O(A)\supset \On{A-1}\supset S_A$. The \On{A} is the group of orthogonal transformations that act on the ``particle-index" space (transformations of nucleon coordinates, $x_{i\alpha} \rightarrow \sum_{j=1}^A x_{j \alpha}\omega_{ji}$, that leave the O(A) scalars $x_\alpha \cdot x_\beta=\sum_{i=1}^Ax_{i\alpha}x_{i\beta}$ invariant for $\alpha,\beta=x,y,z$). This scheme is reviewed in detail in Refs. \cite{Rowe85,Rowe96}.
\On{A-1} is the subgroup of \On{A} which leaves center-of-mass coordinates invariant (note that
center-of-mass coordinates are symmetric with respect to nucleon indices and, therefore, invariant
under $S_A$ permutations) and has as a subgroup the permutation group $S_A$, which permutes the spatial coordinates of a system of $A$ particles.

This scheme underpins the Hyperspherical Harmonics (HH) method. The HH method is a variational method where the trial function is written as an expansion on the HH basis \cite{Borel14}, 
which was first applied to  the Helium atom \cite{Gronwall37}
and to the $^3$H ground state \cite{Clapp49} with central and tensor forces, 
and further advances have been developed in Refs. \cite{Simonov66, NeudachinS69}.
The hyperspherical harmonics are the $A$-body generalization of the spherical harmonics $Y_{lm}$, and nuclear orbital wave functions can be expressed as products of functions of a global radius (hyper-radius) $\rho^2 =\sum_{i\alpha}\eta_{i \alpha}^2$ [$\vec \eta_i$ are the $A-1$ Jacobi coordinates] and \On{3A - 3} hyperspherical harmonics.
The symmetrization of the hyperspherical harmonics presents one of the main difficulties of the method. 
For $A>4$ systems, a direct symmetrization becomes impractical and more sophisticated approaches are adopted, such as the construction of HH based on $\On{3A-3}\supset \SO{3} \times \On{A-1}$ (e.g., see \cite{Filippov73,VanagasK73} and the extensive reviews of the \On{A-1}-based theories \cite{Vanagas76, Filippov78}).
An efficient technique has been developed in Ref. \cite{Barnea97}
to solve the reduction problem
from $\On{A-1} \supset S_A$. Using central $NN$ interactions,  this technique has allowed applications of the HH method  for more
than four fermions, namely, for the binding energies of  $^6$Li,  $^8$Be, and $^{12}$C  \cite{BarneaLO99},
while an alternative non-recursive approach, based on an HH expansion in terms of the Slater-determinant basis of the HO shell model, 
has been applied to $^{3-7}$H and $^{4-10}$He isotopes  \cite{Timofeyuk02, Timofeyuk04}.
Similarly to the NCSM with effective interactions, the Effective Interaction for Hyperspherical Harmonics (EIHH) method has been developed \cite{BarneaLO01}, with the unitary transformation applied with realistic potentials at the two-/three-body level. 
The complications with the antisymmetrization problem for the HH functions, however, have limited the model applications to $A < 8$ \cite{LeidemannO13}.
A fully converged {\it ab initio} result, achieved in the framework of the  hyperspherical harmonics  method, of $24.23$ MeV for the $^4$He binding energy with the realistic AV14 potential \cite{WiringaSA84}
has been reported in Ref. \cite{Kievsky08}.  Further details on the HH method and its applications can be found in the review \cite{LeidemannO13}.

As discussed in Ref. \cite{Rowe96}, although the \SpR{3} and \On{A - 1} approaches to a shell-model theory are
complementary and compatible with one another, they play fundamentally different roles. The groups \On{A - 1} $\supset S_A$ are symmetry groups. They
provide good quantum numbers, which, by duality, define an \SpR{3} irrep, and which are shared by
every state within an \SpR{3}  irrep.  The \On{A - 1}  describes the intrinsic component of the many-particle wave function (particle distribution), while the complementary \SpR{3} wave function describes the collective component (deformation of the nuclear system) \cite{Rowe85}. In addition, while a general \On{A - 1}  transformation does not leave the nuclear Hilbert space invariant (the nuclear Hilbert
space contains only part of an \On{A - 1} representation space, the part with the correct permutation
symmetry), the  \SpR{3} representation space lies completely within the nuclear Hilbert space \cite{Rowe96}.

\subsection{Wigner \SU{4} supermultiplet and alpha-clustering in nuclei \label{su4}}
The spin-isospin degrees of freedom of $A$ nucleons, complementary to the 3-dimensional coordinate space degrees of freedom as shown in the classification (\ref{phasespace}),  are described by \SU{4}, called the Wigner supermultiplet group \cite{Wigner37}, with \Un{4}=\SU{4}$\times$ \Un{1}. It ``rotates" the four nucleon degrees of freedom as four components of an \SU{4} multiplet, $\ket{1}=\ket{p\uparrow}$, $\ket{2}=\ket{n\uparrow}$, $\ket{3}=\ket{p\downarrow}$, and $\ket{4}=\ket{n \downarrow}$. The supermultiplet classification  is detailed in Ref. \cite{HechtP69}, with basis states further labelled by the spin (S)
and isospin (T) quantum numbers of the reduction chain, $\SU{4} \supset \SU{2}_S \times \SU{2}_T$.  Isospin is not  exactly conserved in nuclei but is  only slightly broken and can be treated as an exact symmetry. 

As discussed in the review article \cite{Lee09}, \SU{4} symmetry arises naturally in the limit of large number of colors \cite{KaplanS96,KaplanM97,CalleC08}.
In this limit, one can view the symmetry as arising from a combinatorial enhancement
of interaction terms which are spin and flavor independent \cite{KaplanS96, CalleC08}.
Furthermore, recent lattice
QCD simulations have shown that SU$(4)$ symmetry becomes increasingly accurate
at heavier quark masses \cite{Beane13, BeaneCDOPS15}.
The low-energy nuclear interactions show an approximate \SU{4} symmetry in the $S$-wave
scattering channels. While the Coulomb interaction and one-pion exchange interaction break this
 \SU{4} symmetry, the  short-distance part of the $S$-wave nucleon-nucleon interactions obey the symmetry
rather well \cite{MehenSW99, EpelbaumMGE02}.  

For nuclear systems, early applications have identified cases where the \SU{4} symmetry is approximately valid (e.g., see \cite{FranziniR63}). In Ref. \cite{HechtD74}, an important  measure for the goodness/symmetry-breaking of the Wigner supermultiplet symmetry has been introduced, based on spectral distribution theory (SDT, discussed in Sec. \ref{sdt_NN}) \cite{French66, FrenchR71,ChangFT71}. In this study, the \SU{4}  symmetry-breaking has been investigated for the $A = 25$ $sd$-shell  nuclei, which by the complementary nature of the space and spin-isospin symmetries,   has provided a simple measure of the amount of mixing to be expected between states of different space symmetry. This study has pointed to the dominance of the  highest spatial symmetry, while mixing of a few other irreps have been suggested to be linked to collective degrees of freedom. Namely, a dominant \SU{3} irrep can belong to different \Un{\Omega_\eta} irreps (see Table \ref{UNtoSU3}), which as a result mix \cite{HechtD74}. Similarly, some \SU{4} fragmentation has been recently observed for $^{11}$B within the {\it ab initio}  no-core shell-model framework  \cite{Johnson16}.
Medium- to heavy-mass nuclei are found to exhibit significant fragmentation of the wave function over many \SU{4} irreps  (e.g., 
 \cite{VogelO93,FrazierBMZ97}), with results suggesting evidence of coherent quasi-dynamical symmetry within a rotational band \cite{Johnson16}.

%%%%%%%LEFT
\subsubsection{Symmetry-guided techniques in the \textit{ab initio} lattice effective field theory \label{latticeEFT}}
The importance of an approximate \SU{4} symmetry of the low-energy nucleon-nucleon interactions  is explored in \textit{ab initio} lattice
simulations for nuclear systems using chiral effective field theory  \cite{EpelbaumKLM11}. While these simulations are not based on the shell-model theory, we briefly review the symmetry-guided techniques used in this approach.

Indeed, there is experimental evidence that some predictions that one can derive
from SU$(4)$ symmetry  are well satisfied by the spectrum of light nuclei
\cite{ChenDS04,Lee07} (for an extensive review, see \cite{Lee09}).  In Ref. \cite{Lee07},
a general theorem on spectral convexity with respect to particle number $A$ for $2k$ degenerate components of fermions has been derived that only assumes that the  interactions are governed by an \SU{2k}-invariant two-body potential with  negative-definite Fourier transform. It shows that the ground state of any fermionic system with such potentials  is in a $2k$-particle clustering phase [for \SU{4}, it implies an $\alpha$ clustering phase] and obeys a set of spectral convexity bounds.
Indeed, as shown in \cite{Lee07}, all of the \SU{4} convexity constraints are satisfied for nuclei up to $A=20$.  Furthermore, recent lattice simulations have also shown alpha cluster structures consistent with an \SU{4}-clustering phase in $^{12}$C  \cite{EpelbaumKLM11, EpelbaumKLLM12, EpelbaumKLLM13},
$^{16}$O \cite{EpelbaumKLLMR14}, and other $sd$-shell alpha-like nuclei \cite{LahdeEKDMR14}.
These results give further evidence that an approximate description of light nuclei may be possible using an attractive \SU{4}-symmetric potential. 
As noted in \cite{Lee07}, these results do not imply that Monte Carlo simulations of nucleons using chiral effective theory can be performed without sign or phase oscillations, but they suggest that the simulations are possible with only relatively mild cancelations, given the approximate \SU{4} symmetry
and attractive interactions at low energies.

The projection Monte Carlo method with auxiliary fields has been used to study low-energy nucleons in chiral effective
field theory \cite{BorasoyEKLM07, BorasoyEKLM08}. 
A two-step approach is used: a pionless \SU{4}-symmetric transfer matrix acts as an approximate and inexpensive low-energy filter at the beginning and end time steps; for time steps in the midsection, the
full leading-order (LO) transfer matrix is used and next-to-leading-order (NLO) operators are evaluated perturbatively by insertion
at the middle time step. 
The pionless \SU{4}-symmetric transfer matrix is computationally inexpensive because it requires only one auxiliary field and, more importantly, the path
integral is strictly positive for any even number of nucleons~\cite{ChenLS04}.
Although there is no positivity guarantee for odd numbers of nucleons, sign
oscillations are suppressed in odd-$A$ systems when they are only one
particle or one hole away from an even system with no sign oscillations.
Recently a technique called symmetry-sign extrapolation (similar to an extrapolation technique used in shell
model Monte Carlo calculations~\cite{AlhassidDKLO94,KooninDL97}) has been developed, which uses the approximate SU(4) symmetry of
the nuclear interaction to control the sign oscillations without introducing
uncontrolled systematic errors \cite{LahdeTLLMEKR15}. 
For further details and applications of the method, see the reviews \cite{Lee09,Lee16}.
 %%%%%%%%%%%

\subsubsection{Cluster model}
%SU(3)$\times$SU(4) and Sp(3,R)$\times$SU(4): Alpha-cluster systems/reactions\\
%Work by Hecht, Suzuki, Draayer

Cluster models, which can offer a unified theory of structure and reactions~\cite{WildermuthT77},  assume a formation of substructure systems, typically, $\alpha$ clusters. The latter figure prominently in the decay of heavy nuclei or low-lying $0^+$ states in $A=4,8,12,16,20,\dots $ nuclei. The physical significance of $\alpha$-cluster models is 
related to the fact that the $\alpha$-particle is tightly bound. Indeed, in its lowest-energy configuration,  it is a [4] spatial configuration, corresponding to an \SU{3} scalar,  $(\lambda\,\mu)=(0\,0)$ (spherical deformation) and an \SU{4} scalar, a single  $(0\,0\,0)$ \SU{4} irrep, with $S=0$ and $T=0$. 
As discussed above, these simple 2p-2n localized configurations have been shown  to emerge in nuclear modeling  in the framework of the {\it ab initio} lattice EFT  \cite{EpelbaumKLM11, EpelbaumKLLM12, EpelbaumKLLM13,EpelbaumKLLMR14}  (Sec. \ref{latticeEFT}) and in the NCSpM no-core symplectic shell model \cite{DreyfussLTDB13} with no {\it a priori} cluster assumption (Sec. \ref{NCSpM}).  Another interesting approach, e.g., the stochastic variational method with a correlated Gaussian basis makes it possible  to describe, with no {\it a priori} cluster ansatz, both localized cluster states and shell-model like states \cite{SuzukiV98,MitroyBHSACSKBV13}.  

While the cluster model has been mainly applied to the two-cluster case with a cluster ansatz, the extensions of the model to incorporate microscopic clusters and multi-clusters have considerably advanced over the last two decades (for applications, including studies of light exotic nuclei, see the book \cite{SuzukiLYV03}).
Remarkable progress has been made in recent years in the development of approaches from first principles to scattering and nuclear reactions (see, e.g., \cite{NollettPWCH07, HagenDHP07,NavratilQSB09,ElhatisariLRE15}).  In particular,  the shell-model based $NN$-informed NCSM/RGM \cite{QuaglioniN09,NavratilQSB09} has achieved successful descriptions with applications to fusion reactions and astrophysics  \cite{NavratilQ12,RedondoQNH14,HupinQN14}. The method  combines a microscopic cluster technique, the resonating-group method (RGM) \cite{WildermuthT77}, with the {\it ab initio} NCSM \cite{NavratilVB00}  -- it empowers the NCSM with the capability to simultaneously describe both bound and scattering states in light nuclei, while preserving  the Pauli exclusion principle and translational invariance; it also extends the RGM to utilize realistic interactions and first-principle NCSM wave functions. 
The latest developments in cluster physics  are thoroughly reviewed in Ref.~\cite{HoriuchiIK12}. Here, we focus on the use of symmetry-adapted schemes in microscopic cluster models, as discussed next.

\subsubsection{Resonating-group method (RGM) in the \SU{3} and symplectic schemes}

The nuclear wave function of the cluster model consists of ``cluster-internal" and ``cluster-relative" parts.  In the framework of the  microscopic resonating-group method (RGM) \cite{WildermuthT77}, the internal cluster wave functions can be expressed in terms of the HO shell-model basis assuming a common oscillator constant $\ho$ for all the clusters.  
For a relative motion between the clusters  that is very spatially extended, a shell-model representation of  clustering may require ultra-large model spaces. This makes the use of symmetry-based schemes advantageous.

\noindent
{\bf \SU{3}-scheme RGM.} The wave functions of the cluster system  are obtained by solving the many-body Schr\"odinger equation via an $R$-matrix coupled-channel method \cite{WildermuthT77,BayeB00}. This requires calculations of Hamiltonian  ($\hat O={\mathcal A} \hat H {\mathcal A}$)  and norm ($\hat O={\mathcal A} {\mathcal A}$) kernels, which involve computations of overlaps of the type $\left\langle \Psi'\right|\hat O\left| \Psi \right\rangle$ (${\mathcal A}$ properly takes into account antisymmetrization). 
In the \SU{3}-based RGM framework of Hecht \cite{Hecht77_NPA283}, the ``localized" part of the kernels is reduced to calculating norm and Hamiltonian overlaps between the \SU{3}-scheme RGM basis, which, e.g.,  for two fragments of mass number $f$ and $A-f$ can be written as,
\begin{align}
\mathcal{A}
\{
\{\phi^{(\lambda_{1}\,\mu_{1})S_1T_1}_{f}\times \phi^{(\lambda_{2}\,\mu_{2})S_2T_2}_{A-f}\}^{(\lambda_{c}\,\mu_{c})S_cT_c}
\times 
\chi^{(Q\,0)}
\}
^{(\lambda\,\mu)}_{\kappa (LS)JMTM_T},
\end{align}
where  the $\phi_f$ and $\phi_{A-f}$ are the microscopic wave functions of the fragments and $Q$ is the number of HO quanta of their relative motion. 
This defers the dependence on angular momentum to the very last step in the calculations,  and, in turn, facilitates  quick calculations. As emphasized and shown in Refs. \cite{Hecht77_NPA283,HechtS82}, the main advantage arises from the fact that the  norm overlaps (both direct and exchange terms) are diagonal in this basis and that one can avoid the complications of embedding the angular momentum. Another important feature is that once the overlaps are calculated in lab-frame coordinates, the translationally-invariant overlaps can be straightforwardly calculated using an  U($A$)$\times$ U(3) approach, which is especially suitable for the \SU{3}-coupled wave functions \cite{Hecht77_NPA283}.
Applications of the model to the intermediate-mass region  typically  employ leading \SU{3} configurations in the cluster wave functions and Gaussian-like interactions, and have successfully calculated  $\alpha$ and $^8$Be cluster amplitudes, spectroscopic amplitudes for heavy-fragment clusters, and sub-Coulomb $^{12}$C+$^{12}$C resonances \cite{HechtRSZ81,HechtB82,SuzukiH82}. 

\noindent
{\bf \SpR{3}-scheme RGM.} The first calculations of the ``no-core shell model with continuum" type with simple $NN$ potentials have been carried forward by Suzuki and Hecht \cite{SuzukiH86} for $^8$Be with a Gaussian-like interaction. The model utilizes a mixed symplectic \SpR{3} and microscopic cluster-model basis.  This  unified framework has been made possible by developing methods for calculating overlaps between \SpR{3}-scheme basis states and cluster states, and for evaluating matrix elements of a general translationally invariant two-body interaction \cite{Suzuki86,SuzukiH86, SuzukiH87}. 
Even though first applications have been carried in limited model spaces, consisting of a restricted single \SpR{3} irrep up through $N_{\rm max}=8$, and assuming no excitations of the alpha particles, the results  have indicated that the mixed symplectic-cluster model leads only to slight improvement to the cluster model in the description of the $\alpha+\alpha$ system. Furthermore,  calculations in the  pure symplectic basis have also provided a good description of $^8$Be. This is not surprising since the overlaps between the two bases for the low-lying states of $p$ and $sd$-shell nuclei have been found to be comparatively large in the low-$N\ho$~ subspaces \cite{Suzuki86}.   This approach has been further applied to study the monopole and quadrupole strengths in light nuclei~\cite{Suzuki87, SuzukiH89}. 
It has been also applied to the $\alpha$+$^{12}$C cluster system and found to contain the important shell-model configurations needed to describe low-lying  spectrum of $^{16}$O~\cite{Suzuki76a,Suzuki76b}.

The mixed symplectic-cluster basis approach has provided important insight for symmetry-guided large-scale shell models  that  aim to achieve faster convergence of states that are influenced by the continuum. In particular, some of the most important shell-model configurations can be expressed by exciting the relative-motion degree of freedom of the clusters. For example, the large overlaps between the first excitation ($2\ho$) indicate that the $\alpha$+$^{12}$C cluster basis significantly contains the quadrupole collectivity. 
A careful comparison with experiment, however, indicates  that the cluster model, if the clusters are frozen to their ground states, 
tends to miss some states with simple core excitations, overestimates cluster decay widths, and underestimates $E2$ transition rates \cite{SuzukiLYV03}. 
But  in the cases where clusters have little overlap, the corresponding cluster state strongly deviates from the usual ``shell-model"-like configurations and project onto ultra-large shell-model spaces.  As overlaps between the microscopic cluster and symplectic bases decrease in higher-$N\ho$ subspaces, the mixed-bases approach prove to be advantageous, in which both bases play a complementary role~\cite{HechtB82}.

Recent {\it ab initio} large-scale applications that utilize a mixed shell-model and RGM basis to achieve a faster convergence have been carried forward in the framework of  the no-core shell model with continuum (NCSMC) \cite{BaroniNQ13}. This study focuses on the unbound $^7$He nucleus and its controversial $1/2^-$ resonance. The approach and other successful applications to light nuclei are detailed in Ref. \cite{QuaglioniHCNR15}.

\subsection{Seniority scheme and exact pairing theory }
%SU(2): identical pairing; Sp(2): isovector pairing\\
%Work by Racah, Flowers, Kerman, Richardson, Feng Pan, Dukelsky
With an expanding body of experimental evidence that exposed prominent systematic features of nuclei, such as pairing gaps in energy spectra and even-odd mass difference, pairing correlations have been the focus of various models, including the early algebraic pairing models \cite{Racah,RacahB, Flowers52, Kerman61} and exact pairing models for a shell-model framework \cite{Richardson63a, Richardson63b,Gaudin76,PanD99,DukelskyES01,BalantekinP07,Guan12,DeBaerdemacker12}. 

The pairing problem, which was suggested by Racah \cite{Racah} in atomic physics
as a seniority scheme to describe coupling of identical electrons, was introduced first to
nuclear structure by Jahn and Flowers \cite{Jahn50,Flowers52} to completely classify the
states of the  $j^n$ nuclear configurations. 
Similar type of correlation effects, based on coupling of identical nucleons, were then suggested by Bohr, Mottelson and Pines
\cite{BohrMottelsonPines} to explain the energy gap observed in the spectra of even-even
nuclei and the concept was soon after applied by Belyaev in the first detailed (mean-field)
study of pairing in nuclei in terms of independent quasi-particles \cite{Belyaev59}.
Along with approximate mean field solutions (for a review see, e.g.,  \cite{Goodman79}), the pairing
problem was approached by various group theoretical methods, e.g., the like-pair \SU{2} seniority model (see, e.g., \cite
{Kerman61,PanDraayerOrmand}), the \SO{5}/\Spn{4} model 
for isovector (pp,pn, and nn) pairing
(see, e.g., \cite{Helmers61, FlowersS64, Hecht65, Ginocchio65,SviratchevaGD04}), and  the \SO{8} 
model with the additional isoscalar  pn channel
(see, e.g., \cite{Pang69,EngelPSVD97}). We note that the notation of \Spn{4} corresponds to a notation of \SpR{6} for the symplectic scheme of Sec. \ref{sp3r}, and is the one typically adopted for pairing models.

The seniority scheme focuses on a single-$j$ level with dimension $4\Omega_j=2(2j+1)$. In the conventional seniority scheme of Racah and Flowers
\cite{Racah,Flowers52}, states of a simple configuration $j^n$ comprised of both protons and
neutrons are completely classified according to the reduction chain,
\begin{equation}
\begin{array}{cccccccccc}
\Un{4\Omega_j }       & \supset &\Un{2\Omega_j }  & \supset &  \Spn{2\Omega_j } & \supset  & \SO{3}  & \supset &  \SO{2}   \\
j^n         &     & T & b  & (w,t)    &a  & J & &M
\end{array},
\label{reductionU4Omega}
\end{equation}
where an irrep of \Un{4\Omega_j} is formed by the $n$-particle antisymmetric wave functions
with total isospin $T$, $(w,t)$ label \Spn{2\Omega_j } irreps ($t$ is the isospin of non-paired particles), and $b$ and $a$ are multiplicity labels
\cite{Racah,Flowers52,Hecht65,Ginocchio65}.
 
The ``quasi-spin" approach of Helmers \cite{Helmers61}, on the other hand, yields a
classification scheme with the same quantum numbers as in (\ref{reductionU4Omega}) based on two
parallel group chains starting with a different and ingenious group decomposition of
\Un{4\Omega_j }, namely 
\begin{equation}
\begin{array}{ccclccc}
\Un{4\Omega_j } & \supset &  \Spn{2\Omega_j }       &     & & \times  & \SO{5},      \\ 
    j^n  &&   j^{\nu} &     & & & (w\leftrightarrow \nu ,t;n,b,T)      \\
&& \quad \cup  \quad a &      & &    &     \\
&&\SO{3}   & &   & &     \\
&&J  & &   & &     \\
\end{array}
\label{reductionU4Omega2}
\end{equation}
where the dependence on $n$, $b$ and $T $ is transferred solely to \SO{5}, locally
isomorphic to \Spn{4}. The group chain of \Spn{2\Omega_j } is the one associated with 
conventional seniority but now is completely specified by the simple configuration $j^\nu $,
where $\nu $ is the total seniority number that counts particles not coupled in a $J=0$ pair
and is related to the maximum number $w$ as $w=4\Omega -\nu $.
A
detailed comparison that reveals the power of the \Spn{4} method versus the conventional
seniority spectroscopy can be found in the literature
\cite{Helmers61, FlowersS64,Hecht65,Ginocchio65}. In general, for the complete shell-model space, coupling of \Un{4\Omega_j} irreps should be considered.

Partial conservation of seniority has been recently examined in many-body systems for identical particles in a single-$j$ level \cite{IsackerH14}. Racah's seniority scheme, however, is badly broken
by single-particle energies \cite{Talmi93}. Nonetheless, for non-degenerate single-particle energies exact solutions to the pairing problem have been derived by
Richardson and Gaudin \cite{Richardson63a, Richardson63b, Gaudin76}, with further extensions
based on the algebraic  Bethe ansatz \cite{PanDraayerOrmand,DukelskyES01,DukelskyES02, ZhouLMG02,RomboutsVD04,DukelskyGVDEL06, BalantekinP07}.
For all these algebraic Bethe ansatz approaches, the solutions are provided by a set of highly non-linear Bethe Ansatz Equations (BAEs). While these applications demonstrate that the pairing problem is exactly solvable, solutions of these BAEs are not easy and typically require extensive numerical
work, especially for a large number of levels and valence pairs \cite{RomboutsVD04}. This limits the applicability of the methodology to
relatively small systems. However, it has been shown recently that the set of Gaudin-Richardson equations for the standard pairing case
can be solved relatively easily by using the extended Heine-Stieltjes polynomial approach~\cite{Guan12, Guan14}. 
Since solutions of the standard pairing model can be obtained from zeros of the associated extended Heine-Stieltjes polynomials,
the approach can be applied to study the model with more pairs over a larger number of single-particle levels. In particular, in the studies of Refs. \cite{Guan12, Guan14}, the pairing Hamiltonian includes non-degenerate single-particle energies plus standard pairing and is exactly solvable. It has been shown that the method provides solutions for the ground states of Ca, Ni, and Sn isotopes that closely reproduce experimentally observed pairing gaps. Such models can be essential in incorporating exact pairing correlations into a general theory of self-consistent mean-field type, such as density functional theory (DFT) framework.  The DFT approach generally yields an excellent accounting of binding energies as well as near ground state phenomena across much of the nuclear landscape (see, e.g., the review \cite{DrutFP10}). It can link to an {\it ab initio} foundation to achieve better predictive capabilities across most of the chart of the nuclides.  
%%%

\section{Highly structured orderly patterns from first principles}
\subsection{Low intrinsic spin \label{spin}}
Realistic inter-nucleon interactions break  \SU{2} spin symmetry. Consequently, in nuclear states  all possible intrinsic spin values for a given nucleus may mix. To investigate the spin pattern in low-lying states, we have studied NCSM eigenstates by projection to spin components \cite{DytrychSBDV_PRL07} and SA-NCSM eigenstates, which, as discussed in Sec. \ref{su3}, are expressed in terms of basis states that by construction have definite proton $S_{p}$, neutron $S_{n}$, and total $S$ spin values  \cite{DytrychLMCDVL_PRL12}.
\begin{figure}[t]
\centerline{
\includegraphics[width=0.75\textwidth]{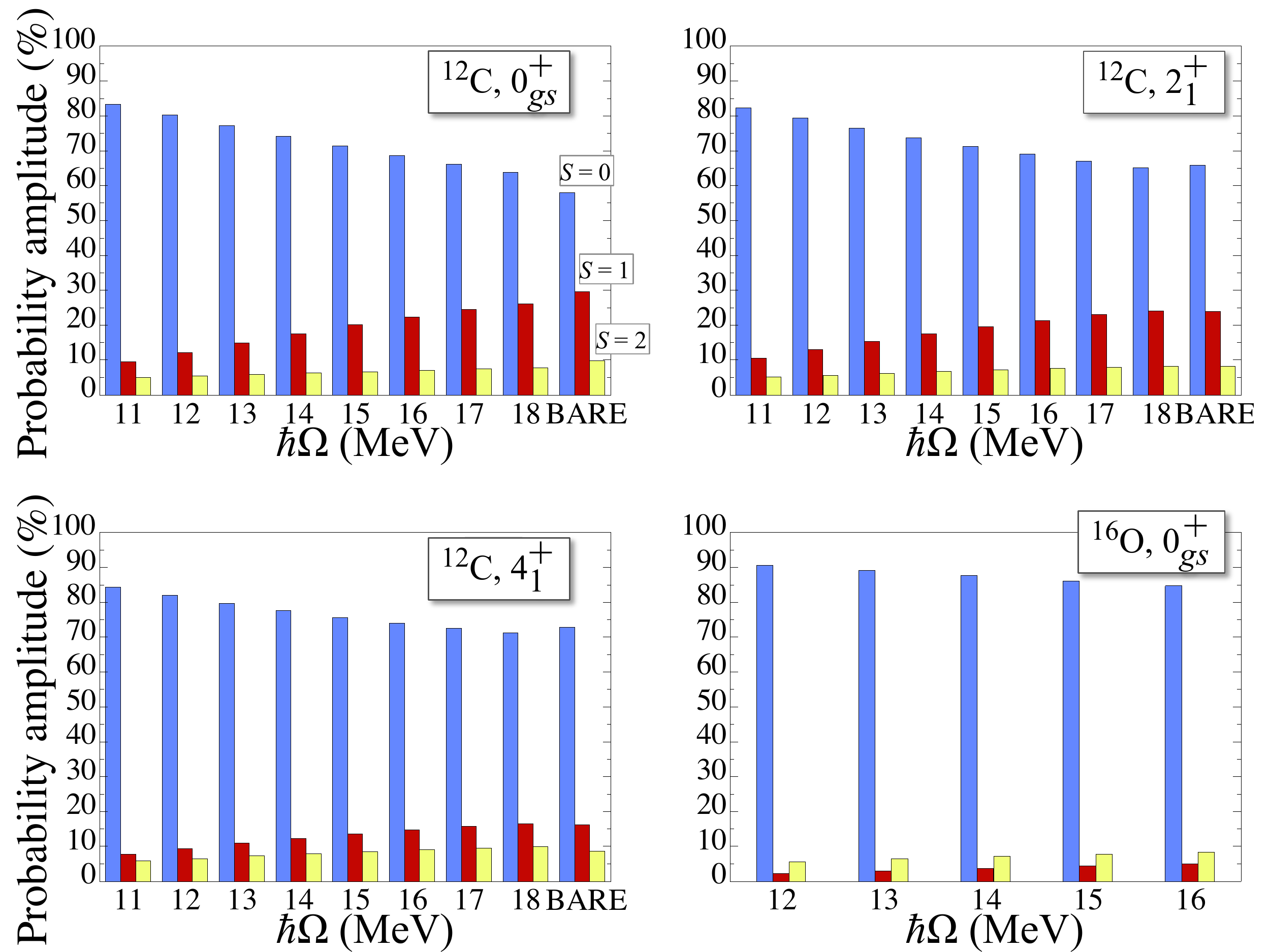}
}
\caption{
Probability distribution of the $S=0$ (blue), $S=1$ (red) and $S=2$ (yellow) components of the $^{12}$C $0^{+}_{gs}$, $2^{+}_1$ and $4^{+}_1$  states and the  $^{16}$O ground state,
calculated with the effective JISP16 interaction for different $\hbar\Omega$ oscillator strengths and with the bare JISP16  interaction for $\hbar\Omega=15$ MeV. Figure adapted from Ref. \cite{DytrychSBDV_PRCa07}.
}
\label{Spin_Mixing}
\end{figure}

For NCSM eigenstates, the spin components and  the corresponding spin probability amplitudes can be determined by the projection operator $P(s_{\min})$,
\begin{equation}
	P(s_{\min})=\prod_{k=s_{\min}}^{S_{\max}}\left(1-\frac{\hat S^{2}}{k\left(k+1\right)}\right),
\end{equation}
where $\hat S$ is the spin operator. To calculate the spin-zero component of an eigenstate, the operator~$P(s_{\min}\!\!=\!\!1)$ is first applied. The resulting $S=0$ component is then removed from the original wave function.  The operator  $P(s_{\min}\!\!=\!\!2)$ is then applied to yield the  $S=1$ component. 
Eventually, the complete spin decomposition is achieved.

%%%%%%%%%%%%%
%
Utilizing this procedure, we have studied the spin decomposition of well-converged NCSM eigenstates in $^{12}$C and $^{16}$O \cite{DytrychSBDV_PRL07,DytrychSBDV_PRCa07}. The NCSM eigenstates employed in this study are obtained with effective
interactions (using the Okubo-Lee-Suzuki procedure) derived from the realistic JISP16 and N$^3$LO $NN$ potentials in the $N_{\max}=6$ model space and
are reasonably well converged.  In addition, calculated
binding energies as well as  other observables for $^{12}$C such as
$B(E2;2^{+}_{1}\!\rightarrow\!0^{+}_{gs}$), $B(M1;1^{+}_{1}\!\rightarrow\!0^{+}_{gs}$),
ground-state proton rms radii and the
$2^{+}_{1}$ quadrupole moment all lie reasonably close to the measured values \cite{NavratilVB00,NavratilO03}.
The analysis of the spin probability amplitudes for various \ho~oscillator strengths  (Fig. \ref{Spin_Mixing}) reveals that spin mixing  follows almost exactly the same pattern in all the low-lying $0^+_{gs}$, $2^+_1$ and $4^+_1$ states, which we will clearly identify as belonging to a rotational band in Sec. \ref{def}. 
The predominance of the $S=0$ component is striking, regardless if effective or bare interactions were used.
\begin{figure}[t]
\centerline{
  \includegraphics[width=.55\textheight]{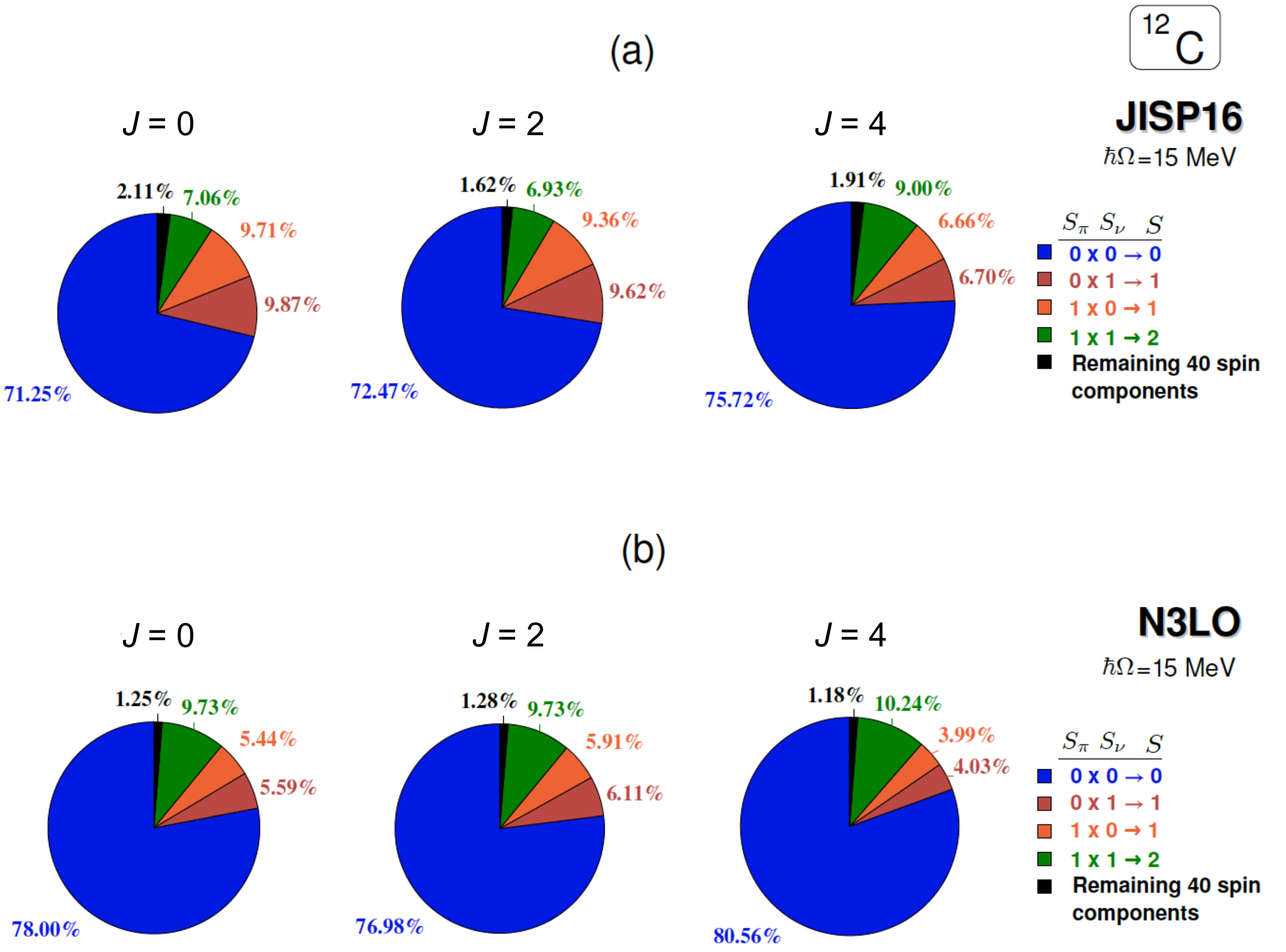}
  }
  \caption{
Intrinsic spin structure of the $J=0^{+}_{\textrm{gs}}, 2^{+}_{1},$ and $4^{+}_{1}$
NCSM states in $^{12}$C obtained using: (a) JISP16, and (b) N$^3$LO effective 
interactions in the $N_{\max}=6$ model space with $\hbar\Omega=15$ MeV. 
}
\label{12C_J024_spin_decomposition}
\end{figure}
\begin{figure}[b!]
\centerline{
  \includegraphics[width =.45\textheight]{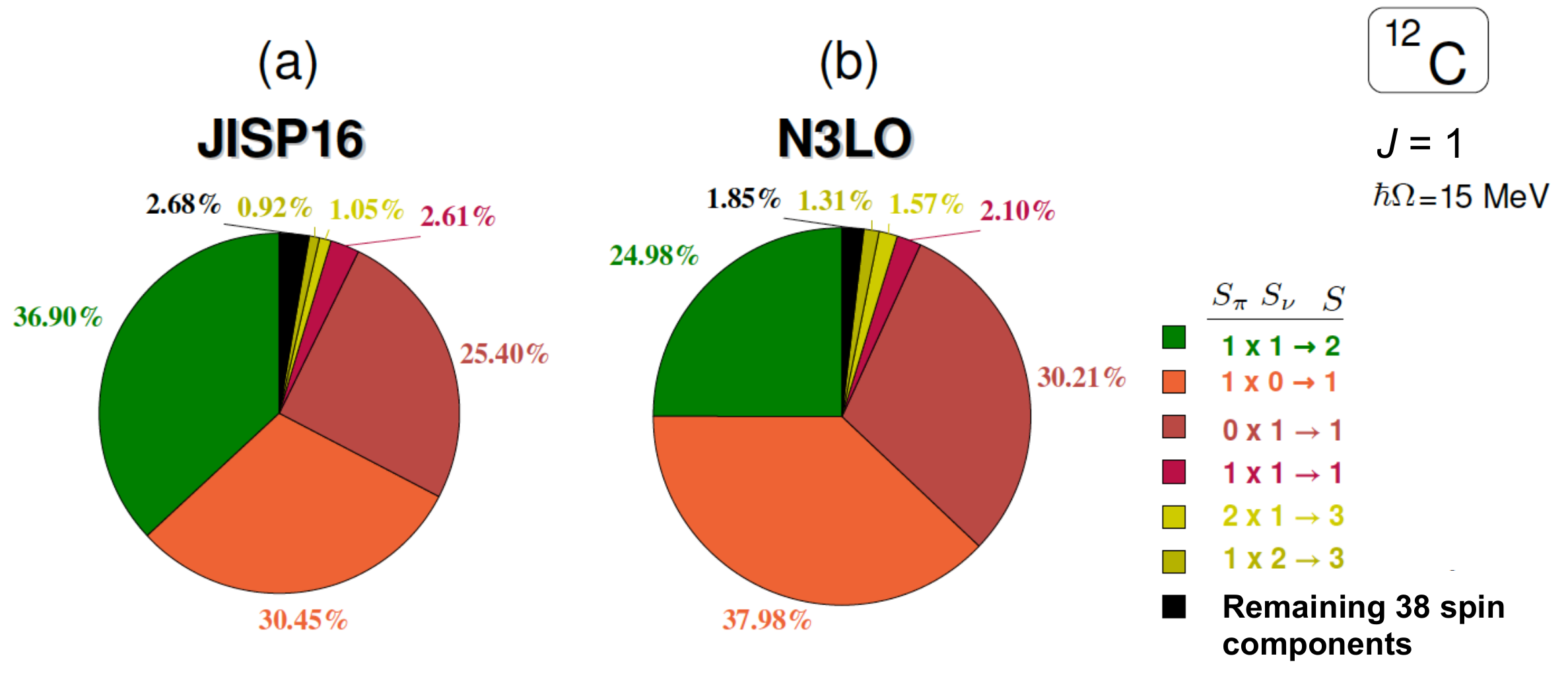}
  }
  \caption{
Intrinsic spin structure of the $J=1^{+}_{1}$ NCSM state in $^{12}$C
obtained using: (a) JISP16 and (b) N$^3$LO effective interactions in the
$N_{\max}=6$ model space with $\hbar\Omega=15$ MeV. 
}
\label{12C_J1_spin_decomposition}
\end{figure}

Considering further the proton and neutron spins \cite{DytrychDSBV09}, we have found that only four configurations with the
lowest intrinsic spins (spin-zero and spin-one) in the proton and neutron sector  contribute  $98\%$ to the states shown in Fig.~\ref{12C_J024_spin_decomposition},
for both the JISP16 and N$^3$LO interactions. The remaining $44$ spin combinations contribute typically less than $2\%$.  Qualitatively,
similar results are obtained for the $1^{+}_{1}$ state, with $4$ lowest-spin 
configurations describing about $95\%$ of the wave function (Fig.~\ref{12C_J1_spin_decomposition}). If the $N_{\rm max} = 6$ model space is restricted to the wave functions with good total angular momentum $J$ and $S_p \le 1$ and $S_n \le 1$, the size of the basis drops by a factor of three relative to the basis which does not impose any restrictions on intrinsic spins. This reduction further improves, albeit slowly, for heavier nuclei and larger model
spaces. 

Low-spin dominance is further confirmed by  {\it ab initio} SA-NCSM calculations with bare JISP16, N$^3$LO, and NNLO$_{\rm opt}$  \cite{Ekstrom13}  interactions \cite{DytrychLMCDVL_PRL12, DytrychDLCL13}. Table \ref{tab:dominantConfig} shows the spin component having the largest contribution to the ground state, which is  more than 50\% up to $\sim90$\% for various $p$-shell and $sd$-shell nuclei.  In addition, supporting the results above, we have found that the SA-NCSM calculated eigenstates project at a 99\% level onto a comparatively
small subset of intrinsic spin combinations.  For instance, the lowest-lying
eigenstates in $^{6}$Li are almost entirely realized in terms of configurations
characterized by the following intrinsic spin $(S_{p}S_{n}S)$ triplets:
$(\frac{3}{2} \frac{3}{2} 3),\,\, (\frac{1}{2} \frac{3}{2}
2),\,\,(\frac{3}{2}\frac{1}{2} 2)$, and $(\frac{1}{2} \frac{1}{2} 1)$, with the
last one carrying over 90\% of each eigenstate \cite{DytrychLMCDVL_PRL12} (see also Fig. \ref{gsStructure} discussed in the next two sections).  Likewise, the same spin
components as in the case of $^{6}$Li are found to dominate the ground state
and the lowest $1^{+}$, $3^{+}$, and $0^{+}$ excited states of $^{8}$B.  The ground state, $2^+_1$ and $4^+_1$  of $^8$Be,
$^6$He, $^{12}$C, $^{16}$O and $^{20}$Ne are all found to be dominated by spin-zero and spin-one proton and neutron spins, with the largest contributions arising from the $(S_p S_n S)$=(000) configurations. 
\begin{table}[th]
\begin{center}
\begin{tabular}{c|ccccc}
\hline
Nucleus &$(S_{p}\,S_{n}\,S)$ & Probability, \% & $ (\lambda_0 \,\mu_0)$ & Probability, \%  & Probability, \% \\
&&  &  & $ (\lambda_0 \,\mu_0)$ & $(\lambda \,\mu)$* \\
\hline
 $^{6}$Li & $(\frac{1}{2}\, \frac{1}{2}\, 1)$ & $93.24 $ & $(2\, 0)$ & 57.36 & $93.11$ \\
 $^{8}$B  & $(\frac{1}{2}\, \frac{1}{2}\, 1)$ & $85.58 $ & $(2\, 1)$ & 56.50 &$82.32 $ \\
 $^{8}$Be & $(0\, 0\, 0)$ & $85.21 $ & $(4\, 0)$ & 55.92 &$85.06 $\\
 $^{12}$C & $(0\, 0\, 0)$ & $55.60 $ & $(0\, 4)$ & 44.10 &$49.03 $ \\
 & [$(0\, 1\, 1), (1\, 0\, 1)$] & [$29.19 $] & [$(1\, 2)$] & [18.63] & [22.52] \\
 $^{16}$O  & $(0\, 0\, 0)$ & $78.42 $ & $(0\, 0)$ & 60.59 & $77.33 $ \\ 
$^{20}$Ne  & $(0\, 0\, 0)$ & $79.73 $ & $(8\, 0)$ & 43.93 & $79.30 $ \\
\hline
\end{tabular}\\
*All $(\lambda \,\mu)$ built over $ (\lambda_0 \,\mu_0)$ according to the rule of Eq.~(\ref{eq:Sp2RSelection}).
\caption{
Probability amplitudes  of the dominant $(S_{p}\,S_{n}\,S)$ spin configurations and the corresponding dominant nuclear deformation $(\lambda_0 \,\mu_0)$  for the ground state of selected nuclei calculated in the {\it ab initio} $N_{\rm max}=8-14$ SA-NCSM with the bare JISP16 interaction (NNLO$_{\rm opt}$is used for $^{20}$Ne) for \ho=20 MeV. 
The second most important contribution is shown for $^{12}$C -- note that the combined $(0\, 0\, 0),(0\, 1\, 1),(1\, 0\, 1)$ contribution for $^{12}$C yields $84.79\%$ for the third column, as well as $62.73\%$ and $71.55\%$ for the last two columns. Note that the SA-NCSM calculations for $^{20}$Ne are performed  in the challenging \Nmax{2}{10} model space of 13 shells of, e.g., $51\times 10^6$ basis states for $6^+$ -- compare to the inaccessible NCSM  space of  $4.4\times 10^{11}$ dimensionality required for the corresponding $N_{\rm max}=10$  fixed-$J$  calculations or of $1\times 10^{12}$ for the conventional $M$ scheme.
}
\label{tab:dominantConfig}
\end{center}
\vspace{-0.50cm}
\end{table}%

A more recent study \cite{Johnson15} using SRG effective interactions in the NCSM  has arrived at the same conclusions, while, in addition, it points to a similar spin pattern in odd-$A$ $p$-shell nuclei.

All of these studies support the dominance of the lowest proton/neutron intrinsic spins in light and intermediate-mass nuclei, followed by low-spin configurations,  as evident from first principles.

\subsection{Large quadrupole deformation \label{def}}
\begin{figure}[t]
\includegraphics[width=0.45\textwidth]{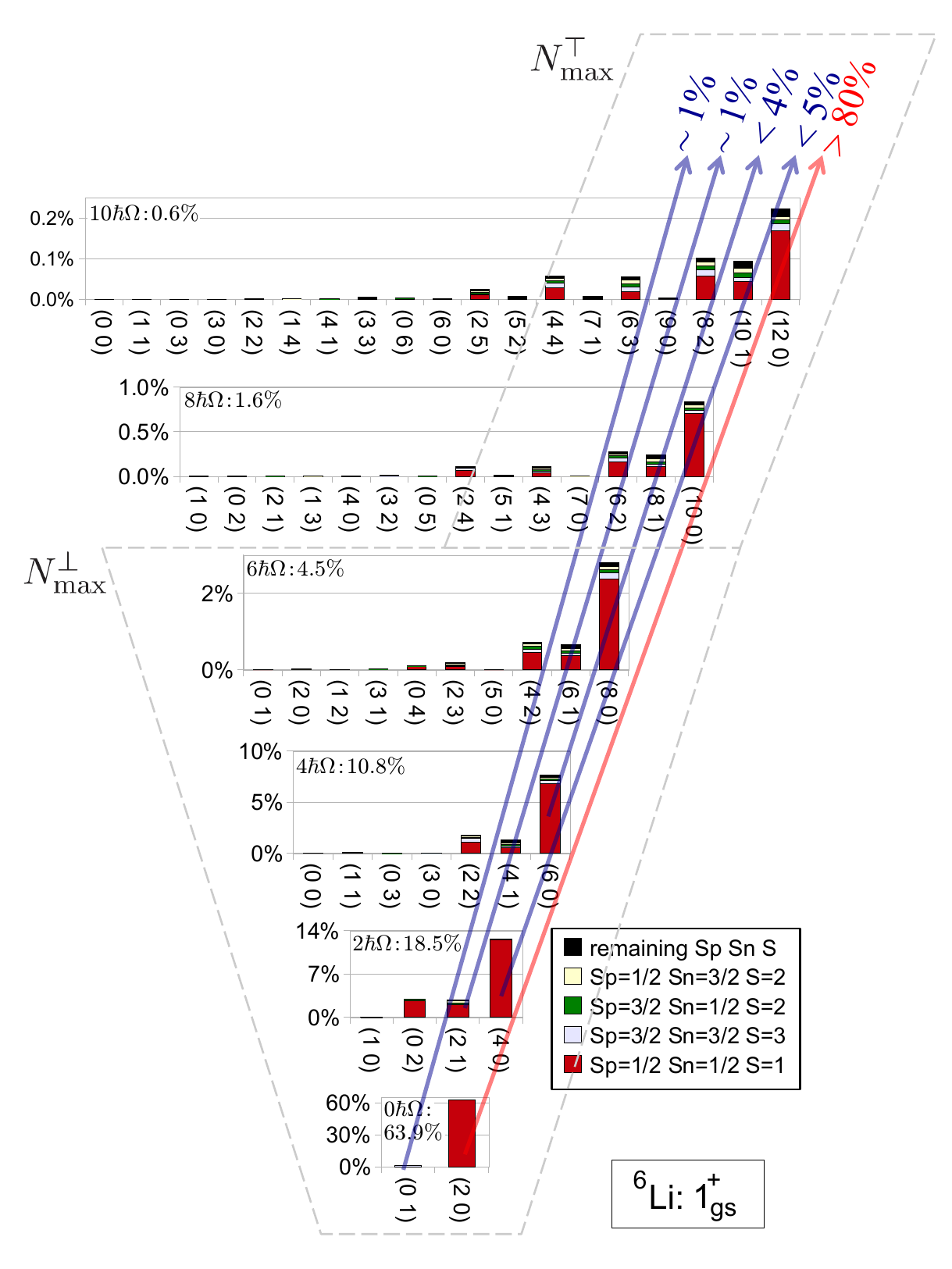} 
\includegraphics[width=0.45\textwidth]{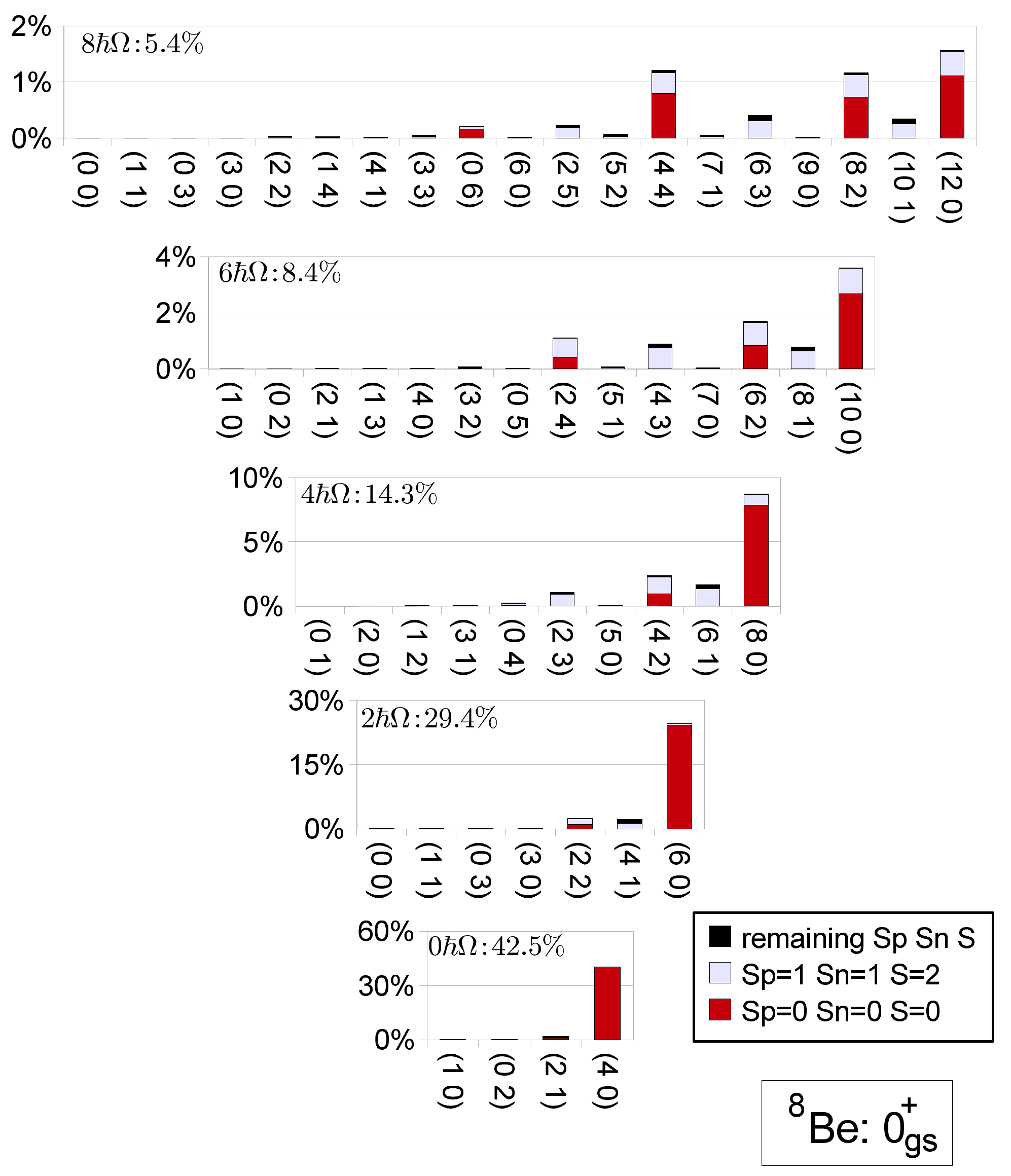} 
\vspace{-0.30cm}
\caption
{
	Probability distributions for proton, neutron, and total intrinsic spin
	components ($S_{p} S_{n} S$) across the Pauli-allowed $(\lambda\,\mu)$
	values (horizontal axis) for the calculated $1^{+}$ ground  state of
	$^{6}$Li obtained for $N_{\max}=10$ and $\hbar\Omega=20$ MeV with the
	JISP16 bare interaction (left) and the $0^{+}$ ground  state of $^{8}$Be
	obtained for $N_{\max}=8$ and $\hbar\Omega=25$ MeV with the N$^3$LO bare
	interaction (right). The total probability for each $N\hbar\Omega$ subspace
	is given in the upper left-hand corner of each histogram.  The concentration of
	strengths to the far right within the histograms demonstrates the
	dominance of collectivity in the calculated eigenstates -- this supports a {\it symmetry-guided} concept (detailed in Sec. \ref{SpConcept}), 
which implies inclusion of the complete 
space up through $N^{\bot}_{\max}$, and a subset of deformation/spin configurations beyond this, up through $N^{\top}_{\max}$ (for the example illustrated in the figure, a selected space includes all possible  configurations within $N^{\bot}_{\max}=6$ and only selected configurations in the $8\ho$, $10\ho$, etc., up to an $N^{\top}_{\max}$ cutoff). The projection onto symplectic vertical slices (with probability $\ge 1\%$) is schematically illustrated for $^{6}$Li by arrows 
and clearly reveals the preponderance of a single  symplectic irrep (vertical cone). Figure adapted from Ref. \cite{DytrychLMCDVL_PRL12}.
}
\label{gsStructure}
\vspace{-0.5cm}
\end{figure}

Within the {\it ab initio} SA-NCSM  framework, it is possible to explore the microscopic nature of the most important collective configurations. In Refs. \cite{DytrychLMCDVL_PRL12, DytrychSBDV_PRL07, DytrychDLCL13},  we have analyzed the probability distribution across Pauli-allowed $(S_{p} S_{n} S)$ and $(\lambda\,\mu)$ configurations of the four lowest-lying isospin-zero $1^+_{\rm gs}$, $3^+_1$,
$2^+_1$, and $1^+_2$ states of $^6$Li; $0^+_{gs}$, $2^+_1$ and $4^+_1$ of $^{8}$Be, $^{6}$He
and $^{12}$C; the lowest $1^{+}$, $3^{+}$, and $0^{+}$ excited states of
$^{8}$B; and the ground state of $^{16}$O.  Results for the ground state of
$^{6}$Li and $^{8}$Be, obtained with the bare JISP16 and chiral N$^3$LO
interactions, respectively, are shown in Fig.~\ref{gsStructure}.  

The results show that the mixing of $(\lambda\,\mu)$ quantum numbers, induced
by the \SU{3} symmetry breaking terms of realistic interactions, exhibits a
remarkably simple pattern.  One of its key features is the preponderance of a
single $0\hbar\Omega$ \SU{3} irrep.  This so-called leading irrep, according to
the established geometrical interpretation of \SU{3} labels
$(\lambda\,\mu)$~\cite{RosensteelR77b,LeschberD87,CastanosDL88}, is characterized by the largest value of the
intrinsic quadrupole deformation. For instance, the low-lying states of
$^{6}$Li, $1^+_{\rm gs}$, $3^+_1$, $2^+_1$, and $1^+_2$,  project predominantly onto the prolate $0\hbar\Omega$ \SU{3}
irrep $(2\,0)$, as illustrated in Fig.~\ref{gsStructure} for the ground
state.  Furthermore, the dominance of the same dominant deformation within these states clearly identifies them as members of a rotational band, that is, different rotations of the same deformation (we will come back to this feature, as illustrated in Table \ref{TABLE_15MeV} for $^{12}$C). Indeed, according to Elliott's rule, for $(2\,0)$, $L=0$ and $2$, which couples with spin-1 to yield $J=1^2,2,3$. 
In addition to $^{6}$Li, Table \ref{tab:dominantConfig} shows the leading $0\hbar\Omega$ \SU{3} irrep for  the ground-state rotational-band states in $^{8}$B, $^{8}$Be, $^{12}$C, and $^{16}$O, which is   (2 1), (4 0), (0 4), and (0 0), respectively (associated with triaxial,
prolate, oblate, and spherical shapes, respectively). 
The clear dominance of the most deformed $0\hbar\Omega$ configuration within low-lying states of light and intermediate-mass nuclei (Table \ref{tab:dominantConfig}) indicates that the quadrupole-quadrupole interaction of the
Elliott \SU{3} model of nuclear rotations \cite{Elliott58,Elliott58b} is realized naturally within an {\textit{ab initio}} framework.

These results corroborate earlier observations \cite{Wiringa06} based on a simple but useful guide, which involves  counting of
the number of interacting pairs in different spin-isospin  states for a given spatial symmetry and provides an estimate for
the overall binding due to one-pion exchange. The ordering of states, estimated in this way, has been found to closely agree with the results of {\it ab initio} Variational Monte Carlo (VMC)  \cite{PieperWC04} calculations. The outcomes have revealed that the lowest-energy configuration for $p$-shell nuclei with nucleons in the $s$ and $p$ shells is of the highest spatial symmetry. This, indeed, as discussed in Sec. \ref{su3}, contains the largest deformation. For example, in the study of Ref. \cite{Wiringa06}, for $^6$Li, the four states referenced in the previous paragraph, have been identified as dominated by $^3$S[4\,2] ($1^+_{\rm gs}$) and  $^3$D[4\,2] ($3^+_1$, $2^+_1$, and $1^+_2$), where [4\,2] contains the \SU{3} (2 0) (see Sec. \ref{su3}), the preponderant configuration revealed in the corresponding SA-NCSM wave functions with  $\sim 87\%$ of the $1^+_{\rm gs}$ state in  $L=0$  \cite{DytrychLMCDVL_PRL12}.

\subsection{Symplectic symmetry from first principles \label{symplectic}}
In the studies referenced above \cite{DytrychLMCDVL_PRL12} and \cite{DytrychSBDV_PRL07}, the existence of a new approximate symmetry in light nuclei, the symplectic \SpR{3} symmetry, and hence their propensity towards development of collectivity, has been confirmed from first principles with no {\em a priori} symmetry assumptions. 
\begin{figure}[t]
\begin{center}
\centerline
{
\includegraphics[width=0.37\textwidth]{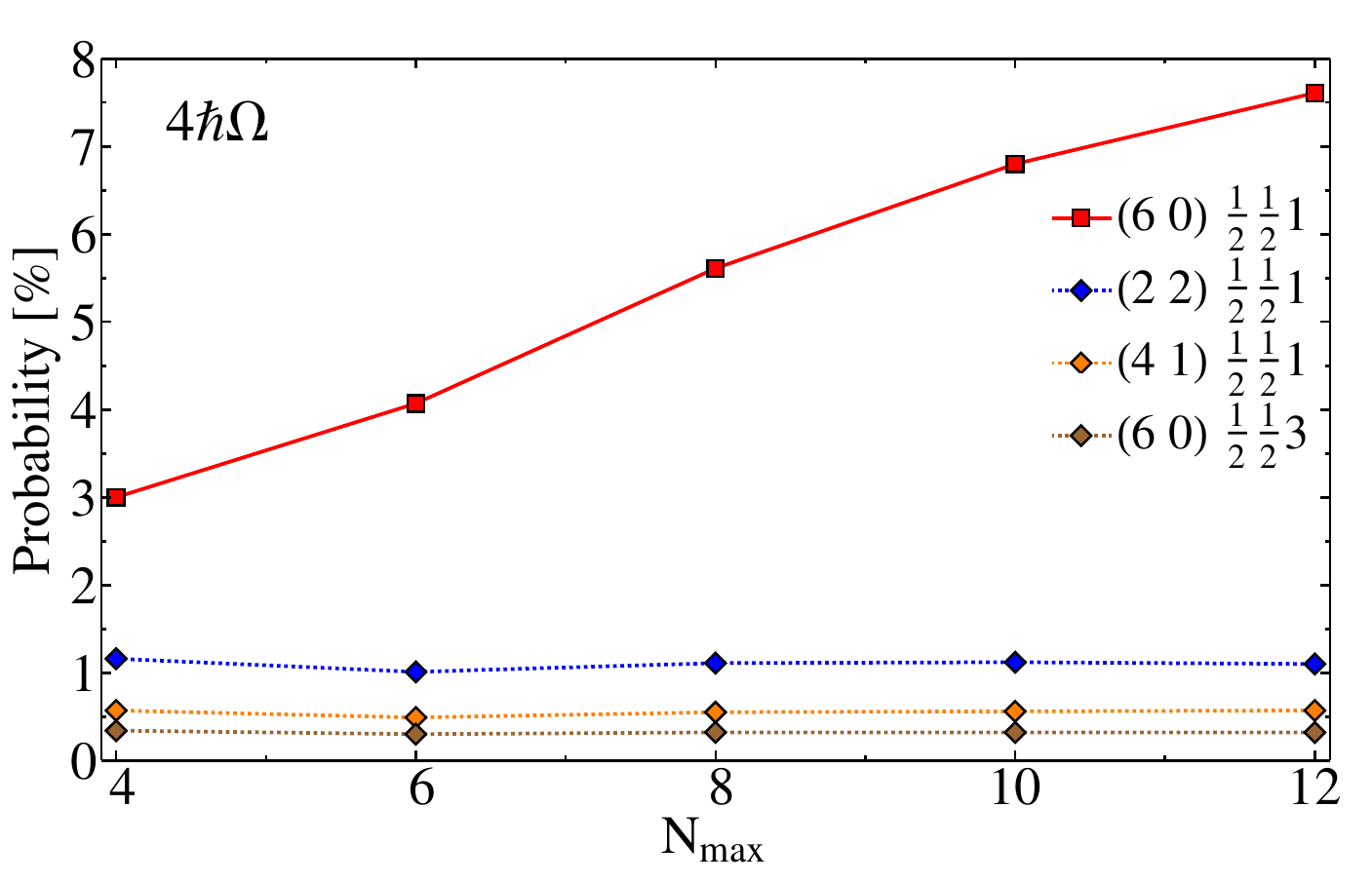}
\includegraphics[width= 0.40\textwidth]{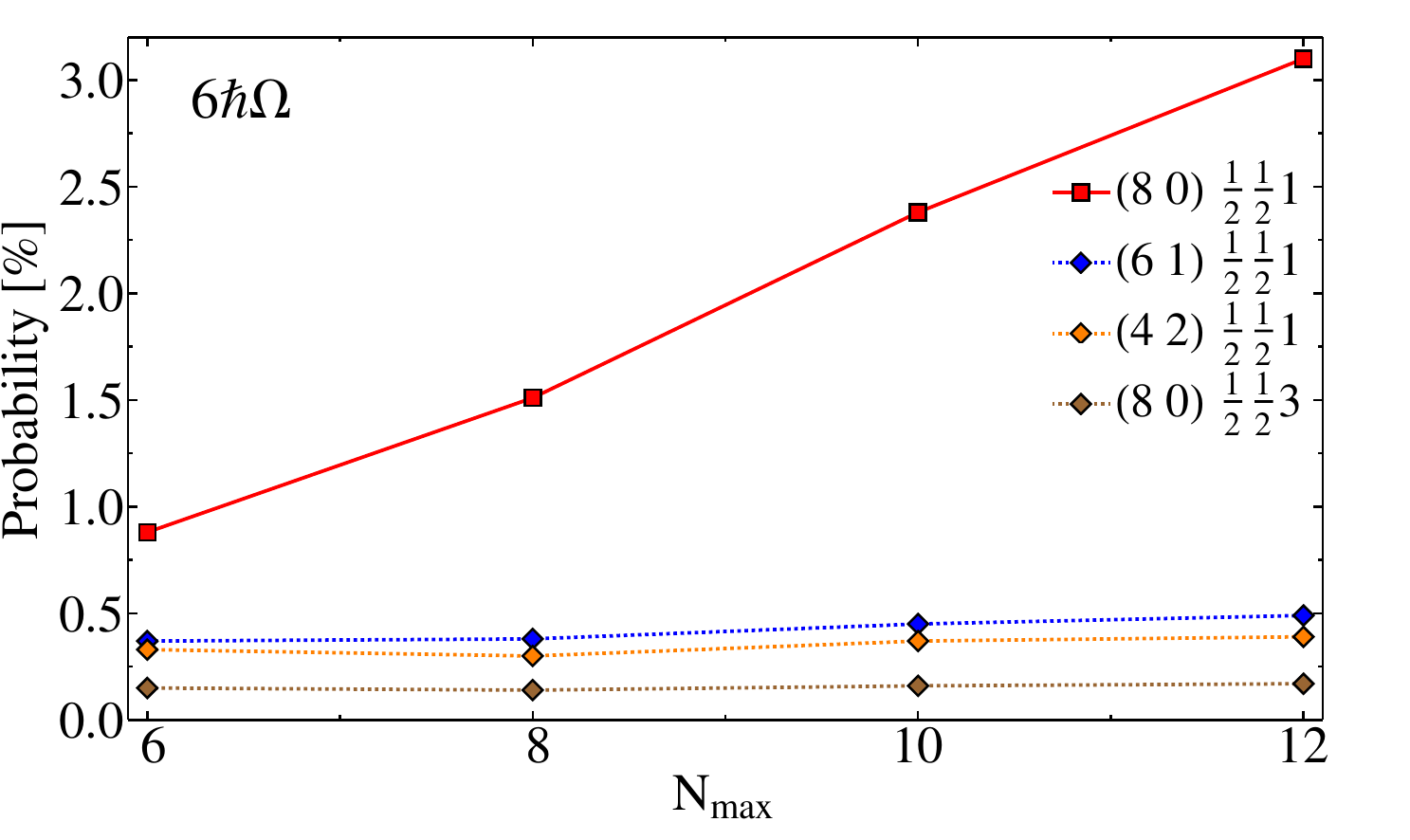} 
}
\centerline{ ({{a}}) \hspace{7cm} ({{b}}) }
\centerline{
\includegraphics[width= 0.40\textwidth]{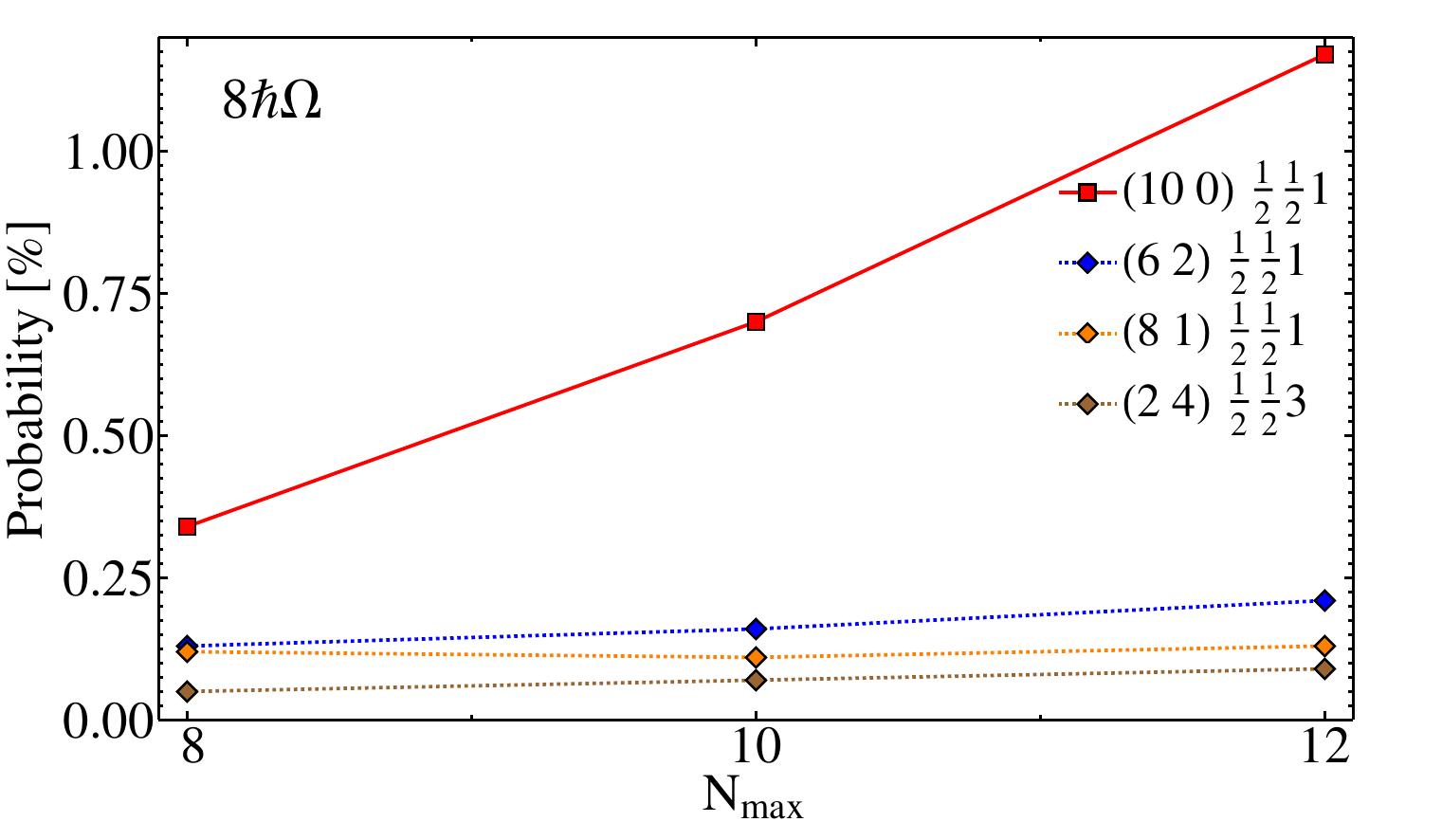}
\includegraphics[width= 0.40\textwidth]{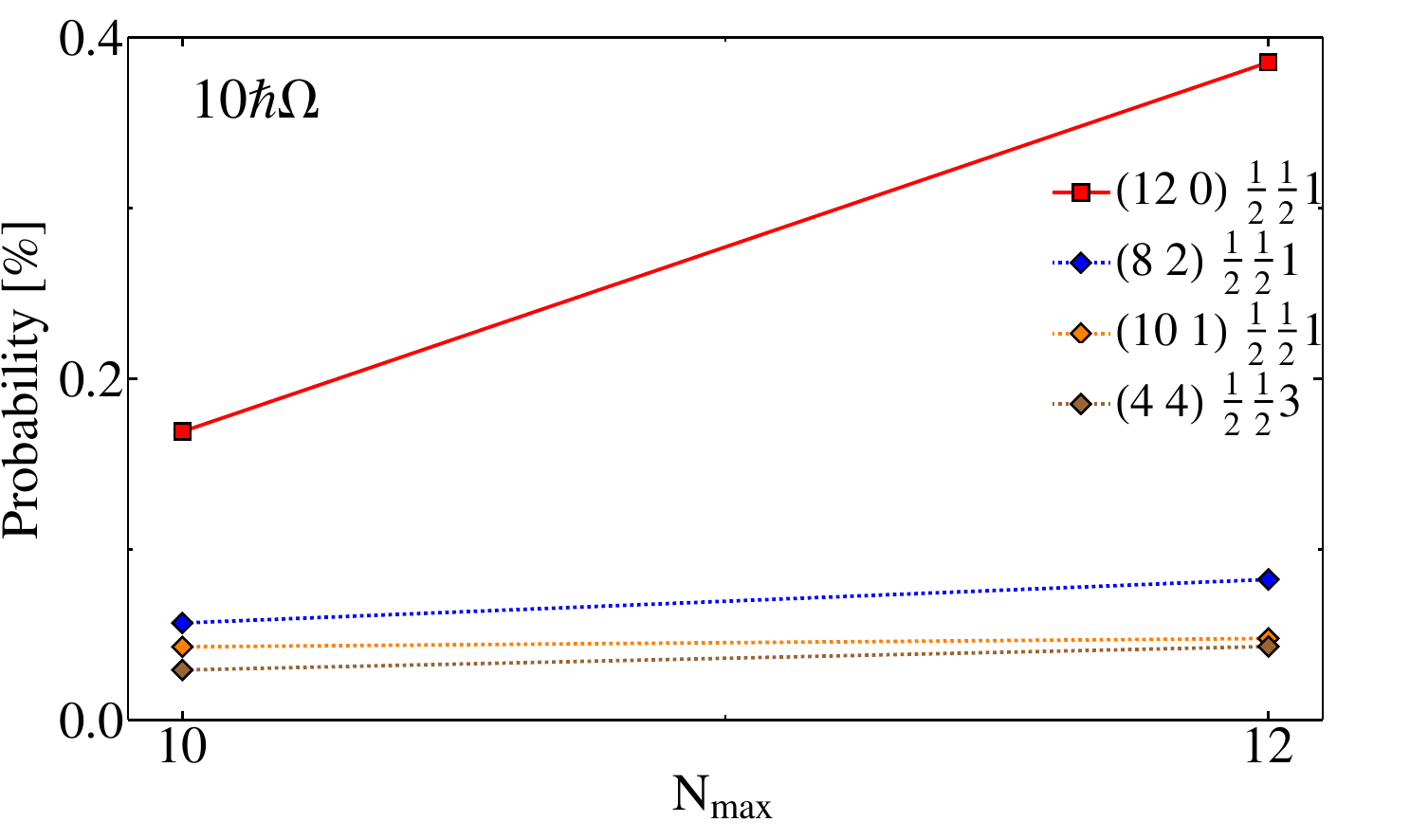}
}
\centerline{ ({{c}}) \hspace{7cm} ({{d}}) }
\caption{
\label{fig:projections}
Probability amplitudes  of the most important $(\lambda\,\mu)$ $S_{p}S_{n}S$ configurations
within various $\hbar\Omega$ components of the $^6$Li ground state, which is calculated in  the SA-NCSM for increasing $N_{\max}$ model space cutoffs, and for the bare JISP16 and \ho=20 MeV.
(a) $4\hbar\Omega$ configurations, (b) $6\hbar\Omega$ configurations, (c) $8\hbar\Omega$ configurations, and (d) $10\hbar\Omega$ configurations. For $N_{\max}=10$, these amplitudes are the ones shown in Fig. \ref{gsStructure}, left. Note that the $S_{p}S_{n}S=$ \half \half 1 $(6\,0)$, $(8\,0)$, $(10\,0)$, and $(12\,0)$ (red) are the stretched states over the most dominant 0\ho~ (2\,0) \SU{3} irrep and they exhibit a comparatively substantial increase in larger model spaces.
}
\end{center}
\end{figure}

The symplectic  symmetry structure was identified in well-converged \textit{ab initio} wave functions for $^{6}$Li (odd-odd), $^{8}$Be (even-even),
$^{6}$He (halo), $^{12}$C (oblate), and $^{16}$O (spherical) nuclei  using realistic nucleon-nucleon ($NN$)
interactions, the bare JISP16  and N$^3$LO potentials, as well as their effective counterparts.  Indeed, the SA-NCSM framework exposes a remarkably simple physical feature that is
typically masked in other \textit{ab initio} approaches; in particular, the
emergence, without {\it a priori} constraints, of  simple orderly patterns that
favor spatial configurations with strong quadrupole deformation and  low
intrinsic spin values (Fig. \ref{gsStructure}; see also Fig. \ref{enSpectrumC12}b for the {\it ab initio} $^{12}$C ground state). This figure illustrates a feature common to all
the low-energy solutions considered; namely, a highly structured and regular
mix of intrinsic spins and \SU{3} spatial quantum numbers that, furthermore,
does not seem to depend on the particular choice of a realistic $NN$ potential.  This feature, once exposed and understood, can in turn
be used to guide a truncation and augmentation of model spaces to ensure that 
important properties 
of atomic nuclei, like enhanced $B(E2)$ strengths,
nucleon cluster substructures, and others important in reactions, are
appropriately accommodated in future \textit{ab initio} studies. 

Moreover, the analysis reveals the preponderance of only a few symplectic \SpR{3} irreps (see arrows in Fig. \ref{gsStructure} that schematically represent symplectic irreps). E.g., for $^6$Li, the dominant \SU{3} basis states belong to a single $(2\, 0)$ symplectic irrep, which comprises more than 80\% of the ground-state wave function, with total of only  5 \SpR{3} irreps realizing more than 90\% of the state.  This clearly reflects the presence of an underlying symplectic \SpR{3} symmetry of microscopic nuclear collective motion \cite{RosensteelR77,Rowe85}.
 \begin{figure}[b!]
\centerline{
\includegraphics[width=0.9\textwidth]{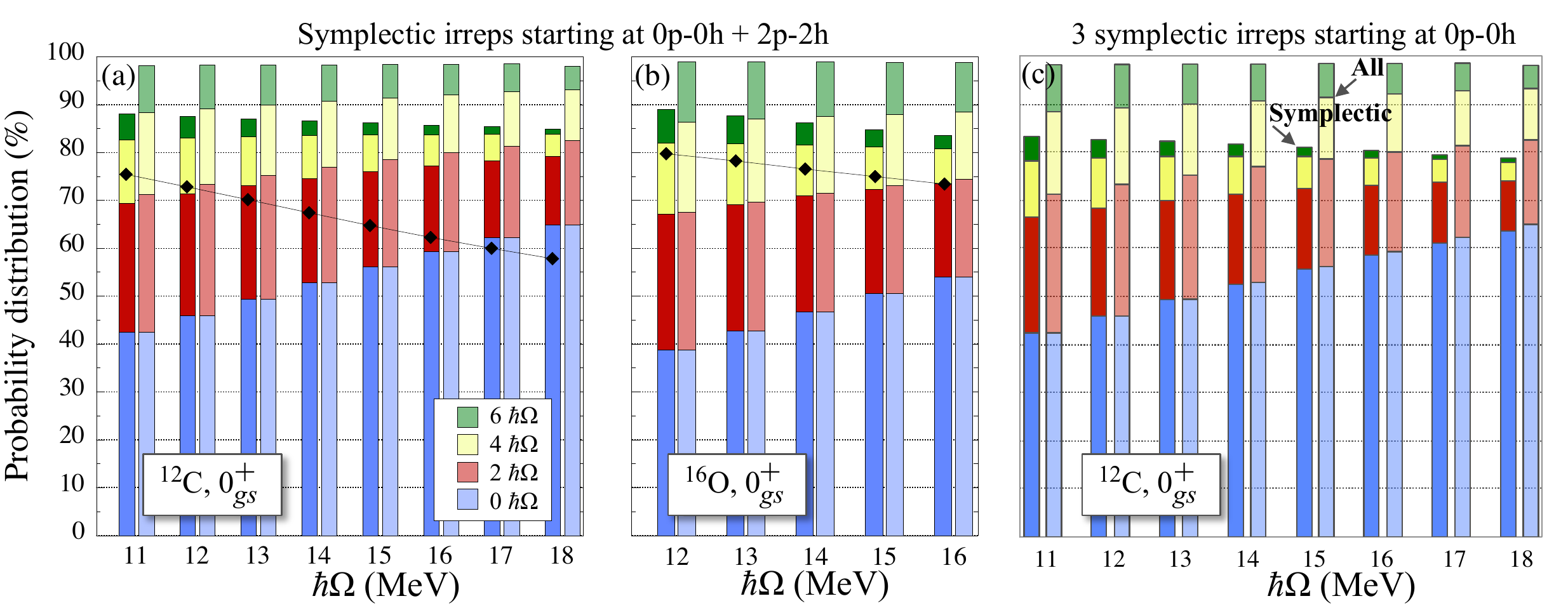}
}
\caption{
Probability distribution 
over $0$\ho~ (blue, lowest) to $6$\ho~ (green, highest) subspaces
for the $N_{\rm max}=6$ NCSM ground state (right bars) of (a) $^{12}$C and (b) $^{16}$O, or equally all $N_{\rm max}=6$ symplectic irreps, and for the
case of the most dominant  \ph{0} + 2\ho~\ph{2} \SpR{3} irreps 
(left bars)  together with the leading \SpR{3} irrep contribution (black diamonds), $(0~4)$ for $^{12}$C and $(0~0)$ for $^{16}$O, as a function of
the \ho~ oscillator strength.
(c) For comparison, the same results without the \ph{2} contribution, that is, for the three \ph{0} \SpR{3} irreps  for $^{12}$C are shown (similarly for the $2^+_1$ and $4^+_1$ states \cite{DytrychSBDV_PRCa07}). Note that a ``2\ho~\ph{2} \SpR{3} irrep"  refers to an irrep that consists of symplectic excitations, driven by $A^{(2\,0)}$ as described in Sec. \ref{sp3r}, over a 2\ho~\ph{2} bandhead of two particles up a shell  (e.g., see Fig. \ref{sp3Rpicture}, 3d vertical cone), and includes configurations that are inaccessible by symplectic excitations built on the \ph{0} bandheads (e.g., see Fig. \ref{sp3Rpicture}, 1st and 2nd vertical cones). Figure adapted from Refs. \cite{DytrychSBDV_PRL07,DytrychSBDV_PRCa07}.
}
\label{C12prblty_vs_hw}
\end{figure}

In addition, the dominant \SU{3} basis states at each
$N\hbar\Omega$ subspace ($N=0, 2, 4,\dots$) are typically those with
$(\lambda\, \mu)$ quantum numbers given by
\begin{equation}
\lambda + 2\mu = \lambda_{0} + 2\mu_{0} + N	
\label{eq:Sp2RSelection}
\end{equation}
where $\lambda_{0}$ and $\mu_{0}$ denote labels of the leading \SU{3} irrep 
in the $0\hbar\Omega$ ($N=0$) subspace [in general, this implies  $\lambda + 2\mu = \lambda_{\sigma} + 2\mu_{\sigma} + N-N_\sigma$, for a $(\lambda_{\sigma}\,\mu_{\sigma})$ leading irrep in an $N_\sigma$ subspace]. 
Clearly, this regular pattern of \SU{3} configurations, which contribute largely to the low-lying states (Table \ref{tab:dominantConfig}),
further supports the presence of symplectic \SpR{3} symmetry. This can be seen from
the fact that $(\lambda\, \mu)$ configurations that satisfy
condition~(\ref{eq:Sp2RSelection}) can be determined from the leading \SU{3}
irrep $(\lambda_{0}\, \mu_{0})$ through a successive application of a specific
subset of the \SpR{3} symplectic $2\hbar\Omega$ raising operators, Eq. (\ref{sp3RgenA}).  This subset
is composed of the three operators, ${A}_{zz},{A}_{zx}$, and ${A}_{xx}$ (expressed in Cartesian coordinates), that distribute two oscillator quanta in $z$ and $x$
directions, but none in $y$ direction, thereby inducing \SU{3} configurations
with ever-increasing intrinsic quadrupole deformation.  These three operators
are the generators of the \SpR{2} subgroup~\cite{PetersonH80}, according to $\SpR{3}\supset \SpR{2}$, and give rise to deformed configurations
that are energetically favored by an attractive quadrupole-quadrupole
interaction~\cite{Rowe85}.  

Furthermore, there is an apparent hierarchy among states that fulfill
condition~(\ref{eq:Sp2RSelection}).  In particular, the $N\hbar\Omega$
configurations with $(\lambda_{0}\!+\!N\,\,\mu_{0})$, the so-called stretched
states, carry a noticeably higher probability than the others. For instance,
the $(2\!+\!N\,\,0)$ stretched states contribute at the 85\% level to the
ground state of $^{6}$Li, as can be readily seen in
Fig.~\ref{gsStructure}.
Moreover, the dominance of the stretched states is rapidly increasing with
increasing $N_{\max}$, as illustrated in
Fig~\ref{fig:projections}.  The sequence of the stretched states is formed by
consecutive applications of $\hat{A}_{zz}$ over the leading \SU{3} irrep.  The $\hat{A}_{zz}$ operator is the generator of
$\SpR{1}$  subgroup according to $\SpR{3}\supset\SpR{2}\supset\SpR{1}$.
This translates into distributing $N$ oscillator quanta along the direction of
the $z$-axis only and hence rendering the largest possible deformation.  The
important role of the stretched configurations for the description of the
rotational bands in $N=Z$ even-even nuclei has been recognized heretofore using a
simple microscopic Hamiltonian~\cite{Arickx76}. 
\begin{table}[b!]
\caption{Probability distribution of {\it ab initio} eigenstates for $^{12}$C and $^{16}$O across the 
dominant \ph{0} and 2\ho~\ph{2} \SpR{3} irreps, \ho=15 MeV, as compared to the complete $N_{\rm max}=6$ space (``All").\label{TABLE_15MeV}}
\begin{center}
{\small 
\begin{tabular}{crrrrr} 
 & $0\hbar\Omega$ & $2\hbar\Omega$ & $4\hbar\Omega$ & $6\hbar\Omega$ & Total \\
\hline
\multicolumn{6}{c}{$^{12}$C, $0^+_{gs}$} \\
\hline
$(0\;4)S=0$     & $46.26$ & $12.58$ & $4.76$  & $1.24$  & $64.84$\\
                   $(1\;2)S=1$     & $4.80$  & $2.02$  & $0.92$  & $0.38$  & $8.12$\\
   	                $(1\;2)S=1$ & $4.72$  & $1.99$  & $0.91$  & $0.37$  & $7.99$ \\
             2\ho~\ph{2}            &         & $3.46$  & $1.02$  & $0.35$  & $4.83$\\
\cline{2-6}
                   Total selected      & $55.78$& $20.05$& $7.61$ &  $2.34$  & $85.78$ \\
  All          & $56.18$ & $22.40$ & $12.81$ &  $7.00$  & $98.38$ \\
\hline
\multicolumn{6}{c}{$^{12}$C, $2^+_1$} \\
\hline
 $(0\;4)S=0$ & $46.80$ & $12.41$ & $4.55$  & $1.19$     &  $64.95$\\
                    $(1\;2)S=1$ & $4.84$  & $1.77$  & $0.78$  &  $0.30$    &  $7.69$ \\
                    $(1\;2)S=1$ & $4.69$  & $1.72$  & $0.76$  &  $0.30$    &  $7.47$ \\
                    2\ho~\ph{2} &         & $3.28$  & $1.04$  &  $0.38$    &  $4.70$\\
\cline{2-6}
         Total  selected   & $56.33$ & $19.18$ & $7.13$ &  $2.17$  & $84.81$ \\
   All       & $56.18$ & $21.79$ & $12.73$ &  $7.28$ & $98.43$ \\
\hline
\multicolumn{6}{c}{$^{12}$C, $4^+_1$} \\
\hline
 $(0\;4)S=0$ & $51.45$ & $12.11$ & $4.18$   & $1.04$     & $68.78$\\
                    $(1\;2)S=1$ & $3.04$  & $0.95$  & $0.40$   & $0.15$     & $4.54$ \\
                    $(1\;2)S=1$ & $3.01$  & $0.94$   & $0.39$  & $0.15$     & $4.49$ \\
                   2\ho~\ph{2}  &         & $3.23$   & $1.16$  & $0.39$     & $4.78$\\
\cline{2-6}
           Total selected & $57.50$ & $17.23$ & $6.13$  &  $1.73$  & $82.59$ \\
  All        & $57.64$ & $20.34$ & $12.59$ &  $7.66$  & $98.23$ \\
\hline
\multicolumn{6}{c}{$^{16}$O, $0^+_{gs}$} \\
\hline
 $(0\;0)S=0$ & $50.53$ & $15.87$ & $6.32$ & $2.30$ & $75.02$ \\
           2\ho~\ph{2} & & $5.99$ & $2.52$ & $1.32$ & $9.83$ \\
\cline{2-6}
                  Total  selected     & $50.53$ & $21.86$  & $8.84$  &  $3.62$ & $84.85$ \\
  All                     & $50.53$ & $22.58$ & $14.91$ &  $10.81$ & $98.83$ \\
\end{tabular}
}
\end{center}
\end{table}

%%%%%%%%%%%%%%%%%%%%%
This is consistent with our earlier findings of a clear symplectic \SpR{3} structure with the same pattern
(\ref{eq:Sp2RSelection}), as unveiled  in \textit{ab initio} eigensolutions for $^{12}$C and $^{16}$O~\cite{DytrychSBDV_PRL07}.
These eigenstates, determined within the framework of  the no-core shell model using the JISP16  realistic interaction, have been found to typically project 
at the 85-90\% level onto a few  symplectic vertical slices, starting at the most deformed \ph{0} and 2\ho~\ph{2} configurations, that span 
only a small fraction of the complete model space (Fig. \ref{C12prblty_vs_hw}a\&b and Table~\ref{TABLE_15MeV}). The results are nearly independent of  whether the bare or renormalized effective interactions are used in the analysis and reveal a
clear dominance, for any \ho, of the \ph{0} \SpR{3} irreps  (Fig. \ref{C12prblty_vs_hw}c and Table~\ref{TABLE_15MeV}).  In particular, for the $0^+_{gs}$ and the lowest $2^+$ and $4^+$ states in $^{12}$C,  there are nonnegligible overlaps for only 3 of the 13 \ph{0} \SpR{3} irreps.
Moreover, these 3 irreps are the same for these states with nearly $J$-independent contributions, thereby pointing to a clear rotational structure. 
In addition, the largest contribution comes from the \SpR{3} irrep built on the most deformed 0\ho~configuration (Fig. \ref{C12prblty_vs_hw}, black diamonds).

\begin{figure}[th]
\centerline{
\includegraphics[width=0.75\textwidth]{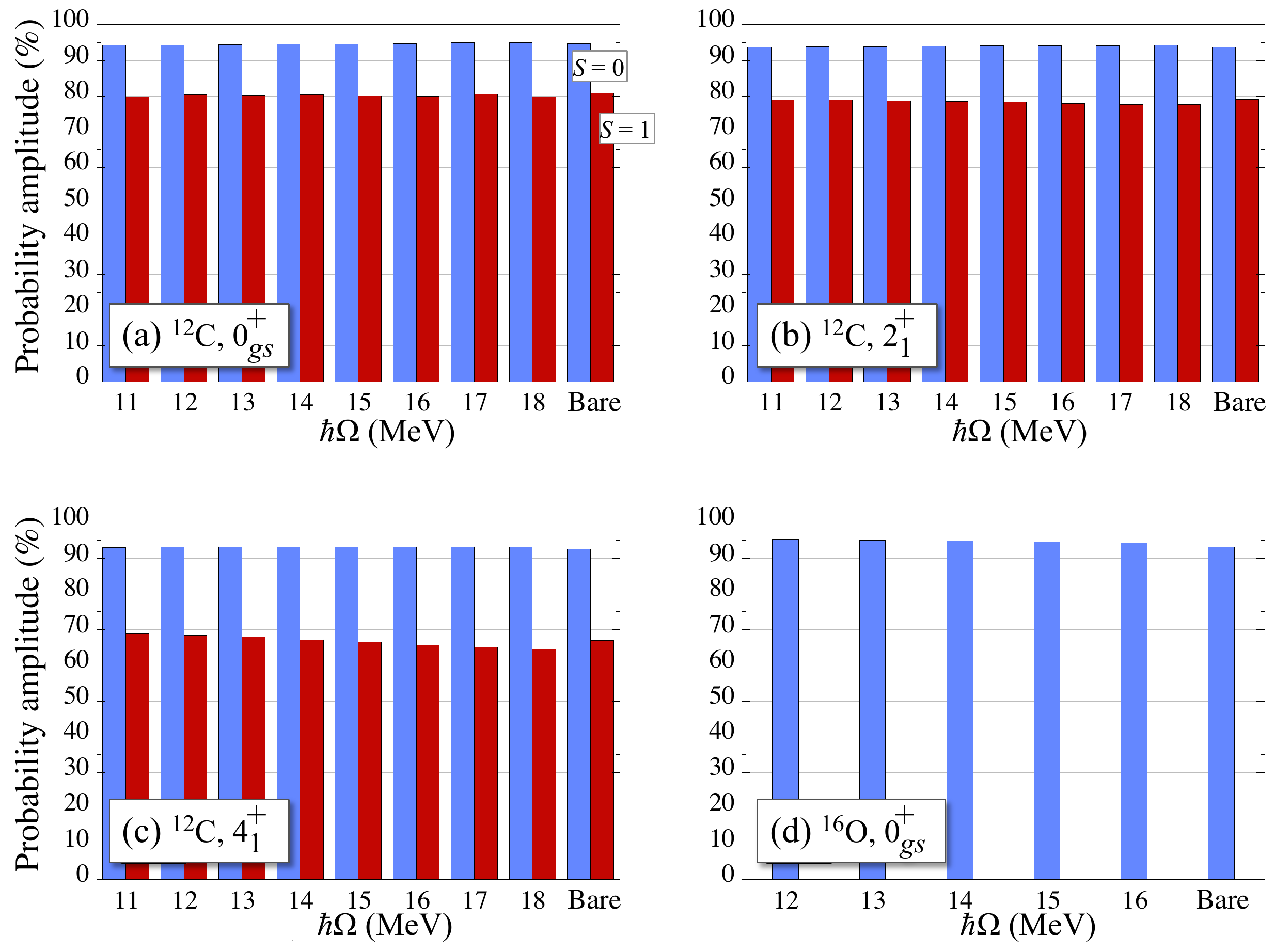}
}
\caption{
Projection of the most dominant \ph{0}+\ph{2} $S=0$ (blue, left bars) and $S=1$ (red, right bars) \SpR{3} irreps onto the
corresponding spin components of the NSCM wave functions for (a) $0^{+}_{gs}$, (b)
$2^{+}_1$, and (c)
$4^{+}_1$ in
$^{12}$C and (d) $0^{+}_{gs}$ in $^{16}$O, for the effective JISP16 interaction for different
$\hbar\Omega$ oscillator strengths and for the bare JISP16 interaction (for $\ho=15$ MeV). Figure adapted from Ref. \cite{DytrychSBDV_PRL07}.
}
\label{Overlaps_SpinScaled}
\end{figure}

A striking property of the low-lying eigenstates is revealed when their spin-0 and spin-1
components  are examined, namely, the \SpR{3} symmetry within each spin component and hence the geometry of the
nucleon system being described is nearly independent of the \ho~oscillator strength and regardless of whether the bare or the
effective interactions are used (Fig.~\ref{Overlaps_SpinScaled}).
The symplectic structure is preserved, only the \SpR{3} irrep contributions change, as illustrated in Fig. \ref{C12prblty_vs_hw}, because the $S=0$ ($S=1$) part of the NCSM eigenstates decrease (increase) towards higher \ho~ frequencies, as shown in Fig. \ref{Spin_Mixing}.  This suggests that the  Okubo-Lee-Suzuki transformation, which effectively compensates for the finite space truncation by  renormalization of
the bare interaction, does not affect the \SpR{3} symmetry structure of the spatial part of the
wave functions. Hence, the symplectic structure detected in the analysis for
smaller model spaces is what would emerge in NSCM evaluations with a sufficiently large
model space to justify use of the bare interaction. 

\begin{figure}[th]
\centerline{
\includegraphics[width=0.75\textwidth]{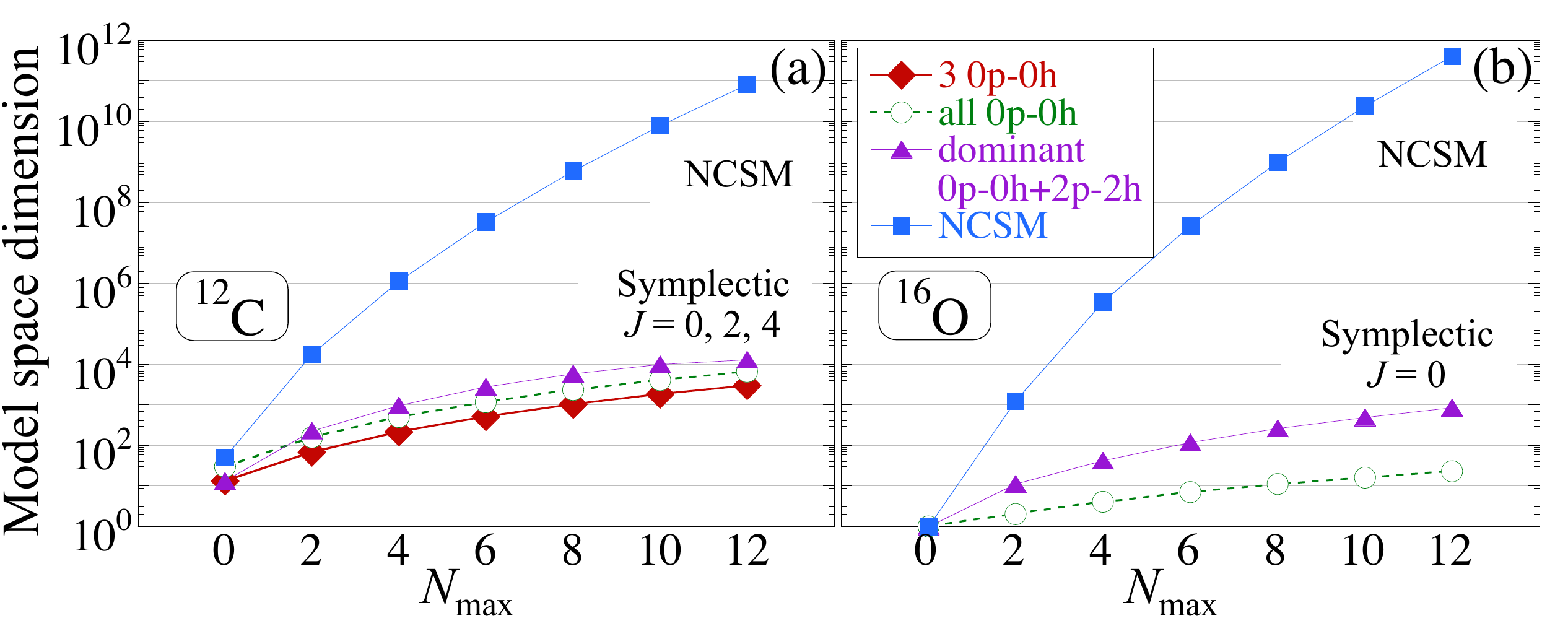}
}
\caption{
NCSM model space dimension as a function of the $N_{\rm max}$ cutoff, as compared to that of the \SpR{3} subspace considered in Fig. \ref{C12prblty_vs_hw}: (a) $J=0,2,$ and
$4$ for
$^{12}$C, and (b) $J=0$ for
$^{16}$O. Figure adapted from Ref. \cite{DytrychSBDV_PRL07}.
}
\label{dimMdlSpace}
\end{figure}

The typical dimension of a symplectic irrep in
the $N_{\rm max}=6$ space is on the order of $10^{2}$ as compared to
$10^{7}$ for the complete NCSM $M$-scheme basis.
As $N_{\max}$ increases the dimension of the  $J=0,2,$ and
$4$ \SpR{3}-scheme model space  built on the \ph{0} \SpR{3} irreps for $^{12}$C
grows very slowly  and remains a small fraction of the complete model
space even when the most dominant 2\ho~\ph{2} \SpR{3} irreps are included (Fig. \ref{dimMdlSpace}a). The space reduction is even
more dramatic in the case of $^{16}$O (Fig.
\ref{dimMdlSpace}b). This means that a space spanned by a set of symplectic basis
states  is computationally manageable even when they extend to large $N_{\rm max}$.

\subsection{Dominant \SU{3} modes in bare and effective $NN$  interactions \label{sdt_NN}}
The nucleon-nucleon interaction itself possesses a remarkable  \SU{3} structure \cite{LauneyDDSD15}. This is observed in the decomposition of the $NN$ interaction into $\SU{3}\times\SU{2}_{S_0}\times\SU{2}_{T_0}$ tensors (isoscalar  interactions will be henceforth considered). This is analogous to the unitary transformation of a $V_{NN}$ two-body interaction represented in an $M$-scheme  HO basis to a $JT$-scheme basis, which renders $V_{NN}$ as only one $\SU{2}_{J_0}\times\SU{2}_{T_0}$  tensor of rank $J_0=0$ and $T_0=0$ (a scalar with respect to rotations in coordinate and isospin space).
For \SU{3} interaction tensors, the $(\lambda_0\,\mu_0)=(0\, 0)$ scalar  does not mix nuclear deformation in analogy to the isoscalar part of an interaction that does not mix isospin values. In addition, the $(\lambda_0\,  \mu_0)$ interaction components with $\lambda_0= \mu_0$ are almost diagonal, that is, connect configurations within a few shells, while  interaction components with a large difference $|\lambda_0-\mu_0|$ typically couple to high momenta.
\begin{figure}[b!]
\begin{center}
\includegraphics[width=0.33\textwidth]{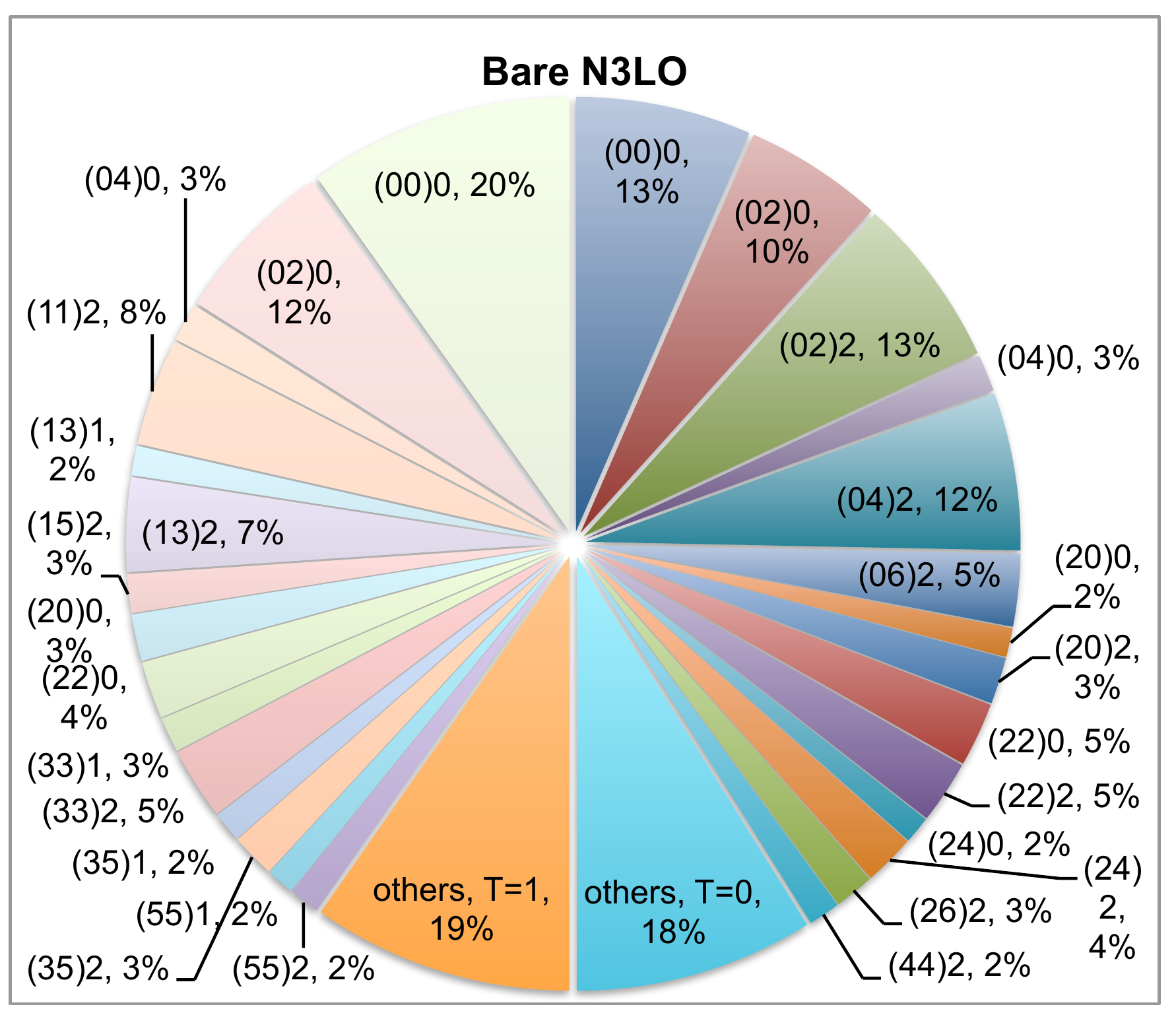}
\includegraphics[width=0.33\textwidth]{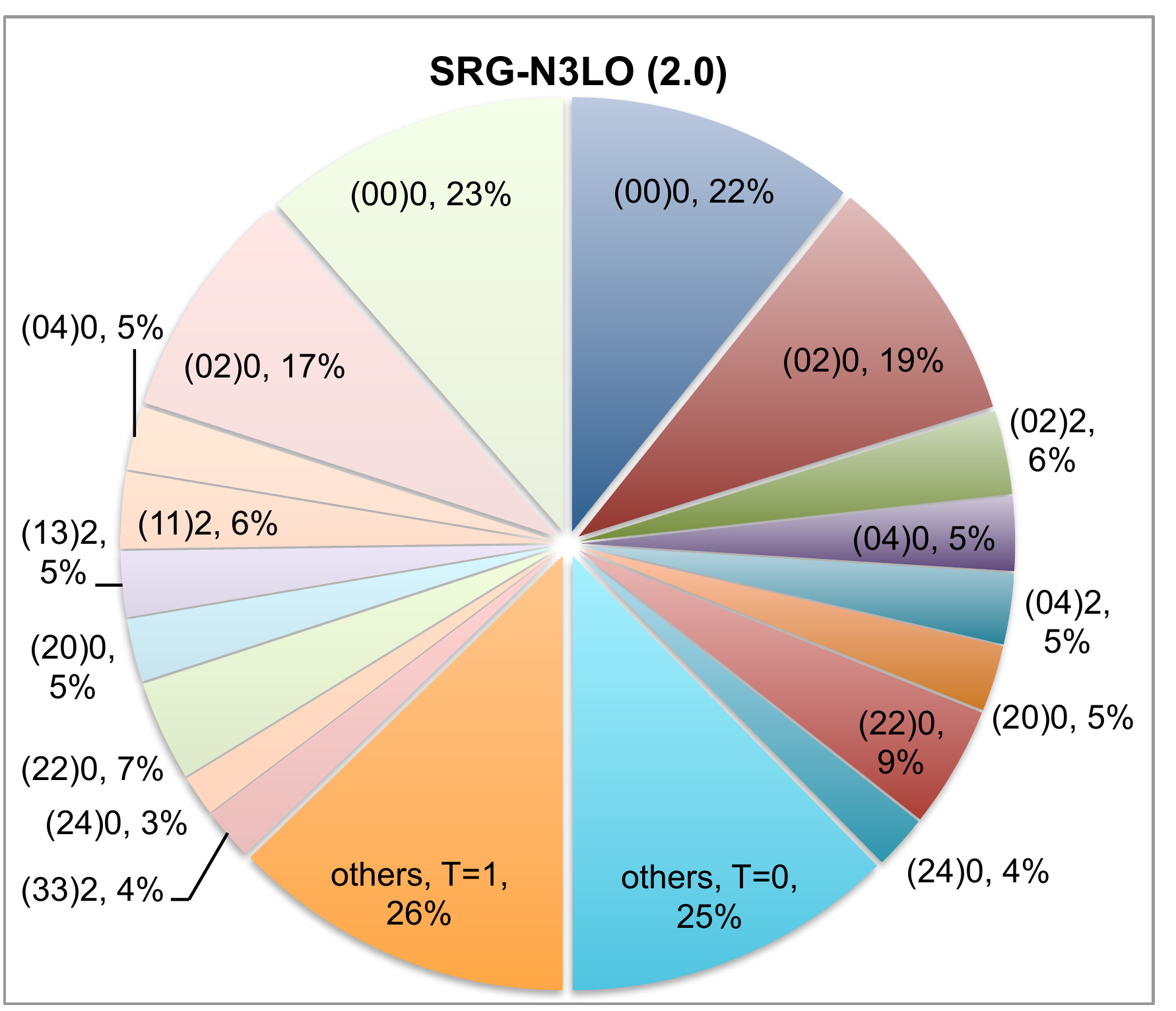}\\
(a) \hspace{2.3in} (b) \\
\includegraphics[width=0.33\textwidth]{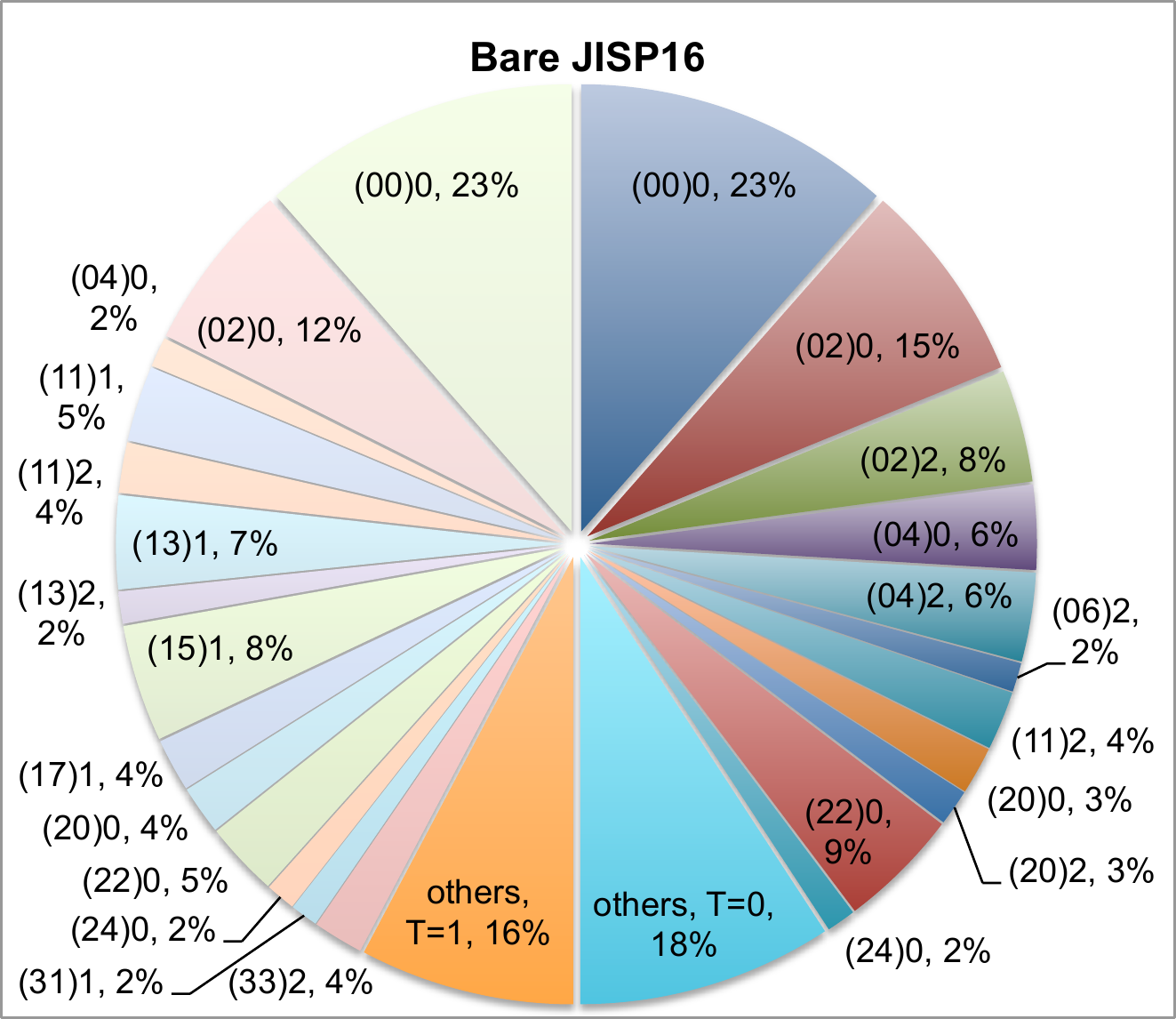}
\includegraphics[width=0.33\textwidth]{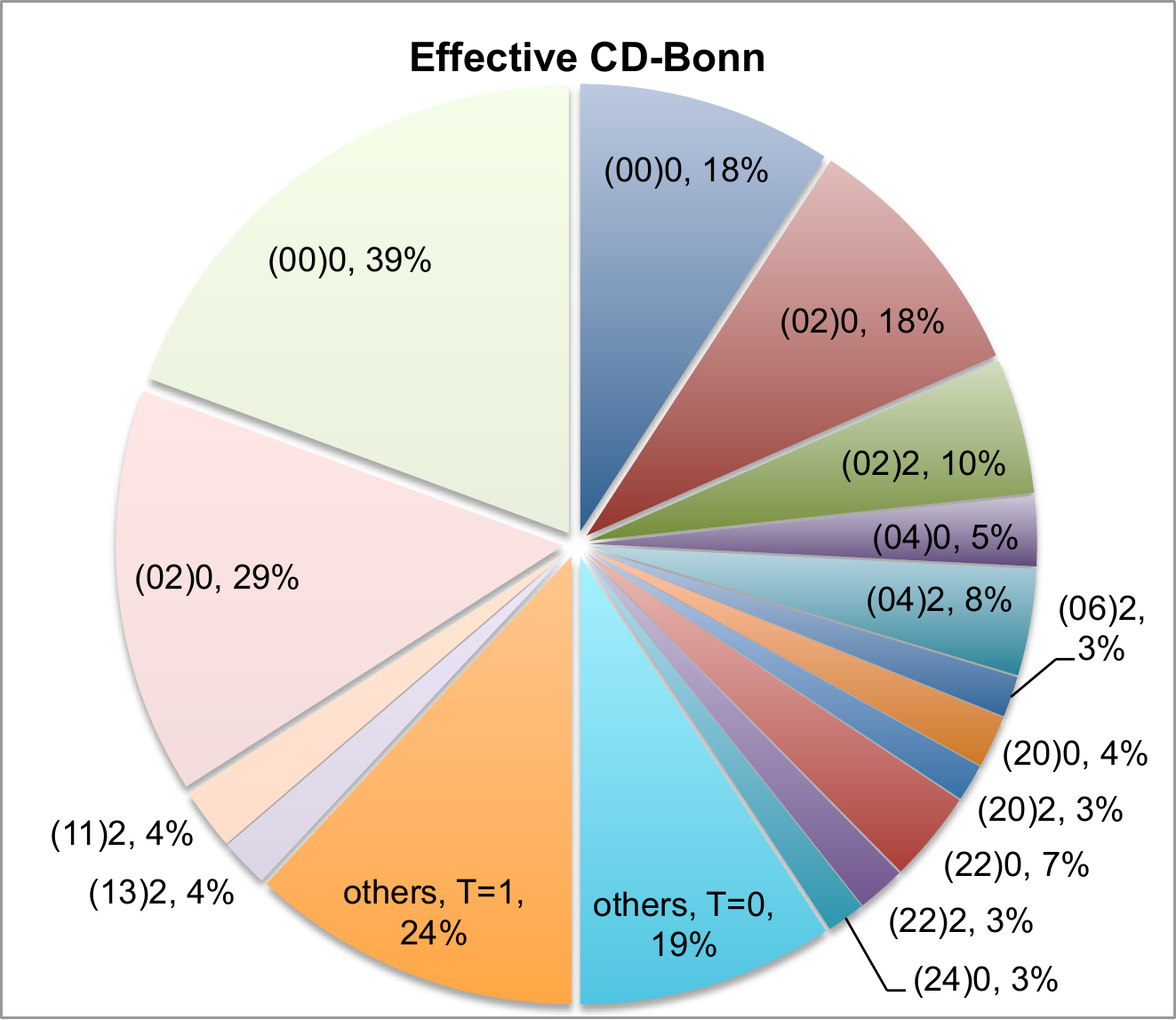}\\
(c) \hspace{2.3in} (d) 
\end{center}
\caption{\label{bareJISP16hw15_Vsu3}
Relative strengths of the $T=1$ (left half) and $T=0$ (right half) interaction tensors labeled by $(\lambda_0\,\mu_0)S_0$  for  $N_{\max}=6$ for $p$-shell nuclei. The contributions of the conjugate $(\mu_0\, \lambda_0)$ tensors are not shown, but are equal to the ones of their $(\lambda_0\,\mu_0)$ counterparts. (a) Bare N$^3$LO interaction \cite{EntemM03} for $\hbar\Omega=11$ MeV, and (b) its renormalized counterpart using Similarity Renormalization Group (SRG) \cite{BognerFP07} with a cutoff of $\lambda_c=2$ fm$^{-1}$. (c) Bare JISP16 interaction \cite{ShirokovMZVW07} for $\hbar\Omega=15$ MeV. (d) Effective interaction, based on the bare CD-Bonn interaction \cite{Machleidt01} for $\hbar\Omega=15$ MeV and using the Okubo-Lee-Suzuki (OLS) renormalization technique \cite{LeeSuzuki80}.
}
\end{figure}

To study the dominant pieces of various $NN$ interactions and their similarity, we utilize tools developed in spectral distribution theory (SDT) \cite{French66,FrenchR71,ChangFT71,HechtD74}. Specifically, we employ second-order energy moments widely used as measures of the overall strength of an interaction (norm of a many-body Hamiltonian matrix $H$) and its similarity to other interactions (inner product of two Hamiltonian matrices $H$ and $H'$). As is well-known, the smaller the norm, the weaker the interaction (the more compressed the energy spectrum of $H$), while a zero inner product indicates lack of any overlap between $H$ and $H'$ or no similarity  \cite{FrenchR71}.    It is worth noting that, while not utilized in this study, SDT is actually a many-body microscopic approach. It  originated as an alternative microscopic approach to the conventional shell model technique. The efficacy of the theory stems from the fact that typically low-order energy moments dominate the many-particle spectroscopy as a result of leading surviving features of the underlying microscopic interaction. Convergence to the shell-model results improves as higher-order energy moments are taken into account or  toward the limit of many particles occupying a much larger available single-particle space. The theory provides the means to calculate important average contributions, nuclear level densities,  degree of symmetry violation such as parity/time-reversal violation, nuclear structure and reactions, quantum chaos measures, as well as to understand dominant features of realistic $NN$ interactions (see, e.g., \cite{SviratchevaDV08}) and the effect of SRG-induced  three-body forces \cite{LauneyDD12} (see the book \cite{KotaH10} on SDT and its applications).  

By examining the norm of each \SU{3} component of $V_{NN}$,  we find a dominance of the $(0\, 0)$ scalar part, independent of the  $NN$ realistic interaction (Fig. \ref{bareJISP16hw15_Vsu3}). It is  followed by  spin-zero $(0\, 2)$, and its conjugate $(2\, 0)$, and by  $(0\, 4)+(4\, 0)$. These \SU{3} modes are the ones that also appear in the kinetic energy, the monopole operator, as well as the $Q\cdot Q$ interaction.  Other dominant modes are the spin-2  $(0\, 2)+(2\, 0)$, as well as $(1\, 1)$, which can be linked to the tensor force. Additional important tensors include  $(1\, 1)$, $(2\, 2)$, $(3\, 3)$, and etc., which typically dominate for the pairing interaction or contact term. These features, we find, repeat for various realistic bare interactions (Fig. \ref{bareJISP16hw15_Vsu3}a and c) and, more notably, are further  enhanced for their renormalized counterparts (Fig. \ref{bareJISP16hw15_Vsu3}b and d) \cite{LauneyDDSD15}.

\section{Understanding emergent collectivity  from first principles with SA-NCSM \label{appSANCSM}}

\subsection{Symmetry-guided concept \label{SpConcept}}
\begin{figure}[th]
\centerline{
\includegraphics[width=0.85\textwidth]{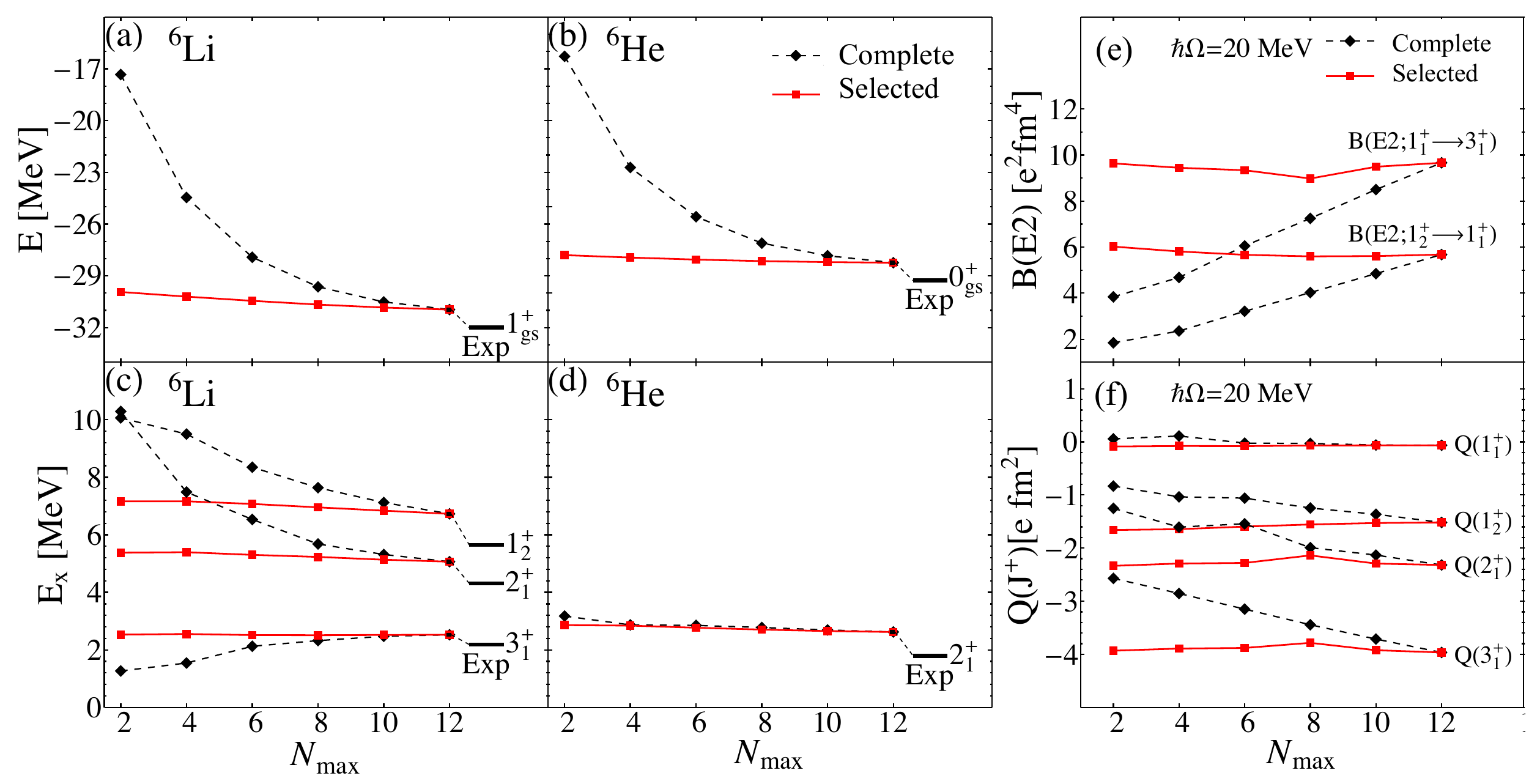}
}
\caption{
Ground-state and excited energies of $^{6}$Li and $^{6}$He, (a)-(d), together with (e) $E2$ transition probabilities and (f) quadrupole moments  for $^{6}$Li, shown for the complete $N_{\max}$ model spaces (dashed black curves) and for the
 $\langle N_{\max}\rangle 12$ SA-NCSM selected  model spaces
(solid red lines).  Results shown are for JISP16 and $\hbar\Omega=20$ MeV.  
Note the relatively large changes in the results when the complete space is increased from $N_{\max}=2$ to $N_{\max}=12$
as compared to the nearly constant $\langle N_{\max} \rangle12$ SA-NCSM outcomes. Figure adapted from Ref. \cite{DytrychLMCDVL_PRL12}.
}
\label{6LiT0Spectrum}
\end{figure}

As discussed above, the low-lying states in light nuclei exhibit orderly patterns, as illustrated  in Fig. \ref{gsStructure}, that favor spatial configurations 
with strong quadrupole deformation and complementary low intrinsic 
spin values,  a picture that is consistent with the nuclear symplectic model. 
This  suggests a symmetry-guided basis selection philosophy  that allows the SA-NCSM to obtain results in much smaller spaces that are nearly indistinguishable from the complete basis counterparts. Specifically, the outcome supports a symmetry-guided concept \cite{DytrychLMCDVL_PRL12}, a key feature of the SA-NCSM, namely, the relevant model space can be systematically selected, using a quantified cutoff, starting from the largest deformation/lowest spin within a low-$N\ho$ subspace and associated symplectic excitations thereof (right sector of Fig. \ref{gsStructure}), and including ever smaller deformation until convergence of results is achieved. Specifically, 
one can take advantage of dominant symmetries to relax and refine the 
definition of the SA-NCSM model space, which for the NCSM is fixed by 
simply specifying the $N_{\max}$ cutoff.  SA-NCSM model spaces can be characterized by a pair of numbers, 
$\langle N^{\bot}_{\max} \rangle N^{\top}_{\max}$ (schematically illustrated in Fig. \ref{gsStructure}), which implies inclusion of the full 
space up through $N^{\bot}_{\max}$, and a subset of deformation/spin configurations beyond this, up through $N^{\top}_{\max}$.
The validity of the symmetry-guided concept can be illustrated with SA-NCSM results 
for $p$-shell nuclei obtained in model spaces, which are expanded beyond a
complete $N^{\bot}_{\max}$ space ($N^{\bot}_{\max}=2,4,\dots,10$) by using a relatively few dominant intrinsic
spin and deformation components  up through $N^{\top}_{\max}=12$  (Fig. \ref{6LiT0Spectrum}). 
Clearly, the results indicate that the observables obtained in the much smaller symmetry-guided selected spaces are excellent approximations to the corresponding $N_{\max}=12$ complete-space counterparts. 
A crucial advantage of this  symmetry-guided concept is that SA-NCSM can carry out  investigations beyond the current $N_{\max}$ NCSM limits.
Within this context, it is important to emphasize again that for model spaces truncated according to
$(\lambda\,\mu)$ irreps and spins $(S_{p}S_{n}S)$, the spurious
center-of-mass motion is factored out exactly~\cite{Verhaar60,Hecht71,Millener92}, 
which represents an important advantage of this scheme.

\begin{figure}[th]
\begin{center}
\includegraphics[width=0.48\columnwidth]{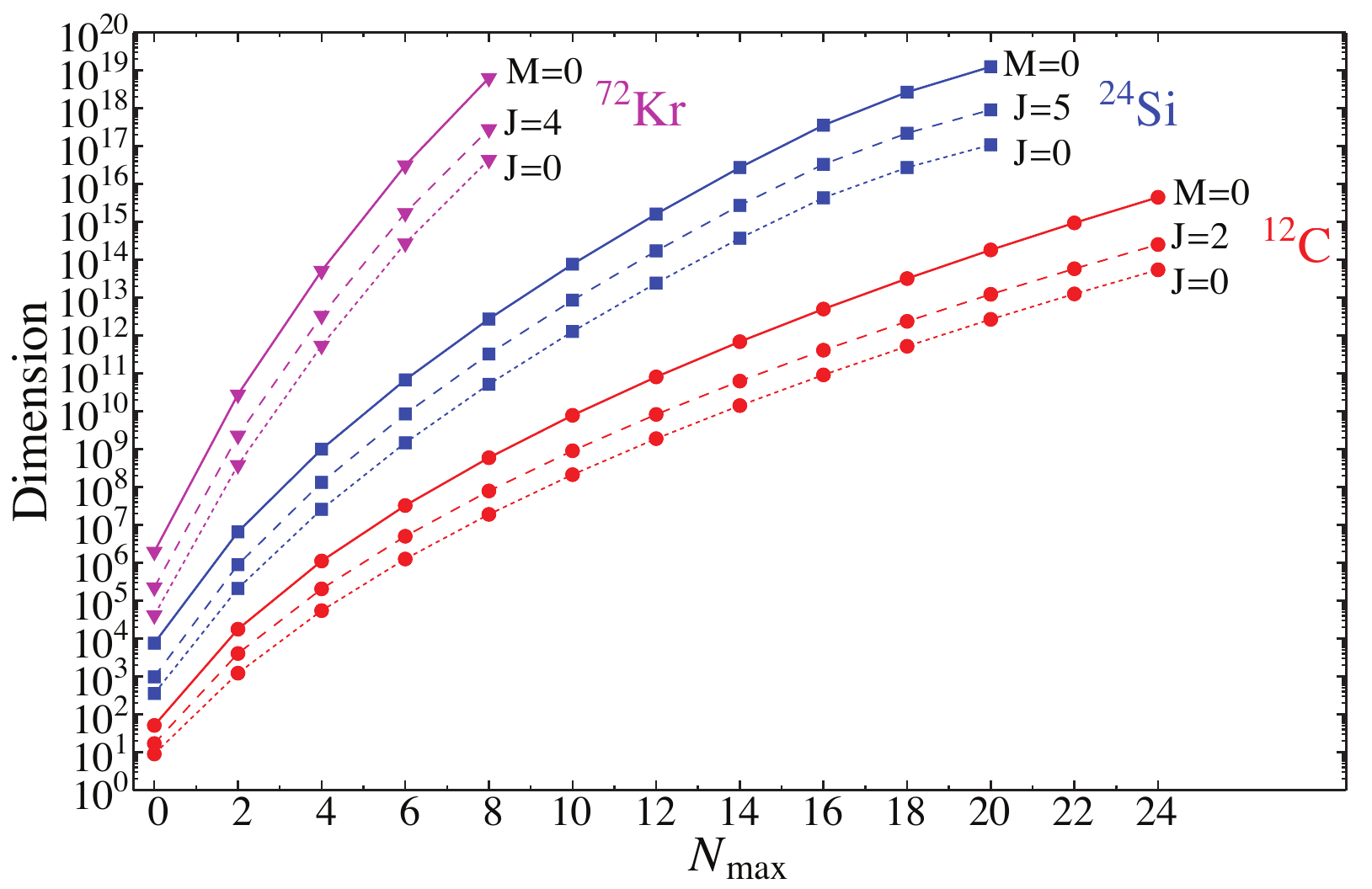}
\includegraphics[width=0.48\columnwidth]{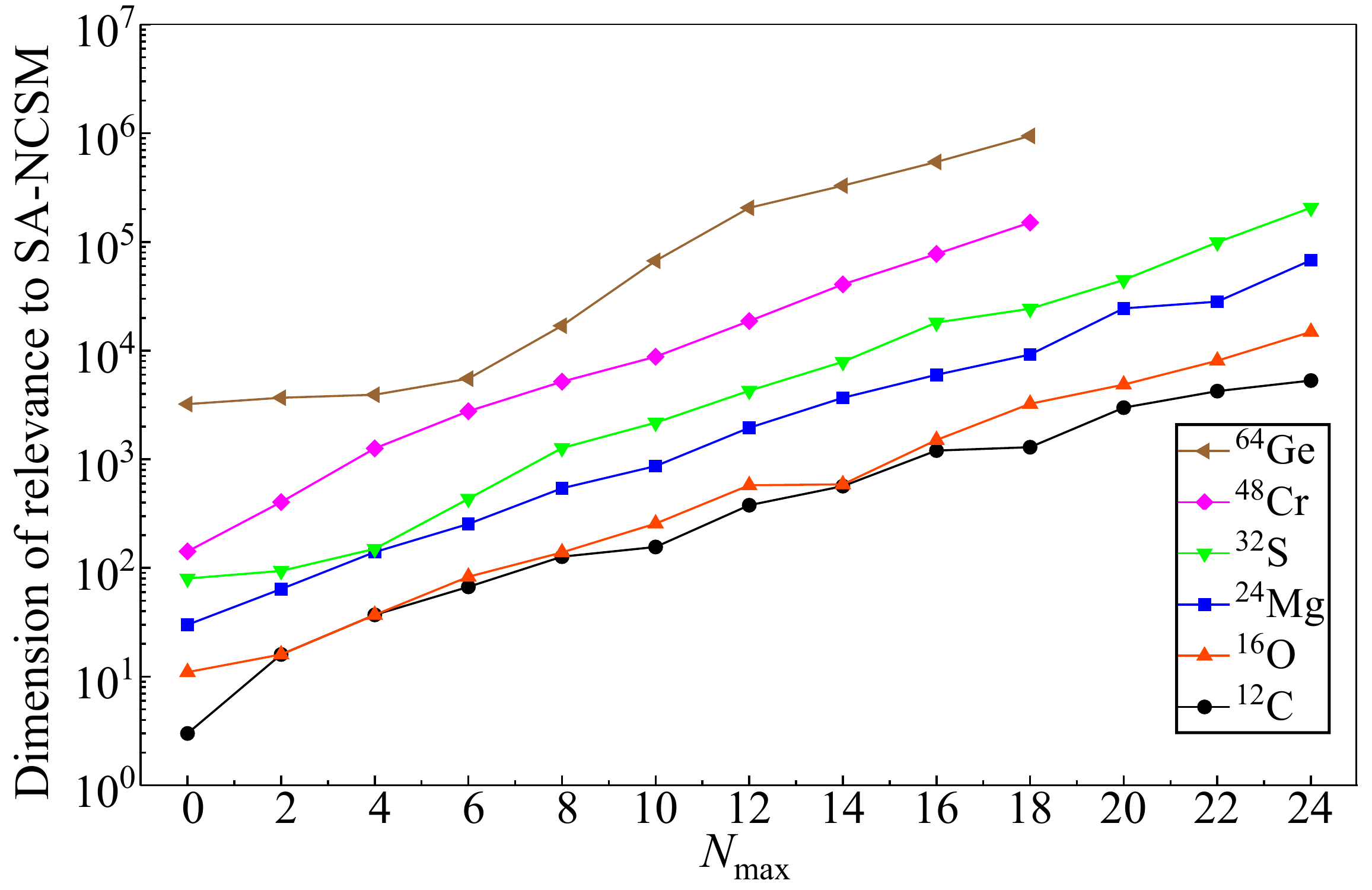}\\
(a) \hspace{3.2in}(b)
\caption{
(a)  Dimensions of positive-parity model spaces as functions of $N_{\rm max}$ for selected nuclei for the $M$ scheme (solid curves) and the $J$ scheme (dashed and dotted curves).
(b) Dimension of relevance to the SA-NCSM, namely, number of  many-nucleon single-shell  basis configurations that generate the SA-NCSM model space. The increase in particle number is shown for selected nuclei in $p$, $sd$, and $pf$ shells. 
}
\label{fig:dims}
\end{center}
\end{figure}

The number of basis states used, e.g., for each $^6$Li state, is
only about 10-12\% for $\langle 2\rangle 12$, $\langle 4\rangle 12$, $\langle
6\rangle 12$, 14\% for $\langle8\rangle 12$, and 30\% for $\langle 10\rangle
12$ as compared to the number for the complete $N_{\max}=12$ SA-NCSM model
space.  To add a degree of computational  specificity to
this, the runtime of the SA-NCSM code exhibits a quadratic
dependence on the number of  $(\lambda\,\mu)$ and  $(S_p S_n S)$ irreps --
there are $1.74\times10^{6}$ such irreps for the complete $N_{\max}$ = 12 model
space of $^{6}$Li  (see Fig. \ref{fig:dims} for selected nuclei), while only 8\%-30\% of these are
retained in the selected space. The net
result is that calculations in the selected spaces
require one to two orders of magnitude less
computational time than SA-NCSM calculations for the complete $N_{\max}=12$
space. 
(Going into detail, which undoubtedly depends on the current  code implementation and available computational resources, model spaces up to 15 active HO shells for intermediate-mass nuclei with dimensions ranging to $10^8$ basis states presently require several hours, while utilizing 22,425 Cray XE6 compute nodes on the Blue Waters system.)
Indeed, with the help of HPC resources, 
the use of such basis spaces in {\it ab initio} studies is manageable as well as expandable; that is, one expects to be able to extend the reach of our SA-NCSM scheme from applications that are doable today to the larger spaces and heavier nuclear systems of tomorrow, utilizing at each stage the accuracy and predictive power of the {\it ab initio} approach. 

%%%%%%%%%%%%%%%%%%
%Li-6, He-6, Be-8, C-12

\subsection{Spectral properties of  light and intermediate-mass nuclei \label{SANCSMspectra}}
In Ref. \cite{DytrychLMCDVL_PRL12}, {\it ab initio} SA-NCSM spectral properties of  $p$-shell nuclei  are presented  for selected  model spaces that expand up to $N_{\rm max}=12$.   Energies and physical observables are calculated with a Coulomb plus JISP16 $NN$ interaction for $\hbar\Omega$ values ranging from $17.5$ up to $25$ MeV, along with the
Gloeckner-Lawson prescription~\cite{Lawson74} for elimination of spurious
center-of-mass excitations. 
\begin{figure}[th]
\begin{center}
\includegraphics[width=0.75\textwidth]{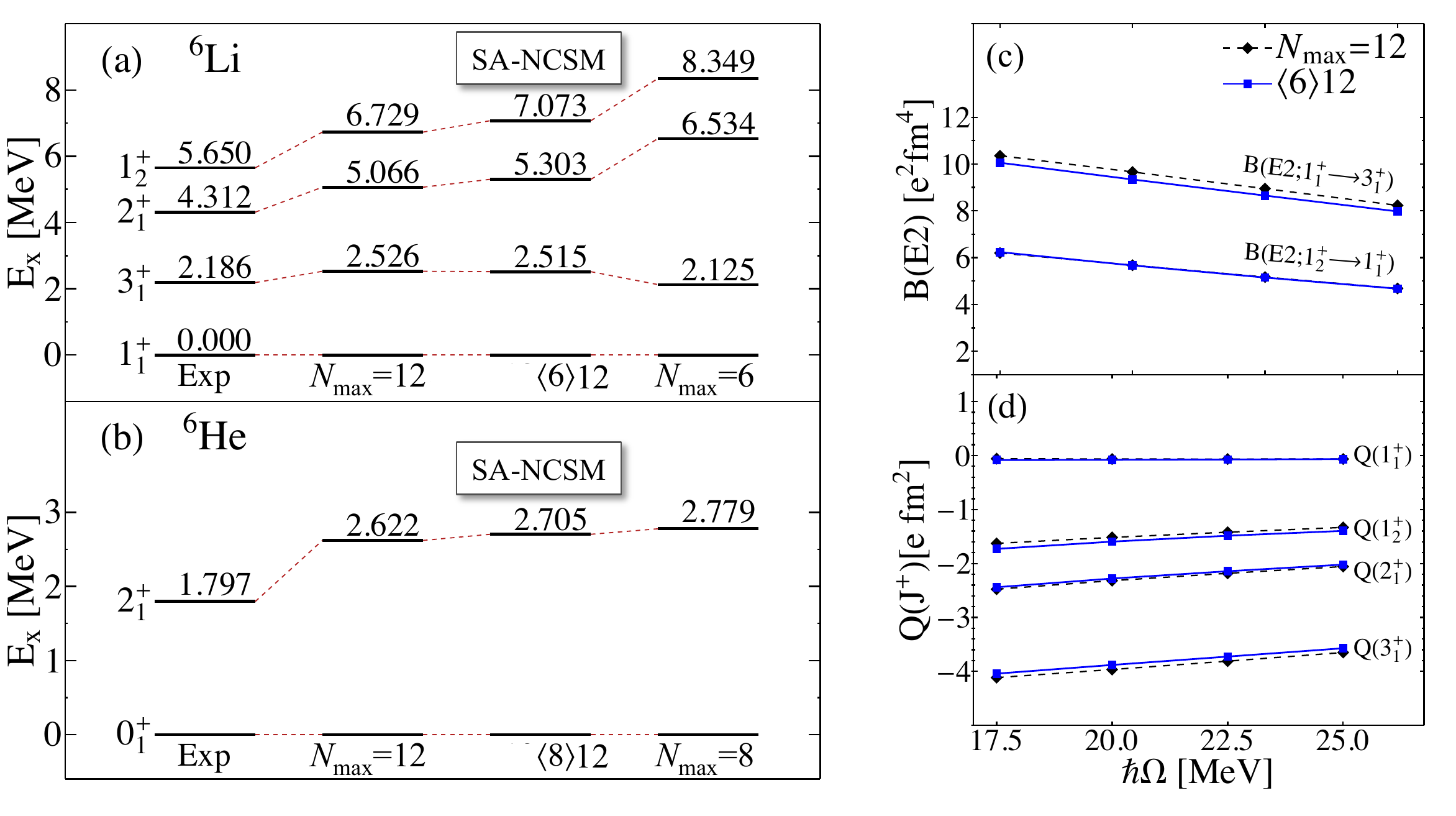}
\caption{
	Experimental (from Ref. \cite{Tilley2002}) and theoretical excitation energies for  (a) 
	$^{6}$Li and (b) $^{6}$He. The theoretical
	results are for JISP16 and $\hbar\Omega=20$ MeV in the complete $N_{\max}=6$ and $12$ spaces and in selected model  spaces. 
(c) Electric quadrupole transition probabilities and (d) quadrupole moments for $^{6}$Li, calculated using the JISP16 interaction without using effective charges, 
 as a function of $\hbar\Omega$  for the complete $N_{\max}=12$ space (dashed black lines) and $\langle6\rangle 12$
selected space (solid blue lines).  
Experimentally, $B(E2;1^{+}_{1}\rightarrow
3^{+}_{1})=25.6(20)\,e^2\mathrm{fm}^4$ \cite{Tilley2002}. Figure adapted from \cite{DytrychLMCDVL_PRL12}.
}
\label{fig:BE2Q2}
\end{center}
\end{figure}

The ground-state binding energies represent from 98\% up to 98.7\% of the complete-space binding
energy in the case of $^{6}$Li, and reach over 99\% for  $^{6}$He (Fig.~\ref{6LiT0Spectrum}).
Furthermore, the excitation energies differ only by 11 keV to a few hundred keV
from the corresponding complete-space results, and the agreement with known experimental data is reasonable over a broad range
of $\hbar\Omega$ values (Fig.~\ref{fig:BE2Q2}).

% \mu
% Expt.: \mu of p = 2.793 \mu_N, \mu of n = 1.913 \mu_N => p-n=0.88 \mu_N
As illustrated in Table~\ref{tab:muRMS}, the magnetic dipole moments for $^{6}$Li agree to
within 0.3\%.  
In addition, the ground-state magnetic dipole moment agrees with the  experimental value and turns out to be very close to the sum of the magnetic
moments of the neutron and the proton.
Qualitatively similar agreement is achieved for $\mu(2^{+}_{1})$ of $^{6}$He.
The results suggest that it may suffice to include all low-lying $\hbar\Omega$
states up to a fixed limit, e.g. $N^{\bot}_{\max}=6$ for $^{6}$Li and $N^{\bot}_{\max}=8$ for
$^{6}$He, to account for the most important correlations that contribute to the
magnetic dipole moment. 
\begin{table}[t]
\caption{
Selected observables for the two lowest-lying states of  $^{6}$He and $^{6}$Li obtained in
the complete $N_{\max}=12$ space and selected model subspaces forJISP16 and $\hbar\Omega=20$ MeV. The experimental values for
the $^{6}$Li  $1^+_{gs}$  are  measured to be  $Q (1^+) =
-0.0818(17)\,e\mathrm{fm}^2$ and $\mu=+0.822$ $\mu_{N}$ \cite{Tilley2002}.
\label{tab:muRMS}}
\begin{center}
\begin{tabular}{l|ccc|ccc}
\hline
& \multicolumn{3}{c|}{$^{6}$He} & \multicolumn{3}{c}{$^{6}$Li} \\
& & $N_{\max}=12$ & $\langle$8$\rangle$12 & & $N_{\max}=12$ & $\langle$6$\rangle$12 \\
\hline
$B(E2;2^{+}_{1}\rightarrow 0^{+}_{1})$, $e^2$fm$^4$	&		&	0.181	&	0.184	&		&		&		\\
$Q$, $e$fm$^2$	&	$2^{+}_{1}$	&	-0.69	&	-0.711	&	$1^{+}_{1}$	&	$-0.064$	&	$-0.08$	\\
$\mu$, $\mu_{N}$  	&	$2^{+}_{1}$	&	-0.873	&	-0.817	&	 $1^{+}_{1}$ 	&	0.838	&	0.839	\\
	&		&		&		&	 $3^{+}_{1}$ 	&	1.866	&	1.866	\\
$r_m$, fm	&	$2^{+}_{1}$    	&	2.153	&	2.141	&	 $1^{+}_{1}$ 	&	2.119	&	2.106	\\
	&	$0^{+}_{1}$     	&	2.113	&	2.11	&	 $3^{+}_{1}$ 	&	2.063	&	2.044	\\
	\hline
\end{tabular}
\end{center}
\end{table}
\begin{table}[b!]
\caption{Observables  for several model spaces, namely, excitation energies $E$,
electric quadrupole moments $Q$, together with point-particle matter rms radii  $r_m$ for selected low-lying states (including the ground state, $gs$) in $^{12}$C. The observables are calculated for \ho=20 MeV  using the bare JISP16 interaction and compared to the experiment \cite{ASelove90} (``Expt."). The SA-NCSM  results are obtained in a reduced $\Nmax{6}{8}$ model space  with a complete space up to 6\ho. Two selection patterns, ``A"  and ``B", are shown. The fraction of the many-body basis dimension used in the calculations as compared to the complete $N_{\rm max}=8$ $M$-scheme dimension is also specified. 
}
\begin{center}
\begin{tabular}{c|llll}
\hline
        & $\Nmax{6}{8}$-A & $\Nmax{6}{8}$-B &  $N_{\rm max}=8$ &  Expt.\\
      Model Space Dimensionality  &   3.8\%$^{\text a}$  &   9.2\%$^{\text b}$   &   100\% &         \\
      \hline
$E_{2^+_1}$ (MeV)	&	5.253	&	4.644	&	4.685	&	4.439	\\
$E_{1^+_1}$ (MeV)	&		&	14.199	&	14.161	&	12.71	\\
$E_{4^+_1}$ (MeV)	&	17.132	&	16.324	&	16.255	&	14.083	\\
$r_m (0^+_{gs})$ (fm)	&	2.007	&	2.005	&	2.003	&	2.43(2)$^{\text c}$	\\
$r_m (2^+_1)$	(fm) &	2.027	&	2.023	&	2.024	&	N/A	\\
$r_m (4^+_1)$	(fm) &	2.058	&	2.055	&	2.061	&	N/A	\\
$Q_{2^+_1}$	($e$ fm$^2$) &	3.712	&	3.735	&	3.741	&	 +6(3)	\\
$Q_{4^+_1}$	($e$ fm$^2$) &	4.826	&	4.845	&	4.864	&	N/A	\\
\hline
\end{tabular}
\\
$^{\text a}$ Model space for all $0^+,\,2^+$, and $4^+$ states in $^{12}$C.\\
$^{\text b}$ Model space for all $0^+,\,1^+,\,2^+$, and $4^+$ states in $^{12}$C.\\
$^{\text c}$ Ref. \cite{Tanihata85}\\
\end{center}
\label{12C_observables}
\end{table}

% B(E2), Q, rms 
To explore how close one comes to reproducing the important long-range
correlations of the  complete 
$N_{\max}=12$ space in terms of nuclear collective
excitations within the symmetry-guided 
spaces under consideration,
we compare observables that are sensitive to the tails
of the wave functions; specifically, the point-particle rms matter radii, the
electric quadrupole moments and the reduced electromagnetic $B(E2)$ transition
strengths.  
As Table~\ref{tab:muRMS} clearly shows, the complete-space
results for these observables are remarkably well reproduced by the SA-NCSM for $^{6}$He
in the restricted 12$\langle$8$\rangle$ space.  Similarly, the
12$\langle$6$\rangle$ eigensolutions for $^{6}$Li yield results for 
$B(E2)$ strengths and quadrupole moments that
track very closely with their complete
$N_{\max}=12$ space counterparts for all
values of $\hbar\Omega$ (Fig.~\ref{fig:BE2Q2}). 
The $B(E2)$ strengths almost double upon increasing the model space from
$N_{\max}=6$ to $N_{\max}=12$.
While  larger model spaces may be needed to achieve convergence, the close correlation between the $N_{\max}=12$
and 12$\langle$6$\rangle$ results is 
nevertheless impressive. 
In addition to being in agreement, the results  reproduce the challenging sign and magnitude of the
ground-state quadrupole moment (Table~\ref{tab:muRMS}) that is measured to be $Q (1^+) =
-0.0818(17)\,e\mathrm{fm}^2$ \cite{Tilley2002}. The sign can be easily understood in terms of the \SU{3} structure of the $^{6}$Li  ground-state rotational band, which is dominated foremost by a $0\ho(2\,0)$ configuration (of a prolate shape), followed by  high-\ho~ configurations  of prolate deformation, which, as mentioned earlier, implies a negative  $Q$ value.
Finally, the results for the rms matter radii of $^{6}$Li, listed in
Table~\ref{tab:muRMS}, 
agree to within 1\%.
The differences between selected-space and complete-space results are found
insensitive to the choice of $\hbar\Omega$ and appear sufficiently small as to
be inconsequential relative to the residual dependencies on $\hbar\Omega$ and
on $N_{\max}$.

For $^{12}$C \cite{LauneyDDTFLDMVB13}, we construct two SU(3)-based selection model spaces, based on an algorithm that first includes  largest-deformation/lowest-spin configurations and symplectic configurations thereof according to the rule (\ref{eq:Sp2RSelection}), followed by configurations of increasing spin and decreasing deformation values. These are referred to as  ``A" (a smaller set of basis states) of dimensions $2.8\times10^6$ (all $0^+$ states), $10.2\times10^6$ (all $2^+$ states), and $9.4\times10^6$ (all $4^+$ states), and ``B" (an extended set) of dimensions $4.0\times10^6$ (all $0^+$ states), $16.3\times10^6$ (all $2^+$ states), $20.3\times10^6$ (all $4^+$ states) and $14.4\times10^6$ (all $1^+$ states). Indeed, these sizes realize only $0.5\%$ to $3.5\%$ of the corresponding complete $N_{\rm max}=8$ model space. 
Table \ref{12C_observables} reveals that by employing the drastically reduced model space, $\Nmax{6}{8}$-B, the SA-NCSM isospin-zero $2^+_1$, $1^+_1$, and $4^+_1$ excitation energies  are found to deviate from the corresponding $N_{\rm max}=8$ results \cite{Maris12}  only by 0.9\%, 0.3\%, and 0.4\%, respectively.
These states  lie remarkably close to the complete-space counterparts even when the smaller $\Nmax{6}{8}$-A  SA-NCSM model space is utilized.  
In addition, radii and quadrupole moments shown in Table \ref{12C_observables} are well reproduced by the  SA-NCSM calculations in both selected spaces. This indicates that these observables are not sensitive to the fine-tuning of the selected space and only a manageable number of  symmetry-adapted configurations of the 8\ho~ subspace appears sufficient for their description.

\begin{figure}[h]
\begin{center}
 \includegraphics[width =0.55 \columnwidth]{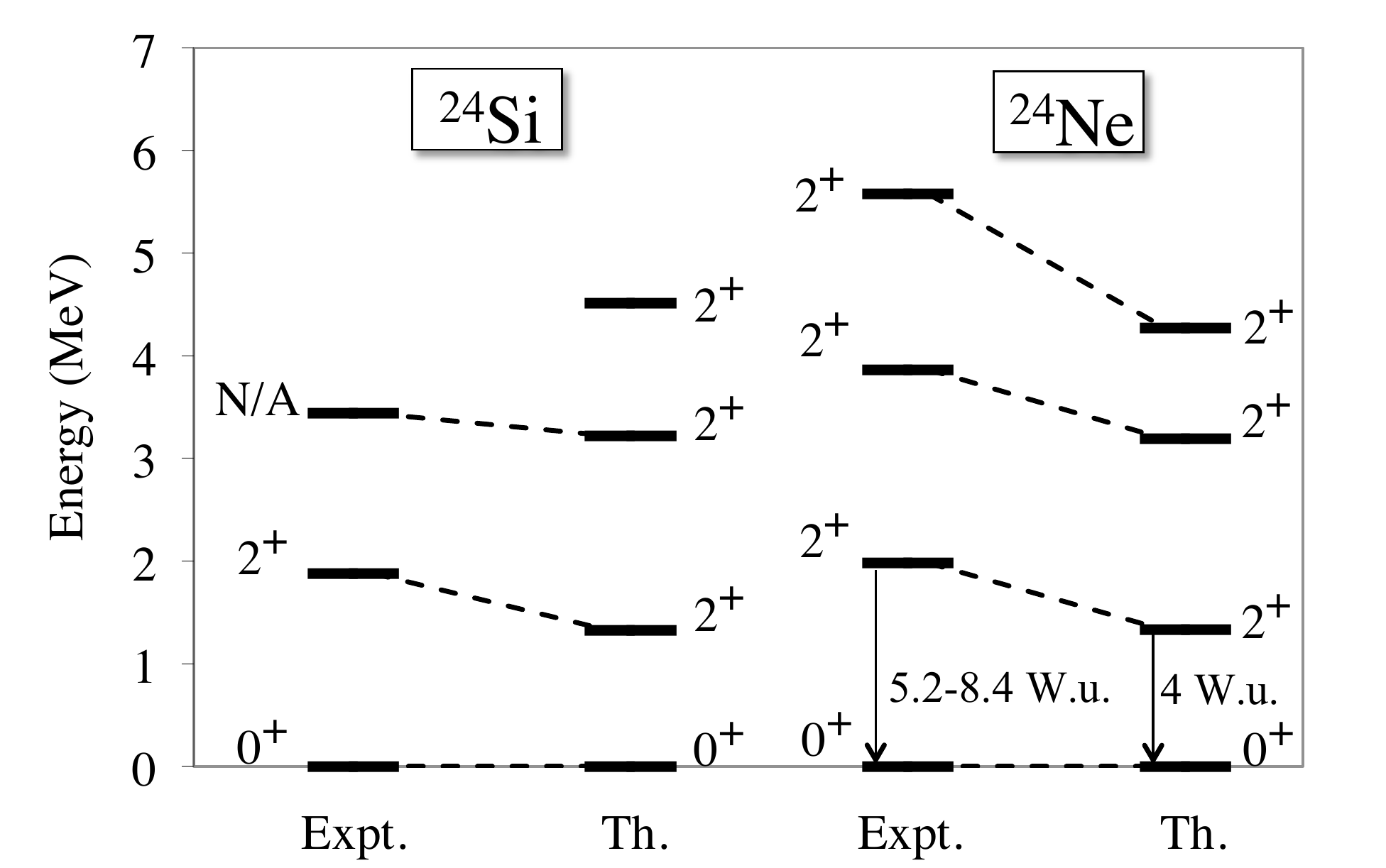}
\caption{
First SA-NCSM calculations with realistic interactions for open-shell intermediate-mass nuclei, $^{24}$Si and $^{24}$Ne, 
	obtained for a selected $N_{\max}=\langle 2 \rangle 6$ model space (that is, the complete space up through 2\ho~ and selected 4\ho~ and 6\ho~ subspaces with $3.5\times 10^6$ dimensionality  -- compare to the currently inaccessible $N_{\max}=6$ complete model space of $\sim 10^{11}$ dimensionality,  as shown in Fig. \ref{fig:dims}a, required for the corresponding $M$-scheme  NCSM  calculations). Calculations are performed for $\hbar\Omega=15$ MeV and using the SRG-N$^3$LO $NN$
	interaction for an SRG cutoff $\lambda_c=2$ fm$^{-1}$. Figure adapted from Ref. \cite{DraayerDLDL_Cocoyoc14}.
}
\label{sdNuclei}
\vspace{-0.3in}
\end{center}
\end{figure}

\vspace{22pt}
 {\it Ab initio} investigations of open-shell nuclei in the $sd$-shell region are now feasible and  the first calculations for $^{24}$Si and its mirror nucleus $^{24}$Ne (Fig. \ref{sdNuclei}) have been achieved in the framework of the SA-NCSM with SRG-evolved (to $\lambda_c=2$ fm$^{-1}$) chiral N$^3$LO interactions in an $N_{\max}=\langle 2 \rangle 6$  symmetry-selected model space   of $3.5\times 10^6$ dimensionality  (dimensionality of the corresponding $J=0$ and $J=2$ $N_{\rm max}=6$ complete model space  is $8\times 10^9$) \cite{McGrawHill14,DraayerDLDL_Cocoyoc14}. While structural properties of the short-lived $^{24}$Si are difficult to measure, $E2$ transitions are experimentally available for $^{24}$Ne and the SA-NCSM result for $B(E2;2^{+}_{1}\rightarrow 0^{+}_{1})$ agrees with the experimental value without using effective charges (note that conclusive results will require the exploration of bare $3N$ chiral interactions, along with the  SRG-induced many-body components of the interaction and observables).  {\it Ab initio} results up to $N_{\rm max}=12$ for $sd$-shell nuclei are detailed in Ref. \cite{Dytrych_sd_15}.

\subsection{Electron scattering for light nuclei}
In the SA-NCSM, the impact of the symmetry-guided space selection on  the charge density components for the  ground state of $^6$Li  in momentum space is studied, including the effect of  higher shells \cite{DytrychHLDMVLO14}, by  investigating the electron scattering charge form factor for momentum transfers up to $q \sim 4$ fm$^{-1}$.
The results demonstrate that  this symmetry-adapted framework can achieve significantly reduced dimensions for equivalent large shell-model spaces while retaining the accuracy of the form factor for   any momentum transfer.
This confirms the previous outcomes  for selected spectroscopy observables in light nuclei, as discussed in Sec. \ref{SANCSMspectra}. 

\begin{figure}[t]
\includegraphics[width=0.49\textwidth]{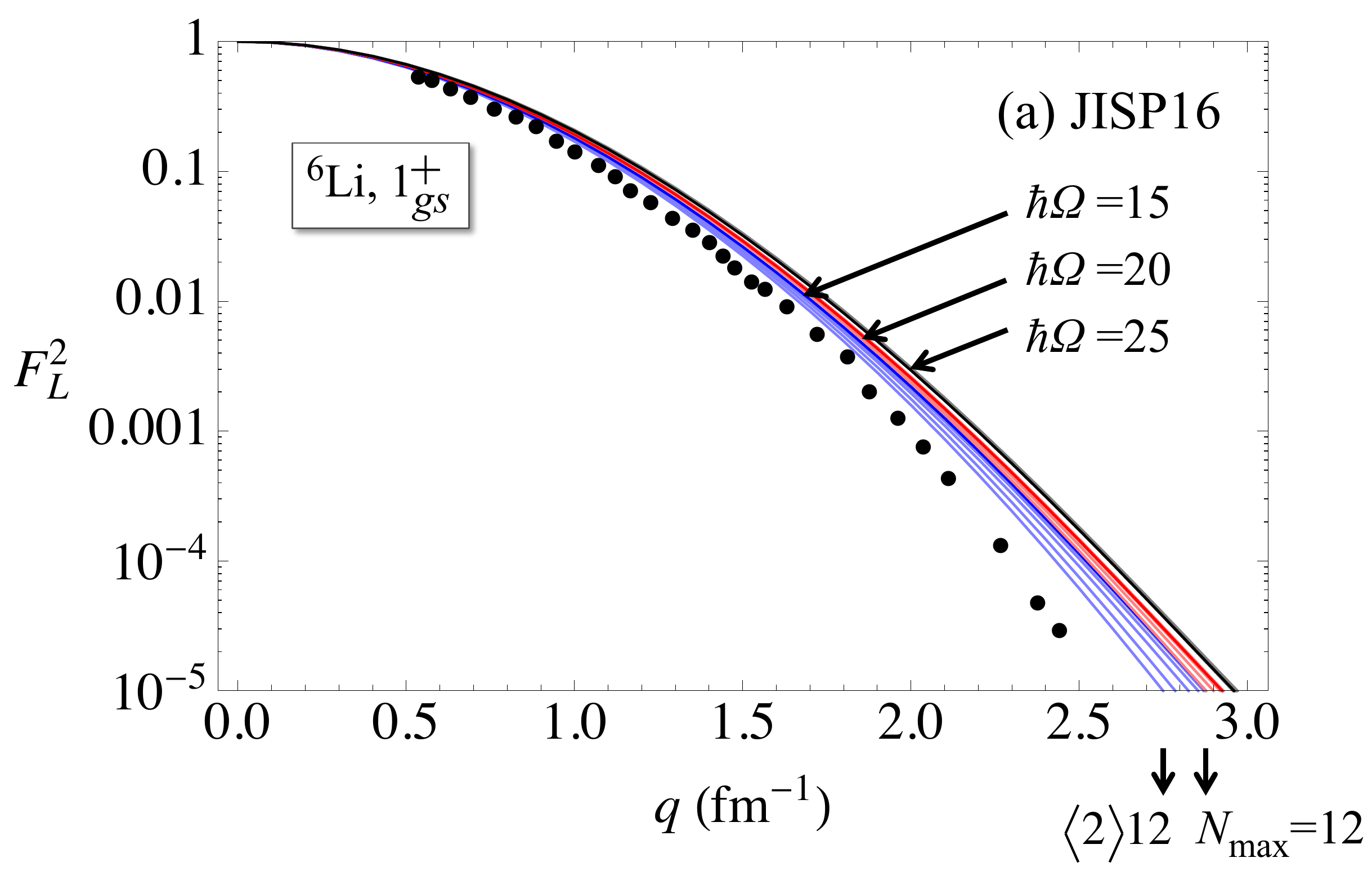}
\includegraphics[width=0.49\textwidth]{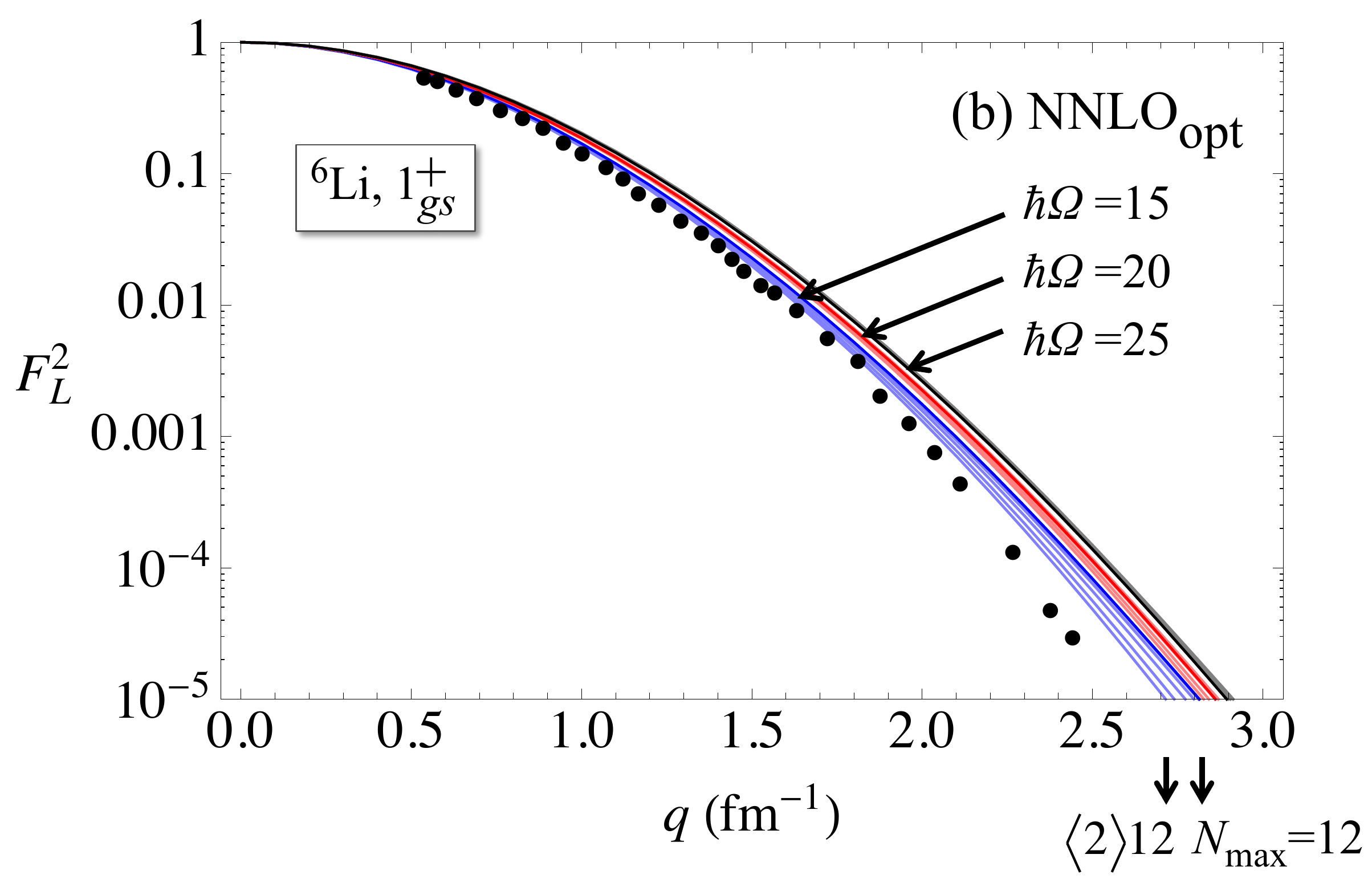} 
\caption
{
Longitudinal $C0$ electron scattering form factors $F_L^2$  for the SA-NCSM $1^{+}$ ground  state of $^{6}$Li calculated in the complete $N_{\max}=12$ space (darker colors) and the \SU{3}-selected spaces, \Nmax{2}{12}, \Nmax{4}{12}, \Nmax{6}{12}, \Nmax{8}{12}, and \Nmax{10}{12} (lighter colors), for selected $\hbar\Omega$ values and for (a) the bare JISP16 interaction, as well as for (b) the bare  NNLO$_{\rm opt}$ interaction. Deviations due to the \SU{3}-based space selection are indicated by the curve thickness.  All form factors are corrected for the center-of-mass motion and the finite-size effect of the nucleon. Experimental data are taken from Ref. \cite{LiSWY71}. Figure adapted from Ref. \cite{DytrychHLDMVLO14}.
}
\label{FLC0_Li6}
\end{figure}

In particular, Fig. \ref{FLC0_Li6} shows longitudinal electron scattering form factors for the ground  state of $^{6}$Li in the framework of the {\it ab initio} SA-NCSM  for several \SU{3}-selected spaces, \Nmax{2}{12}, \Nmax{4}{12}, \Nmax{6}{12}, \Nmax{8}{12}, \Nmax{10}{12}, together with  the complete $N_{\max}=12$ space.  An important result is that in all cases, $\Nmax{6}{12}$ selected-space results are found to be almost identical to the $N_{\rm max}=12$ complete-space counterparts  for any momenta, shown here up to momentum transfer $q \sim 4$ fm$^{-1}$, 
while being reasonably close to experiment. This remains valid for various \ho~ values, as well as when different bare interactions are employed. Deviations in the form factor (and in the one-body densities) as a result of the \SU{3}-based selection of model spaces are found to be only marginal and to decrease  for higher  \ho \cite{DytrychHLDMVLO14}. 

While results using NNLO$_{\rm opt}$ lie slightly closer to experiment, both
interactions show similar patterns with a small dependence on \ho~(Fig.
\ref{FLC0_Li6}).  Furthermore, as one increases $N_{\max}$ (e.g., from
$N_{\max}=8$ to $N_{\max}=12$), SA-NCSM predictions are  reasonably trending
towards experiment (see Fig. 3 of Ref. \cite{DytrychHLDMVLO14}).  We note that
the  $N_{\max}=12$ results continue to deviate from the experimental data for
intermediate momenta, especially for $q \gtrsim 2$ fm$^{-1}$,  where two-body currents become significant for $C0$ as shown by the Variational Monte Carlo (VMC)  \cite{WiringaS98} with the AV18 \cite{WiringaSS95} two-nucleon and
Urbana IX \cite{PudlinerPCW95} three-nucleon  interactions.
Nonetheless, the  low-\ho~ SA-NCSM $F_L^2$ calculations using NNLO$_{\rm opt}$  agree with
the ones of the VMC using AV18/UIX  (without contributions from two-body
currents) for  $q \lesssim 2$ fm$^{-1}$ (e.g., compare Fig. \ref{FLC0_Li6} and Fig. 1 of Ref. \cite{WiringaS98}).  The agreement might be a consequence
of the fact that the NNLO$_{\rm opt}$  is designed to minimize the contribution
due to three-nucleon interactions (similarly, for JISP16).  In order to gain
additional insight into the similarities and differences among the {\it ab
initio} results for $^6$Li, we present in Table \ref{tab:cmp_abinitioModels}
the energies, electromagnetic moments, and point-nucleon  rms radii for
selected states in $^{6}$Li, as calculated in the present SA-NCSM approach with
the JISP16 and NNLO$_{\rm opt}$, and in other {\it ab initio} models, such as the NCSM,
the VMC with AV18/UIX and the Green's function Monte Carlo (GFMC) with
AV18/UIX.  The results presented in Table \ref{tab:cmp_abinitioModels} show
good correlations among the different models with, perhaps, the exception of
the smaller rms radii obtained with JISP16 (SA-NCSM \& NCSM) and the larger magnitude of the
electric quadrupole moment obtained with the VMC. 
\begin{table}[th]
\caption{
$^{6}$Li binding energy (BE), excitation energies ($E$), electric quadrupole ($Q$) and
magnetic dipole ($\mu$) moments, as well as point-nucleon proton ($r_p$) and
matter  ($r_m$) rms radii,
as calculated in the  \Nmax{6}{12} SA-NCSM with the JISP16 $NN$ interaction and
for \ho=20 MeV (taken from Ref.  \cite{DytrychLMCDVL_PRL12}) and compared to
other {\it ab initio} approaches:  the complete $N_{\rm max}=12$ model
space  \cite{DytrychLMCDVL_PRL12} (or NCSM), as
well as VMC and GFMC using AV18/UIX  interaction
(energies taken from Ref. \cite{WiringaS98}; radii and electromagnetic
moments  taken from Ref.  \cite{PudlinerPCPW97}, without contributions from
two-body currents). Experimental results (Expt.) taken from Ref.
\cite{Tilley2002} unless otherwise specified.
}
\begin{center}
\begin{tabular}{l|lllll}
\hline\hline
		&	SA-NCSM 	&	NCSM	 & VMC	&	GFMC	&	Expt.	\\
	\hline
		&		&		&	$1^+_{gs}$	&		&		\\
	BE [MeV]	&	30.445	&	30.951	&	27.0(1)	&	31.2(1)	&	31.99 	\\
	rms $r_p$ [fm]	&	2.112	&	2.125	&	2.46(2) 	&		&	2.43$^a$ 	\\
	rms $r_m$ [fm]	&	2.106	&	2.119	&		&		&	2.35(3)$^b$	\\
	$Q$ [$e$ fm$^2$]	&	-0.08	&	-0.064	&	-0.33(18)	&		&	-0.0818(17)	\\
	$\mu$ [$\mu_N$]	&	0.839	&	0.838	&	0.828(1)	&		&	0.822 	\\
		&		&		&	$3^+$	&		&		\\
	$E$ [MeV]	&	2.515	&	2.526	&	3.0(1)	&	2.7(3)	&	2.186	\\
	rms $r_m$ [fm]	&	2.044	&	2.063	&		&		&		\\
	$Q$ [$e$ fm$^2$]	&	-3.88	&	-3.965	&		&		&		\\
	$\mu$ [$\mu_N$]	&	1.866	&	1.866	&		&		&		\\
		&		&		&	$2^+$	&		&		\\
	$E$ [MeV]	&	5.303	&	5.066	&	4.4(1)	&	4.4(4)	&	4.312	\\
	rms $r_m$ [fm]	&	2.18	&	2.204	&		&		&		\\
	$Q$ [$e$ fm$^2$]	&	-2.279	&	-2.318	&		&		&		\\	
	$\mu$ [$\mu_N$]	&	1.014	&	0.97	&		& & \\
	\hline\hline
\end{tabular}\\
\footnotesize{$^a$ Deduced from the $^{6}$Li charge radius of 2.56(5) fm \cite{LiSWY71}}\\
\footnotesize{$^b$ From Ref. \cite{Tanihata88}} 
\end{center}
\label{tab:cmp_abinitioModels}
\end{table}%

\begin{figure}[th]
\centerline{
\includegraphics[width=1\textwidth]{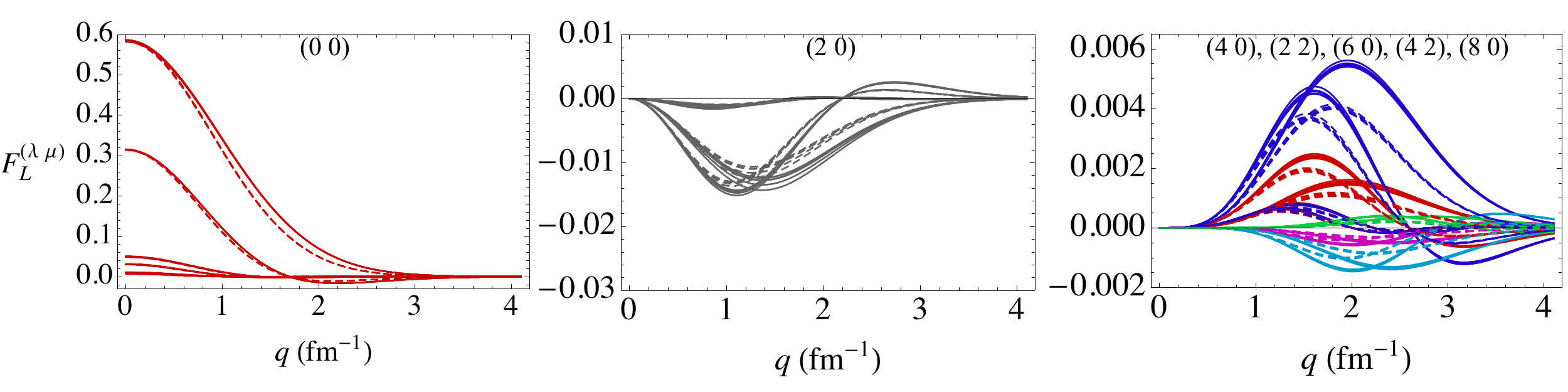} 
}
\caption
{
Most dominant \SU{3} contributions  to $F_L$ for the longitudinal  $C0$ form factor, with (solid) and without (dashed) removing the CM  contribution. \SU{3} contributions are labeled by $(\lambda\,\mu)$, shown at the top of each plot in order of decreasing maximum amplitude. The $N_{\max}=12$ SA-NCSM $1^{+}$ ground  state of
	$^{6}$Li is calculated with the NNLO$_{\rm opt}$ bare interaction and for $\hbar\Omega=20$ MeV. Deviations due to the \SU{3}-based space selection are indicated by the curve thickness.  Note that the vertical axis scale is reduced by an order of magnitude from the left to the right panels.  Figure adapted from Ref. \cite{DytrychHLDMVLO14}.
}
\label{fig:contributions_to_FLcmpCM}
\end{figure}

The largest contribution to the C0 form factor comes from the $(\lambda\,\mu)=(0\,0)$ one-body density and, for all $q$ values, from the $(0\,0)$ contribution to the $F_L$ (Fig. \ref{fig:contributions_to_FLcmpCM}), as a result of the largest density within the $s$, $p$, $sd$, and $pf$ shells. In addition, for all \ho, only the (0\,0)+ (2\,0)/(0\,2) components are found sufficient to reproduce the low-momentum regime of the form factor. The $(4\,0)$,  $(2\,2)$, and $(8\,0)$ components are the ones that are most responsible for larger form-factor values at intermediate momenta. The preponderance of $0\ho(0\,0)$, $2\ho(2\,0)$, \dots, and $8\ho(8\,0)$ together with  $0\ho(2\,2)$ and $2\ho(4\,2)$  (and their conjugates) in the one-body densities and in the  form factor can be recognized as another signature of the \SpR{3} symmetry. Above all, the symmetry-adapted model spaces include the important excitations to higher HO shells as seen in their significant contributions at low- and intermediate-momentum transfers. The outcome further confirms the utility of the SA-NCSM concept for low-lying nuclear states.

\section{Summary and outlook}

We have reviewed  exact and near symmetries of atomic nuclei that have been long recognized and only recently, exposed from first principles and exploited in {\it ab initio} theories. Such symmetries include, for the spatial degrees of freedom, the deformation-related \SU{3} together with \SpR{3} and its complementary \On{A-1}, and, for the spin-isospin degrees of freedom, Wigner's  \SU{4}. We have also presented various symmetry-guided techniques, with a focus on open-core shell-model theory and the use of symmetries in large-scale nuclear simulations that permit symmetry mixing. In such calculations, symmetries are not used to restrict the nuclear dynamics (e.g., limiting to symmetry-preserving interactions or a single irrep, which is often very restrictive), but rather to expose physically relevant degrees of freedom to organize large-scale model spaces. 
We have demonstrated the unique role of symmetries in large-scale applications  of  the {\it ab initio} SA-NCSM, NCSM, HH method, lattice EFT, and GFMC to nuclear dynamics. We have discussed an important new development of the theory, the \SU{3}-based SA-NCSM, that, for the first time, has demonstrated that observed collective phenomena in light nuclei emerge naturally from first-principle {\it ab initio} considerations. The results, supporting experimental evidence, underscore the strong dominance of configurations with large quadrupole  deformation and low intrinsic spins, and symplectic excitations thereof. In addition to this, an overall pattern coherently propagates through states that form a rotational band, as first recognized in Ref. \cite{BahriR00}.  

Within this context, a symmetry-guided concept has been discussed, within the SA-NCSM framework, that accommodates spatially expanded
correlations that are essential to collective features, by
 including complete low-$N\ho$ subspaces (all ``horizontal" mixing) and symmetry-guided selected subspaces up to a very large $N_{\rm max}$ (``vertical cones" built on dominant \SU{3} configurations). We have demonstrated that  this symmetry-guided framework can achieve significantly reduced dimensions for equivalent large shell-model spaces while retaining the accuracy of the corresponding complete-space results; they also achieve a close agreement with experiment, when using the  bare JISP16 $NN$ interaction that minimizes the contribution of the $3N$ interactions \cite{ShirokovMZVW07}. While the $3N$ interactions are secondary in importance to the physics and underpinning symmetries discussed here,  and often secondary to the need for larger model spaces, they play an important role \cite{navratilGVON07} toward reproducing experimental binding energies and the physics of certain nuclear states, such as the ones influenced by the spin-orbit force. Including $3N$ interactions  in the {\it ab initio} SA-NCSM is underway (e.g., the N$^2$LO and forthcoming N$^3$LO $3N$  chiral interactions  \cite{EpelbaumNGKMW02}, or $3N$ interactions  induced during a renormalization procedure). In particular, current efforts focus to derive
and implement a unitary transformation of $M$-scheme or $JT$-scheme $3N$ interaction matrix elements into the \SU{3} scheme. This represents an elaborate task, which involves the computing of a huge number of  \SU{3}  coupling/recoupling coefficients. Fortunately, it can be straightforwardly implemented based on our successful strategy for $NN$ interactions, which has been recently shown to run very efficiently on modern GPU architectures by a symmetry-aware reordering of input matrix elements \cite{DytrychO16}. This greatly reduces the thread divergence in branching conditionals and improves utilization of fast memories. Another important consideration relates to an efficient memory storage of the $3N$ matrix elements, which will greatly benefit from the group-theoretical foundation of the SA-NCSM. For example, Ref. \cite{RothCLB14} has already shown that the use of $JT$-scheme $3N$ interactions reduces memory footprints by up to three orders of
magnitude as compared to the use of their $M$-scheme representation. 
 
The results further anticipate the likely significance of $LS$-coupling and \SU{3} as well as the overarching symplectic
symmetry for an extension of {\it ab initio} methods to the heavier nuclei.  And while medium-mass nuclei are expected to exhibit a stronger \SU{3} mixing in low-$N\ho$ subspaces as a result of the richer subspaces as compared to  light nuclei (e.g., the dimension of the 0\ho~subspace is 4 for the $^6$Li  ground state and  $4.1\times10^4$ for the $^{48}$Cr ground state), the relevant $(\lambda\,\mu)S$ configurations in these subspaces typically remain a fraction of the complete subspaces. It is important to emphasize that such an \SU{3} mixing is readily accounted for in the SA-NCSM, by the complete horizontal span of the SA-NCSM model space, while the vertical selection is guided by symplectic vertical cones built on the most dominant \SU{3} configurations.
And while it is clear that  selected model spaces, without compromising the accuracy of results, can reduce the memory demand in computations as compared to the complete space, utilizing symmetries leads to time-consuming evaluations of the many-body Hamiltonian. Nonetheless,  it transforms a memory-bound unfeasible problem in a complete $N_{\rm max}$ model space into a CPU-bound problem  in a selected $N^{\top}_{\rm max}$ model space, with  $N^{\top}_{\rm max}$ much larger than $N_{\rm max}$, which is attainable on today's petascale architecture. This allows a theory of the symmetry-guided large-scale shell-model type to extend the reach of the standard schemes by exploiting symmetry-guided principles that enable one to include large spatial configurations, and in so doing capture the essence of long-range correlations that often dominate the nuclear landscape.
The findings  reviewed here start to unveil new physics, namely, understanding the mechanism on how simple patterns in nuclei and a diversity of nuclear properties emerge from a fundamental  level.

\section*{Acknowledgements}
We thank D. Rowe, G. Rosensteel, and C. Bahri, together with  the PetaApps Collaboration, J. P. Vary, P. Maris, U. Catalyurek, and M. Sosonkina, for useful discussions. We also acknowledge useful discussions with J. L. Wood, D. Lee, C. W. Johnson, R. B. Wiringa, Y. Suzuki, and F. Pan on various topics covered in this review. We also thank A. C. Dreyfuss, G. K. Tobin, M. C. Ferriss, and  R. B. Baker for providing important results.
This work was supported by the U.S. NSF (OCI-0904874,  ACI -1516338), the U.S. DOE (DE-SC0005248), and Southeastern Universities Research Association (SURA), and the Czech Science Foundation under Grant No. 16-16772S.
This work also benefitted from computing resources provided by Blue Waters, as well as the Louisiana Optical
Network Initiative and high performance computing resources provided by Louisiana State University ({\tt http://www.hpc.lsu.edu}).  A portion
of the computational resources was provided by the National Energy Research
Scientific Computing Center.  T. D. acknowledges support from Michal Pajr and CQK Holding.

\bibliographystyle{siamnew}
\bibliography{lsu_latest,reviewSA_2015}

%\begin{thebibliography}{99}
%\itemsep -2pt 
%\bibitem{faes0} A. Valcarce, A. Buchmann, F. Fern\'andez, and Amand Faessler, \Journal{\PRC} {51}{1480} {1995} 
%\end{thebibliography}
\end{document}